\documentclass[aps,prx,reprint,twocolumn,floatfix,superscriptaddress,amsmath,amssymb,amsfonts,nofootinbib,balancelastpage]{revtex4-2}

\usepackage{url}
\usepackage{bm}
\usepackage{graphicx}
\usepackage{amsmath}
\usepackage{amstext}
\usepackage{amssymb}
\usepackage{amsfonts}
\usepackage{amsbsy}
\usepackage{verbatim}
\usepackage{color}
\usepackage[colorlinks=true, urlcolor=blue, linkcolor=blue, citecolor=blue, pdftex]{hyperref}
\usepackage{floatrow}
\usepackage{float}

\usepackage{gensymb}
\usepackage{textcomp}
\usepackage{enumitem}
\usepackage[version=3]{mhchem}
\usepackage{afterpage}
\usepackage{pbox}
\usepackage{dsfont}
\usepackage{siunitx}
\usepackage{stackengine,wasysym,scalerel}
\usepackage{latexsym}
\usepackage{cancel}
\usepackage{comment}
\usepackage{array, multirow, bigdelim, makecell, booktabs}
\usepackage[misc]{ifsym}
\usepackage[capitalise]{cleveref}
\usepackage{sidecap}
\usepackage[caption=false]{subfig}
\usepackage[dvipsnames]{xcolor}

\usepackage{tikz}

\usepackage{color}
\usepackage{pifont}

\usepackage[normalem]{ulem}

\usepackage{hyperref}
\hypersetup{
	colorlinks=true,
	urlcolor=blue,
	citecolor=blue,
	linkcolor=blue,
	breaklinks
}

\usetikzlibrary{shapes.geometric}

\DeclareRobustCommand{\smalltriangleup}{%
  \mathord{%
  \tikz[baseline=-0.6ex]{
  \node[
    regular polygon,
    regular polygon sides=3,
    draw,
    fill=gray,
    inner sep=0.6pt,
    minimum size=3pt
  ] {};
  }%
  }%
}

\DeclareRobustCommand{\smalltriangledown}{%
  \mathord{%
  \tikz[baseline=-0.6ex]{
  \node[
    regular polygon,
    regular polygon sides=3,
    draw,
    rotate = 180,
    fill=gray,
    inner sep=0.6pt,
    minimum size=3pt
  ] {};
  }%
  }%
}

\DeclareRobustCommand{\smallhexagon}{%
  \mathord{%
  \tikz[baseline=-0.6ex]{
  \node[
    regular polygon,
    regular polygon sides=6,
    draw,
    fill=gray,
    inner sep=0.6pt,
    minimum size=4pt
  ] {};
  }%
  }%
}

\DeclareRobustCommand{\smalllozenge}{%
  \mathord{%
  \tikz[baseline=-0.5ex]{
  \node[
    diamond,
    draw,
    fill=gray,
    rotate=90,
    minimum width=6.4pt,
    minimum height=3.7pt,
    inner sep=0pt
  ] {};
  }%
  }%
}

\allowdisplaybreaks

\begin{document}
\title{Classification and correlation signatures of chiral spin liquids on the pyrochlore lattice}
\author{Chunxiao Liu}
\email{chunxiao.liu@universite-paris-saclay.fr}
\affiliation{Université Paris-Saclay, CNRS, Laboratoire de Physique des Solides, 91405 Orsay, France}

\author{Leon Balents}
\email{balents@spinsandelectrons.com}
\affiliation{Kavli Institute for Theoretical Physics, University of California, Santa Barbara, CA 93106-4030, USA}
\affiliation{Canadian Institute for Advanced Research, Toronto, Ontario, Canada}
\affiliation{French American Center for Theoretical Science, CNRS, KITP, Santa Barbara, CA 93106-4030, USA
}
\author{Yasir Iqbal}
\email[Corresponding author: ]{yiqbal@physics.iitm.ac.in}
\affiliation{Department of Physics, Indian Institute of Technology Madras, Chennai 600036, India}

\date{\today}

\begin{abstract}
We present a systematic classification and variational study of chiral quantum spin liquids on the pyrochlore lattice based on fermionic parton constructions. Focusing on chiral $\mathrm{U(1)}$ and $\mathbb{Z}_2$ spin-liquid {\it Ans\"atze}, we characterize their symmetry properties, flux structures, and low-energy spinon spectra within a projective symmetry group framework, and incorporate gauge fluctuations through Gutzwiller-projected wave functions studied by variational Monte Carlo. From the equal-time spin structure factor, we develop correlation-based diagnostics that distinguish gauge-dominated Coulomb phases from states with substantial matter-field and short-range contributions. Distinct chiral flux sectors, though close in energy, exhibit markedly different degrees of emergent $\mathrm{U(1)}$ gauge-field dominance, reflected in the geometry and contrast of pinch-point singularities. Although these states are not competitive ground states of the nearest-neighbor Heisenberg model, they define a physically meaningful family of proximate chiral phases relevant to extended pyrochlore Hamiltonians.
\end{abstract}

\maketitle

\tableofcontents

\section{Introduction}

The pyrochlore lattice has long served as a paradigmatic platform for exploring key concepts in frustrated magnetism. Coulomb spin liquids~\cite{Hermele-2004}, the order-by-disorder mechanism~\cite{Henley-1989,Reimers-1991a}, the three-dimensional quantum spin liquid paradigm~\cite{Canals-1998,Canals-2000}, and, more recently, higher-rank gauge theories~\cite{PhysRevLett.124.127203}—to name only a few—have all found natural realization within this iconic geometry. Beyond the nearest-neighbor Heisenberg limit, a wide variety of extensions, including further-neighbor exchange interactions, anisotropic spin–orbit–coupled terms, and ring-exchange processes, are now known to substantially enrich the phase structure of pyrochlore magnets; see also Ref.~\cite{Gomez-2025} for a recent broader overview of classical pyrochlore spin-liquid regimes. In parallel, continued progress in materials synthesis has brought renewed attention to pyrochlore compounds in which competing interactions and strong spin–orbit coupling coexist. Despite this broad activity, the potential of the pyrochlore lattice as a platform for hosting three-dimensional \emph{chiral} quantum spin liquids remains essentially unexplored. This naturally raises the question of whether the pyrochlore lattice can once again serve as the stage on which a new paradigm in frustrated magnetism emerges.

A particularly important theme in this broader context is \emph{chirality}. While two-dimensional chiral spin liquids~\cite{Kalmeyer-1987,Wen-1989} have been extensively studied and are now well understood as Chern--Simons topological phases~\cite{Wen-1991}, their three-dimensional counterparts remain largely mysterious. Their field-theoretic descriptions remain comparatively underdeveloped~\cite{Laughlin-1990,Tikofsky-1994,Si-2008,Wu-2023}, the very nature of fully gapped versus algebraic ``3D chiral spin liquids'' remains under active debate, and concrete microscopic realizations remain scarce. From a classical perspective, however, chirality arises quite naturally on the pyrochlore lattice: competing interactions readily stabilize noncoplanar magnetic orders, which may be viewed as classical analogues of chiral quantum phases, suggesting a natural route toward three-dimensional chiral spin liquids through quantum disordering. They are, in particular, characterized by finite scalar spin chirality~\cite{Lapa-2012}, which denotes the pseudoscalar operator associated with an oriented triangular plaquette $(i,j,k)$:
\begin{equation}
    \chi_{ijk}
    =
    \mathbf{S}_i\cdot
    \left(
        \mathbf{S}_j\times\mathbf{S}_k
    \right).
    \label{eq:scalar_chirality_intro}
\end{equation}
It changes sign under reversal of the plaquette orientation, and is odd under time-reversal as well as under orientation-reversing spatial operations such as inversion or reflection, while remaining invariant under their product. On the pyrochlore lattice, the elementary triangular plaquettes relevant below are the triangular faces of the up and down tetrahedra, with the sign of $\chi_{ijk}$ fixed by the chosen orientation of each face. Thus, a chiral spin liquid can break time reversal and/or improper spatial symmetries separately while preserving appropriate combined space-time operations.

The connectivity of the pyrochlore lattice, which can be viewed as alternating layers of kagome and triangular planes, is particularly well suited for stabilizing noncoplanar magnetic orders. These include stacked analogs of the cuboctahedral spin configurations~\cite{Lapa-2012} familiar from the kagome lattice~\cite{Domenge-2005,Domenge-2008,Messio-2011}, novel alternating conic spiral states~\cite{Lapa-2012}, and incommensurate noncoplanar spirals~\cite{Iqbal-2017}. As demonstrated in Ref.~\cite{Lapa-2012}, such states arise already in classical ($S\!\to\!\infty$) Heisenberg models with competing long-range interactions, possibly of mixed ferromagnetic and antiferromagnetic type. This raises the intriguing possibility that for small spin $S$, quantum fluctuations destroy conventional dipolar magnetic order while allowing long-range correlations of the scalar spin chirality to survive~\cite{Iqbal-2017}. The resulting quantum-disordered states break time-reversal symmetry and, when combined with fractionalized excitations, provide a natural setting for three-dimensional chiral spin liquids.

This route of ``melting'' noncoplanar magnetic orders through strong quantum fluctuations has been extensively explored in two-dimensional frustrated magnets, most notably on the kagome lattice, where it has yielded a rich landscape of chiral spin liquids~\cite{Gong-2015,Hu-2015,Wietek-2017,Hickey-2017,Cookmeyer-2021}. These include both gapless and gapped chiral phases with emergent $\mathrm{U(1)}$ or $\mathbb{Z}_2$ gauge fields, with the former often associated with the melting of commensurate spin structures and the latter with incommensurate orders. These developments naturally motivate the search for an analogous classification of chiral quantum phases in three dimensions. The pyrochlore lattice is a particularly natural setting for this extension, combining geometric motifs familiar from kagome physics with genuinely three-dimensional gauge dynamics. In this setting, chirality may coexist with gapless $\mathrm{U(1)}$ gauge fields and Coulombic correlations, and may also allow qualitatively new phenomena--such as three-dimensional topological textures including hopfions~\cite{Liu-2020,Naya-2022}---that have no direct two-dimensional counterpart.

Motivated by these considerations, in this work we construct and study a systematic family of chiral U(1) spin-liquid {\it Ans\"atze} on the pyrochlore lattice using fermionic parton wave functions and variational Monte Carlo, extending earlier fermionic-parton studies of pyrochlore $\mathrm{U(1)}$ and $\mathbb{Z}_2$ spin liquids~\cite{PhysRevB.104.054401,Sahu-2024}. Although our classification encompasses both chiral $\mathrm{U(1)}$ and chiral $\mathbb{Z}_2$ spin liquids, we emphasize U(1) states in the present work because their gapless gauge fields lead to sharp and experimentally relevant signatures in equal-time correlations; for a complementary Schwinger-boson classification of chiral $\mathbb{Z}_2$ spin liquids on the pyrochlore lattice, see Ref.~\cite{Schneider-2022}.

Our focus is not on identifying the lowest-energy state of the nearest-neighbor spin-$\tfrac12$ Heisenberg antiferromagnet, whose ground-state character remains actively debated in the recent literature~\cite{Iqbal-2019,Hagymasi-2021,Astrakhantsev-2021,Hering-2022,Schafer-2023,Pohle-2023,Cheng-2025}, but rather on elucidating how different chiral flux patterns control the emergent gauge structure and correlation signatures of U(1) spin liquids. By analyzing equal-time spin structure factors—particularly the geometry and contrast of pinch-point singularities—we develop correlation-based diagnostics that distinguish gauge-dominated states from those in which matter-field and short-range correlations contribute a substantial background. In this sense, our work should be viewed as a classification of chiral $\mathrm{U(1)}$ spin liquids on the pyrochlore lattice based on emergent gauge structure and correlation fingerprints, rather than solely on variational energetics. Although the variational energies of these {\it Ans\"atze} are not competitive with the best available ground-state estimates for the nearest-neighbor model, they define a narrow and physically meaningful family of proximate phases. We argue that such chiral $\mathrm{U(1)}$ states are natural candidates for more general pyrochlore Hamiltonians with further-neighbor exchange, ring-exchange, or explicit chiral interactions, and that the present study provides a framework for exploring these regimes.

The paper is structured as follows. In Sec.~\ref{symmetries} we discuss the lattice and chiral symmetries of the pyrochlore lattice and introduce the framework of projective symmetry groups. In Sec.~\ref{classification} we present the results of the PSG classification. In Secs.~\ref{mf}--\ref{sec:PT_Z2} we construct quadratic spinon Hamiltonians for the different PSGs and discuss the corresponding mean-field ground states, flux structures, and the applicability of the PT theorem in these three-dimensional CSLs. In Sec.~\ref{sec:NNU(1)} we focus on the study of the 12 nearest-neighbor singlet $\mathrm{U(1)}$ {\it Ans\"{a}tze}, including the eight chiral states. After showing their spectra in Sec.~\ref{sec:spec_12}, we conduct a case study in Sec.~\ref{sec:mfs_and_sfs} of two {\it Ans\"{a}tze} with particularly interesting monopole flux patterns. In Sec.~\ref{sec:vmc} we assess the impact of gauge fluctuations via Gutzwiller-projected wave functions analyzed within a variational Monte Carlo framework, and present the equal-time spin structure factors of the chiral {\it Ans\"atze}. We also obtain their variational energies for the spin-$\tfrac12$ nearest-neighbor Heisenberg antiferromagnet. Section~\ref{outlook} concludes with a summary and outlook.

\section{Symmetries, partons, and fluxes}
\label{symmetries}

\subsection{Lattice, time-reversal, and chiral symmetries}
\label{lattice}

\begin{figure}
\includegraphics[width = 0.88\textwidth]{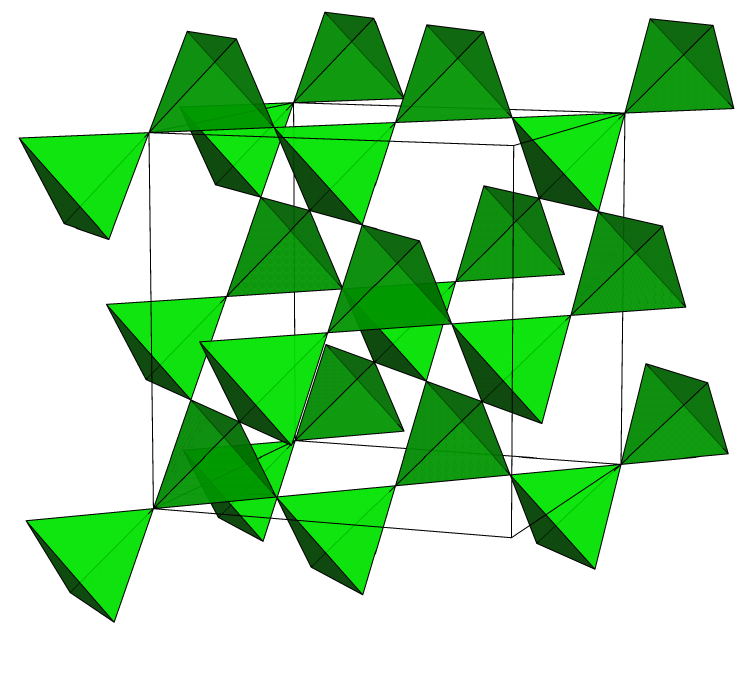}
\caption{Pyrochlore lattice. The sites occupy the vertices of corner-sharing tetrahedra. The conventional cubic unit cell, with edge length $a$, is indicated.}\label{fig:lattice}
\end{figure}

In this section, we introduce the lattice convention and provide a brief description of the symmetries of the pyrochlore lattice. A more detailed
analysis of the pyrochlore symmetry group is presented in Ref.~\cite{PhysRevB.100.075125}.

The pyrochlore lattice is shown in Fig.~\ref{fig:lattice}. It consists of four FCC-type sublattices which
we label by $\mu =0,1,2,3$. We define the following
sublattice-dependent coordinate:
\begin{align}
(r_1,r_2,r_3)_\mu &\equiv \mathbf{r}_\mu \equiv r_1
\mathbf{e}_1+r_2\mathbf{e}_2+r_3\mathbf{e}_3 + \frac{1}{2} \mathbf{e}_\mu\,,
\end{align}
where $\mathbf{e}_0=0$, and
\begin{subequations}\label{fcc_lv}
\begin{align}
\mathbf{e}_1 & =  \frac{a}{2}(\hat{\mathbf{y}}+\hat{\mathbf{z}})\,,\\
\mathbf{e}_2 & =  \frac{a}{2}(\hat{\mathbf{z}}+\hat{\mathbf{x}})\,,\\
\mathbf{e}_3 & =  \frac{a}{2}(\hat{\mathbf{x}}+\hat{\mathbf{y}})
\end{align}
\end{subequations}
are lattice vectors defined in terms of the cubic lattice constant $a$ [see Fig.~\ref{fig:lattice}].

The pyrochlore lattice symmetry is described by the No. 227 space group 
$Fd\overline{3}m$. A convenient choice for the generators of this group is
\begin{equation}\label{gens}
T_1,~~T_2,~~T_3,~~C_3,~~I,~~S,
\end{equation}
where $T_1$, $T_2$, and $T_3$ are translations along the lattice
vectors $\mathbf{e}_1$, $\mathbf{e}_2$, and $\mathbf{e}_3$, respectively, $C_3$ is the threefold rotation around the $[111]$ axis that contains the origin, $I$ the inversion about the origin, and $S$ is a nonsymmorphic screw operation which is the composition
of a twofold rotation around $\mathbf{e}_3$ and a translation by
$\mathbf{e}_3/2$. It is useful to introduce the sixfold rotoreflection $\overline{C}_6 := C_3I$.

A complete set of relations for $Fd\overline{3}m$ can be chosen to be
\begin{subequations}\label{SGtr}
\begin{align}
T_iT_{i+1} T_i^{-1} T_{i+1}^{-1} &= 1\,,\quad i=1,2,3\,, \label{4a}\\
\overline{C}^6_6 &= 1\,,  \\
S^2 T^{-1}_3 &= 1\,, \\
\overline{C}_6 T_i \overline{C}^{-1}_6 T_{i+1} &=1\,,\quad i=1,2,3\,, \label{eq:4d}\\
ST_iS^{-1} T^{-1}_3T_i&= 1\,, \quad i=1,2\,,\\
ST_3 S^{-1} T_3^{-1} &=1\,,\\
(\overline{C}_6S)^4&=1\,, \\
(\overline{C}_6^3S)^2 &=1\,, \label{4h}
\end{align}
\end{subequations}
where it is implicitly understood that $i+3 \equiv i$.

In addition to spatial operations, time-reversal operation $\mathcal{T}$ is also crucial to the description of chiral spin liquids. It
commutes with all space-group operations but acts nontrivially on a spin-$S$ state:
\begin{subequations}\label{trs}
\begin{align}
\mathcal{T}\mathcal{O}\mathcal{T}^{-1}\mathcal{O}^{-1}&=1\,,\quad
\mathcal{O}\in\{T_1,T_2,T_3,\overline{C}_6, S\}\, \label{4j},\\
\mathcal{T}^2 &= (-1)^S\,, \label{4i}
\end{align}
\end{subequations}
Eq.~\eqref{4i} means that half-integer spins carry a projective representation of time-reversal operation. Modulo this subtlety, Eqs.~\eqref{SGtr} and ~\eqref{trs} define a full group of operations which is precisely the No.~1629 magnetic space group (MSG)~\cite{litvin2013magnetic}
\begin{equation}
Fd\overline{3}m1' =Fd\overline{3}m\times\mathbb{Z}_2^{\mathcal{T}}\,,
\end{equation}
where the MSG symbol $Fd\overline{3}m1' $ is used.

Chiral spin liquids by definition break time-reversal symmetry $\mathbb{Z}_2^{\mathcal{T}}$, but may leave certain combinations of time-reversal $\mathcal{T}$ and spatial operation $\mathcal{O}$ invariant. In this work, we explore chiral spin liquids that are \emph{maximally} symmetric, where all spatial operations leave the chiral spin liquid state invariant \emph{up to} time-reversal symmetry. Under this assumption, for any spatial operation $\mathcal{O}$, either itself or the product $\mathcal{O}\mathcal{T}$ (but not both) belongs to the symmetry group of chiral spin liquid, and we term the symmetry operation $\mathcal{O}\mathcal{T}$ \emph{chiral} following Ref.~\cite{Bieri-2015}. Not all spatial operation $\mathcal{O}$ can be transformed into its chiral counterpart: the relation \eqref{4i} of $\mathcal{T}$ restricts chiral operators $\mathcal{O}\mathcal{T}$ to just order-2 elements $\mathcal{O}$ of the generating set \eqref{gens}, namely $\mathcal{O} = I$ or $S$, and this generates all the possible realizations of chiral symmetries of a chiral spin liquid, irrespective of the generating set chosen. We can then label these four types of chiral spin liquids with their associated MSG counterparts~\cite{litvin2013magnetic}:
\begin{itemize}
\item Type $(I,S)$, corresponding to MSG No.~1628. This is the case when full lattice symmetry is preserved but time-reversal symmetry is either not imposed~\cite{PhysRevB.104.054401} or explicitly broken;
\item Type $(I\mathcal{T},S)$, corresponding to MSG No.~1630 (``$Fd'\overline{3}'m$");
\item Type $(I,S\mathcal{T})$, corresponding to MSG No.~1631 (``$Fd\overline{3}m'$");
\item Type $(I\mathcal{T},S\mathcal{T})$, corresponding to MSG No.~1632 (``$Fd'\overline{3}'m'$").
\end{itemize}
These four types of chiral spin liquids are further subject to the constraints of projective symmetry group, which we present in the next section.

\subsection{Projective symmetry group}
\label{psg}

In this section we introduce the main ideas underlying the framework of the projective symmetry group (PSG), with an emphasis on the classification of chiral spin liquids.

We rewrite the spin operator in terms of the Abrikosov fermionic partons
\begin{subequations}\label{Ssss}
\begin{align}
\hat{\mathbf{S}}_{\mathbf{r}_\mu} &= \frac{1}{4}
\mathrm{Tr}(\Psi^\dag_{\mathbf{r}_\mu} \bm{\sigma} \Psi_{\mathbf{r}_\mu})\,,\\
\Psi_{\mathbf{r}_\mu} &= \left(\begin{array}{cc}
f_{\mathbf{r}_\mu,\uparrow} & f^\dag_{\mathbf{r}_\mu,\downarrow}\\
f_{\mathbf{r}_\mu,\downarrow} & -
f^\dag_{r_\mu,\uparrow}\end{array}\right)\,,\label{defpsi}
\end{align}
\end{subequations}
where $\bm{\sigma}=(\sigma^1,\sigma^2,\sigma^3)$ are the Pauli matrices. 
The Abrikosov fermionic partons $f$ are arranged into the $2\times2$ matrix $\Psi$. 
These partons describe the fermionic spinon excitations of interest. 
The parton Hilbert space enlarges the physical spin Hilbert space, which is recovered by imposing the single-occupancy constraint
\begin{equation}
\sum\limits_{\sigma=\uparrow,\downarrow}
f^\dag_{\mathbf{r}_\mu,\sigma}f^{\vphantom\dagger}_{\mathbf{r}_\mu,\sigma} =
1,\quad \forall \, \mathbf{r}_\mu \,.
\end{equation}

The parton representation introduces an SU(2) gauge invariance: any site-dependent SU(2) gauge transformation
\begin{equation}
G \colon \Psi_{\mathbf{r}_\mu} \rightarrow \Psi_{\mathbf{r}_\mu}
W(\mathbf{r}_\mu)\,,\quad W(\mathbf{r}_\mu)\in \text{SU(2)}\,, \label{op-G}
\end{equation}
leaves the physical spins $\hat{\mathbf{S}}_{\mathbf{r}_\mu}$ invariant. 
Because of this gauge redundancy in the parton description, the physical symmetries of the spin liquid act \emph{projectively} on the parton operators. 
Physically distinct parton Hamiltonians cannot be related by gauge transformations and therefore correspond to different projective representations of the physical symmetry. 
The classification of projective symmetry groups (PSGs) thus provides a classification of symmetric quantum spin liquids.

We derive the PSG equations as follows. 
Under a space-group operation $\mathcal{O}$, the spins transform as
\[
\mathcal{O}\colon
\hat{\mathbf{S}}_{\mathbf{r}_\mu}\rightarrow
U_{\mathcal{O}}
\hat{\mathbf{S}}_{\mathcal{O}(\mathbf{r}_\mu)}
U^\dag_{\mathcal{O}}\,,
\]
where $U_{\mathcal{O}}$ is the SU(2) rotation matrix associated with the operation $\mathcal{O}$. 
The partons transform projectively according to
\begin{equation}\label{b_under_o}
\widetilde{\mathcal{O}} = G_{\mathcal{O}} \circ \mathcal{O} \colon
\Psi_{\mathbf{r}_\mu} \rightarrow
U^\dag_{\mathcal{O}}
\Psi_{\mathcal{O}(\mathbf{r}_\mu)}
W_{\mathcal{O}}[\mathcal{O}(\mathbf{r}_\mu)]\,,
\end{equation}
where the symbol ``$\circ$'' indicates that the projective operation
$\widetilde{\mathcal{O}}$ is the composition of the physical symmetry operation $\mathcal{O}$ and the gauge transformation $G_{\mathcal{O}}$.

The same procedure applies to the time-reversal operation $\mathcal{T}$. 
The spins transform under $\mathcal{T}$ as
\[
\hat{\mathbf{S}}_{\mathbf{r}_\mu}
\xrightarrow{\mathcal{T}}
\mathcal{K}^\dag
U_{\mathcal{T}}
\hat{\mathbf{S}}_{\mathbf{r}_\mu}
U^\dag_{\mathcal{T}}
\mathcal{K}\,,
\]
where $U_{\mathcal{T}} = i\sigma^2$ and
$\mathcal{K}=\mathcal{K}^\dag=\mathcal{K}^{-1}$ denotes complex conjugation acting on all operators to its right. 
Using a special property of the SU(2) algebra, the projective action of $\mathcal{T}$ on $\Psi$ can be written in a unitary form~\cite{PhysRevB.104.054401}:
\begin{equation}\label{timereversaldef3}
\widetilde{\mathcal{T}}=G_{\mathcal{T}}\circ \mathcal{T}\colon
\Psi_{\mathbf{r}_\mu}\rightarrow
U_{\mathcal{T}}
\Psi_{\mathbf{r}_\mu}
{W}_{\mathcal{T}}(\mathbf{r}_\mu)\,,
\end{equation}
while the antiunitary nature of $\mathcal{T}$ will manifest itself at the level of the mean-field {\it Ans\"atze} [see Eq.~\eqref{uuu} below].
For any chiral operator $\mathcal{X} = \mathcal{O}\mathcal{T}$, its projective action $\widetilde{\mathcal{X}}$ is simply the product of $\widetilde{\mathcal{O}}$ and $\widetilde{\mathcal{T}}$.

The classification of PSGs generated by the projective operations $\widetilde{\mathcal{X}}$ amounts to finding all gauge-inequivalent solutions for the gauge transformations $G_{\mathcal{X}}$ that are consistent with the symmetry group of the system.
Any group relation of Eq.~\eqref{SGtr} (or its chiral version, obtained by replacing $\mathcal{O}$ with $\mathcal{O}\mathcal{T}$)
\begin{equation}
\mathcal{X}_1 \circ \mathcal{X}_2 \circ \dots = 1\,, \label{grprlt0}
\end{equation}
leads to a gauge-enriched group relation
\begin{equation}\label{grprlt1}
\widetilde{\mathcal{X}}_1 \circ \widetilde{\mathcal{X}}_2 \circ
\dots = (G_{\mathcal{X}_1}\circ \mathcal{X}_1)\circ
(G_{\mathcal{X}_2}\circ \mathcal{X}_2)\circ\dots = \mathcal{G}\,,
\end{equation}
where $\mathcal{G}$ represents pure gauge transformations that leave the parton Hamiltonian invariant, known as the invariant gauge group (IGG). 
The IGG on each site is a subgroup of SU(2), typically $\mathbb{Z}_2$ or $\mathrm{U(1)}$. 
A gauge choice can always be made such that
$\mathcal{G} = e^{i \sigma^3 \chi}$ with some constant $\chi$. 
In this paper we classify both $\mathbb{Z}_2$ and $\mathrm{U(1)}$ spin liquids, corresponding to IGG $= \mathbb{Z}_2$ and $\mathrm{U(1)}$, for which $\chi = \{0,\pi\}$ and $\chi \in [0,2\pi)$, respectively.

Using the general conjugation rule
\begin{equation}\label{OGO}
\mathcal{X}_i \circ G_{\mathcal{X}_j} \circ \mathcal{X}^{-1}_i
\colon \Psi_{\mathbf{r}_\mu} \rightarrow
\Psi_{\mathbf{r}_\mu}W_{\mathcal{X}_j}[\mathcal{X}^{-1}_i(\mathbf{r}_\mu)]\,,
\end{equation}
which follows directly from Eq.~\eqref{b_under_o},
Eq.~\eqref{grprlt1} can be converted to an SU(2) equation
\begin{eqnarray}\label{phaseequationO}
W_{\mathcal{X}_1}(\mathbf{r}_\mu)
W_{\mathcal{X}_2}[\mathcal{X}_1^{-1}(\mathbf{r}_\mu)]
W_{\mathcal{X}_3}\{\mathcal{X}_2^{-1}[\mathcal{X}_1^{-1}(\mathbf{r}_\mu)]\}\dots
= \mathcal{G}\,. \nonumber\\
\end{eqnarray}

The PSG classification is obtained by listing all group relations and solving the corresponding SU(2) equations \eqref{phaseequationO} up to gauge equivalence. 
A general gauge transformation $G$ as in Eq.~\eqref{op-G} transforms the gauge-enriched relations in Eq.~\eqref{grprlt1} according to
\begin{equation}
(G\circ G_{\mathcal{X}_1} \circ \mathcal{X}_1 \circ G^{-1}) \circ
(G\circ G_{\mathcal{X}_2}\circ \mathcal{X}_2\circ G^{-1})\circ\dots
= \mathcal{G}\,,
\end{equation}
which transforms $W_{\mathcal{X}_i} (\mathbf{r}_\mu)$ by
\begin{equation}
W_{\mathcal{X}_i}(\mathbf{r}_\mu)\rightarrow
W(\mathbf{r}_\mu)
W_{\mathcal{X}_i}(\mathbf{r}_\mu)
W^{-1}[\mathcal{X}_i^{-1}(\mathbf{r}_\mu)]\,.
\end{equation}
Such gauge transformations are used to eliminate duplicate PSG solutions that are gauge equivalent to each other, yielding the final PSG classification.

As the group equations for the chiral cases remain the same as for the nonchiral cases, the PSG classification obtained for the latter also applies to the former. 
In particular, the PSGs for each of the chiral types $(I\mathcal{T},S)$, $(I,S\mathcal{T})$, and $(I\mathcal{T},S\mathcal{T})$ are in one-to-one correspondence with the nonchiral PSGs. 
The corresponding chiral and nonchiral classes therefore share the same PSG solutions, while the distinction between unitary and antiunitary symmetry operations appears only at the level of the mean-field {\it Ans\"atze}.

\begin{figure*}
\includegraphics[width = 0.3\textwidth]{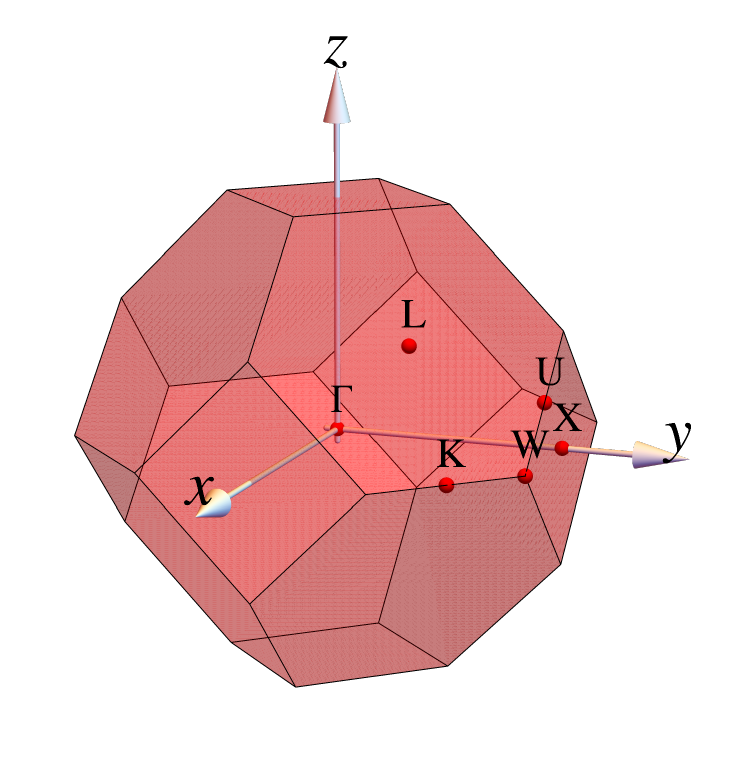}
\includegraphics[width = 0.3\textwidth]{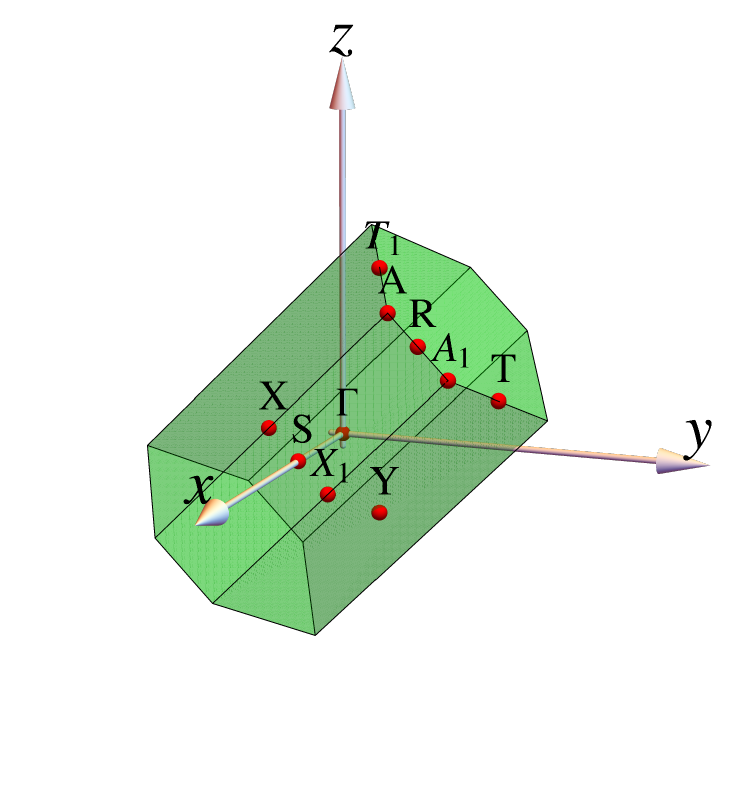}
\includegraphics[width = 0.3\textwidth]{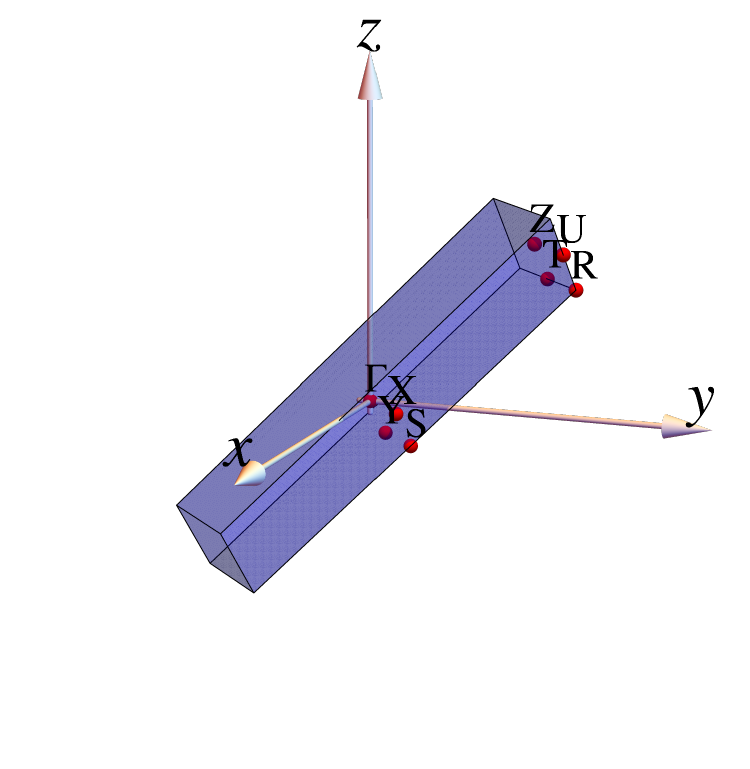}
\caption{Brillouin zones for the (i) 0-flux ($\varepsilon=1$), (ii) $\pi$-flux ($\varepsilon=2$), and (iii) $\pi/2$-flux ($\varepsilon=4$) states. Here, $\varepsilon$ denotes the enlargement factor of the parton Brillouin zone: the parton primitive cell is spanned by $\mathbf e_1$, $\varepsilon \mathbf e_2$, and $\varepsilon \mathbf e_3$, where $\mathbf e_{1,2,3}$ are the FCC lattice vectors defined in Eq.~\eqref{fcc_lv}.}
\label{fig:BZs}
\end{figure*}

\subsection{General mean-field {\it Ans\"atze}}

The most general mean-field Hamiltonian for fermionic spinons is written as
\begin{equation}
H = \sum_{\alpha = 0,x,y,z} H^\alpha,~~~H^\alpha = \sum_{ \mathbf{r}_\mu,\mathbf{r}'_\nu} H^\alpha_{\mathbf{r}_\mu,\mathbf{r}'_\nu}
\end{equation}
with
\begin{equation}\label{H_alpha_term}
H^\alpha_{\mathbf{r}_\mu,\mathbf{r}'_\nu} = \mathrm{Tr}[\sigma^\alpha \Psi_{\mathbf{r}_\mu}u^{(\alpha)}_{\mathbf{r}_\mu,\mathbf{r}'_\nu}\Psi^\dag_{\mathbf{r}'_\nu}]\,,
\end{equation}
where $\sigma^i$, $i=x,y,z$ are the three Pauli matrices, $\sigma^0$ is the identity matrix, and $\Psi_{\mathbf{r}_\mu}$ is defined as in Eq.~\eqref{defpsi}.
Here our convention follows that of Ref.~\cite{PhysRevB.104.054401}: for the bond $\mathbf{r}_\mu \leftarrow \mathbf{r}'_\nu$,
\begin{subequations}
\begin{align}
u^{(0)}_{\mathbf{r}_\mu,\mathbf{r}'_\nu}
&=i\mathrm{Re}t^{(0)}_{\mathbf{r}_\mu,\mathbf{r}'_\nu} 1 - \mathrm{Re}\chi^{(0)}_{\mathbf{r}_\mu,\mathbf{r}'_\nu} \sigma^1 \nonumber \\
&~~~-\mathrm{Im}\chi^{(0)}_{\mathbf{r}_\mu,\mathbf{r}'_\nu} \sigma^2-\mathrm{Im}t^{(0)}_{\mathbf{r}_\mu,\mathbf{r}'_\nu} \sigma^3\,,\\
u^{(i)}_{\mathbf{r}_\mu,\mathbf{r}'_\nu}
&=\mathrm{Re}t^{(i)}_{\mathbf{r}_\mu,\mathbf{r}'_\nu} 1 +i( \mathrm{Re}\chi^{(i)}_{\mathbf{r}_\mu,\mathbf{r}'_\nu} \sigma^1 \nonumber\\
&~~~+\mathrm{Im}\chi^{(i)}_{\mathbf{r}_\mu,\mathbf{r}'_\nu} \sigma^2+\mathrm{Im}t^{(i)}_{\mathbf{r}_\mu,\mathbf{r}'_\nu} \sigma^3)\,,
\end{align}
\end{subequations}
where $i=x,y,z$.
Explicitly, we have
\begin{subequations}\label{simpH}
\begin{align}
H^0_{\mathbf{r}_\mu,\mathbf{r}'_\nu}  
=
&i( t^{(0)}_{\mathbf{r}_\mu,\mathbf{r}'_\nu})^*(f^\dag_{\mathbf{r}_\mu,\uparrow} f_{\mathbf{r}'_\nu,\uparrow}+f^\dag_{\mathbf{r}_\mu,\downarrow}f_{\mathbf{r}'_\nu,\downarrow})\notag\\
&+\chi^{(0)}_{_{\mathbf{r}_\mu,\mathbf{r}'_\nu}} (f^\dag_{\mathbf{r}_\mu,\uparrow}f^\dag_{\mathbf{r}'_\nu,\downarrow}-f^\dag_{\mathbf{r}_\mu,\downarrow} f^\dag_{\mathbf{r}'_\nu,\uparrow})+h.c.\,,\\
H^x_{\mathbf{r}_\mu,\mathbf{r}'_\nu}  
=&
-(t^{(x)}_{\mathbf{r}_\mu,\mathbf{r}'_\nu})^*(f^\dag_{\mathbf{r}_\mu,\downarrow}f_{\mathbf{r}'_\nu,\uparrow}+f^\dag_{\mathbf{r}_\mu,\uparrow} f_{\mathbf{r}'_\nu,\downarrow})\notag\\
&-i\chi^{(x)}_{\mathbf{r}_\mu,\mathbf{r}'_\nu}(f^\dag_{\mathbf{r}_\mu,\uparrow}f^\dag_{\mathbf{r}'_\nu,\uparrow}-f^\dag_{\mathbf{r}_\mu,\downarrow}f^\dag_{\mathbf{r}'_\nu,\downarrow})
+h.c.\,,\\
H^y_{\mathbf{r}_\mu,\mathbf{r}'_\nu}  
=&
-i(t^{(y)}_{\mathbf{r}_\mu,\mathbf{r}'_\nu})^*(f^\dag_{\mathbf{r}_\mu,\downarrow}f_{\mathbf{r}'_\nu,\uparrow}-f^\dag_{\mathbf{r}_\mu,\uparrow}f_{\mathbf{r}'_\nu,\downarrow})\notag\\
&-\chi^{(y)}_{\mathbf{r}_\mu,\mathbf{r}'_\nu}(f^\dag_{\mathbf{r}_\mu,\uparrow}f^\dag_{\mathbf{r}'_\nu,\uparrow}+
f^\dag_{\mathbf{r}_\mu,\downarrow}f^\dag_{\mathbf{r}'_\nu,\downarrow})+h.c.\,,\\
H^z_{\mathbf{r}_\mu,\mathbf{r}'_\nu}  
=&
-(t^{(z)}_{\mathbf{r}_\mu,\mathbf{r}'_\nu})^*(f^\dag_{\mathbf{r}_\mu,\uparrow}f_{\mathbf{r}'_\nu,\uparrow}-f^\dag_{\mathbf{r}_\mu,\downarrow}f_{\mathbf{r}'_\nu,\downarrow})\notag\\
&+i\chi^{(z)}_{\mathbf{r}_\mu,\mathbf{r}'_\nu}(f^\dag_{\mathbf{r}_\mu,\uparrow}f^\dag_{\mathbf{r}'_\nu,\downarrow}+f^\dag_{\mathbf{r}_\mu,\downarrow}f^\dag_{\mathbf{r}'_\nu,\uparrow})+h.c.
\end{align}
\end{subequations}
We introduce $a_h,b_h,c_h,d_h,a_p,b_p,c_p,d_p$ to parametrize the 8 complex hopping and pairing $t^{(0)}$, $t^{(x)}$, $t^{(y)}$, $t^{(z)}$, $\chi^{(0)}$, $\chi^{(x)}$, $\chi^{(y)}$, and $\chi^{(z)}$ of the nearest-neighbor bond $(0,0,0)_0\leftarrow (0,0,0)_1$. For the onsite terms at $\mathbf{r}_\mu = \mathbf{r}'_\nu = (0,0,0)_0$, only four complex parameters $(t^{(0)},\chi^{(x)},\chi^{(y)},\chi^{(z)})$ are nonzero due to fermion anticommutativity and hermiticity of the Hamiltonian, and we parameterize them as $(\alpha_h,\beta_p,\gamma_p,\delta_p)$.

Note that for $\mathrm{U(1)}$ PSG {\it Ans\"atze}, we only have hopping bilinears and no pairing, therefore the parameters with subscript ``$p$'' (hence the $\sigma^1$ and $\sigma^2$ terms) vanish. We then simplify the notation of the hopping parameters by omitting the subscript ``$h$'' and write
\begin{equation}
\begin{aligned}
u^{(0)}_{\mathbf{0}_0,\mathbf{0}_1} &= i\mathrm{Re}a 1-\mathrm{Im}a \sigma^3\,,\\
u^{(x)}_{\mathbf{0}_0,\mathbf{0}_1} &= \mathrm{Re}b 1+i\mathrm{Im}b \sigma^3\,,\\
u^{(y)}_{\mathbf{0}_0,\mathbf{0}_1} &= \mathrm{Re}c 1+i\mathrm{Im}c \sigma^3\,,\\
u^{(z)}_{\mathbf{0}_0,\mathbf{0}_1} &= \mathrm{Re}d 1+i\mathrm{Im}d \sigma^3\,.
\end{aligned}
\end{equation}

For any space-group elements $\mathcal{X}$, the singlet and the triplet sectors transform
as
\begin{equation}\label{uuu}
\begin{aligned}
W_{\mathcal{X}}[\mathcal{X}(\mathbf{r}_\mu)]u^{(0)}_{\mathbf{r}_\mu,\mathbf{r}'_\nu}W^\dag_{\mathcal{X}}[\mathcal{X}(\mathbf{r}'_\nu)] &= (-1)^{c_{\mathcal{X}}}u^{(0)}_{\mathcal{X}(\mathbf{r}_\mu),\mathcal{X}(\mathbf{r}'_\nu)}\,,\\
W_{\mathcal{X}}[\mathcal{X}(\mathbf{r}_\mu)]u^{(i)}_{\mathbf{r}_\mu,\mathbf{r}'_\nu}\mathcal{R}^{\mathcal{X}}_{ij}
W^\dag_{\mathcal{X}}[\mathcal{X}(\mathbf{r}'_\nu)] &=
(-1)^{c_{\mathcal{X}}}u^{(j)}_{\mathcal{X}(\mathbf{r}_\mu),\mathcal{X}(\mathbf{r}'_\nu)}\,,
\end{aligned}
\end{equation}
where $\mathcal{R}^{\mathcal{X}}$ is the $SO(3)$ matrix describing the transformation of the vector $\mathbf{r}$ under $\mathcal{X}$. The new ingredient is the sign factor $(-1)^{c_{\mathcal{X}}}$: when $\mathcal{X}$ is chiral (nonchiral) we have $c_{\mathcal{X}}=1$ ($c_{\mathcal{X}}=0$). The negative sign in the chiral case comes from a special gauge fixing and accounts for the antiunitarity of $\mathcal{X}$.

\subsection{SU(2) flux of parton states\label{sub:fluxes}}

A $\mathrm{U(1)}$ spin-singlet {\it Ansatz} as considered in Refs.~\cite{Burnell-2009, Kim-2008} is parameterized by a single complex hopping $t^{(0)}_{ij}$, where we abbreviate the lattice sites $\mathbf{r}_\mu,\mathbf{r}'_\nu,...$ by $i,j,...$ For the diamond $\smalllozenge$ (i.e., the translation cycles), the triangle $\smalltriangleup/\smalltriangledown$ (on the up and down tetrahedra) and the hexagon $\smallhexagon$ plaquettes (in the kagome planes), one naively defines their fluxes $\phi_{\smalllozenge}$, $\phi_{\smalltriangleup/\smalltriangledown}$ and $\phi_{\smallhexagon}$ to be the phase of the product of nearest-neighbor bonds $t_{ij}$ that border the plaquettes:
$e^{i \phi_p} \sim \prod_{ij \in NN(p)} t^{(0)}_{ij}.$
We will call such defined flux $\phi$ the $\mathrm{U(1)}$ flux, since it is invariant under the $\mathrm{U(1)}$ gauge transformations $f_i\rightarrow e^{i \phi_i} f_i$, $t^{(0)}_{ij}\rightarrow t^{(0)}_{ij} e^{i(\phi_i - \phi_j)}$.

The flux $\phi$ as defined above, however, is not invariant under the projective transformations of PSG, which operates in an enlarged space of SU(2) transformations allowing $f_i$ and $f_i^\dag$ to rotate to each other. Including also the singlet pairings, a spin-singlet {\it Ansatz} has the general form [see Eq.~\eqref{H_alpha_term}]
\begin{equation}
  H^0 = \sum_{\langle ij\rangle}\mathrm{Tr}\left[\Psi_i u^{(0)}_{ij} \Psi^\dag_{j}\right]\,.
\end{equation}
Correspondingly, we define the SU(2) flux $\Phi_p$ for a plaquette $p$ to be
\begin{equation}
F_p\sim e^{i \Phi_p},\quad F_p \equiv \mathrm{Tr}\left[ \prod_{ij \in NN(p)} u^{(0)}_{ij}\right]\,,
\end{equation}
here $F_p$ is defined via the product of the matrix $u^{(0)}_{ij}$ along the NN bonds that enclose the plaquette $p$. It serves as the lattice version of the $SU(2)$ Wilson loop, which is invariant under the SU(2) gauge transformation $\Psi_i\rightarrow \Psi_i W_i$, $u^{(0)}_{ij}\rightarrow W_i u^{(0)}_{ij} W^\dag_j$. The flux $\Phi_p$ is then defined as the phase of $F_p$. 

It is useful to recall how this gauge-invariant parton flux is related to physical spin observables. One should distinguish the mean-field Wilson loop, built from the $c$-number matrices $u^{(0)}_{ij}$, from the corresponding loop operator built from spinon bilinears. For an oriented loop $C=(1,2,\ldots,q)$, the mean-field Wilson matrix is
\begin{equation}
    P^{\mathrm{MF}}_C
    =
    u^{(0)}_{12}u^{(0)}_{23}\cdots u^{(0)}_{q1},
\end{equation}
and $F_p$ is the corresponding gauge-invariant trace for plaquette $p$. The associated operator is obtained by replacing each mean-field link matrix by the singlet spinon-bilinear link operator
\begin{equation}
    \widehat u_{ij}
    =
    \frac{1}{2}
    \begin{pmatrix}
    \bm f_i^\dagger \bm f_j
    &
    \bm f_i^{T}\varepsilon \bm f_j
    \\
    \bm f_j^\dagger \varepsilon \bm f_i^*
    &
    \bm f_j^\dagger \bm f_i
    \end{pmatrix},
    \qquad
    \varepsilon=i\sigma^2 ,
    \label{eq:flux_link_operator}
\end{equation}
where $\bm f_i=(f_{i,\uparrow},f_{i,\downarrow})^T$. Thus $\widehat P_C=\widehat u_{12}\widehat u_{23}\cdots \widehat u_{q1}$. The colons below denote normal ordering with respect to the spinon vacuum. This removes vacuum-contraction terms generated by fermionic anticommutation and yields, after taking the trace, a gauge-invariant spin operator independent of the arbitrary base site of the loop~\cite{Bieri-2016}. For an oriented three-site loop $C=(1,2,3)$, this operator identity reduces to~\cite{Wen-1989,Bieri-2016}
\begin{equation}
    \operatorname{Tr}\!\left[:\widehat{P}_{123}:\right]
    =
    2i\,
    \mathbf{S}_1\cdot
    \left(
        \mathbf{S}_2\times\mathbf{S}_3
    \right),
    \label{eq:triangular_wilson_chirality}
\end{equation}
up to the convention chosen for the orientation of the loop. Thus, for the triangular faces of the pyrochlore tetrahedra, the imaginary part of the triangular Wilson loop is directly tied to the scalar spin chirality. In this precise sense, a nontrivial odd-loop flux in the fermionic parton {\it Ansatz} is the mean-field counterpart of chiral spin correlations. For longer loops, and especially for general $\mathrm{SU}(2)$ gauge fluxes, the corresponding spin operator contains more general multi-spin terms and should not be identified na\"ively with a simple additive $\mathrm{U}(1)$ flux.

We make the following remarks:
\begin{itemize}
\item The Wilson loop $F_p$ may vanish in some cases. In such cases, the SU(2) flux $\Phi_p$ is not well-defined, which we write as $\Phi_p = \forall$.

\item When $\Phi_p$ is well-defined, the special form of $u^{(0)}$ (an imaginary factor times an SU(2) matrix) constrains $F_{\smalllozenge}$ and $F_{\smallhexagon}$ to be real and $F_{\smalltriangleup/\smalltriangledown}$ to be purely imaginary. Consequently, the only possible values of $\Phi_{\smalllozenge}$ and $\Phi_{\smallhexagon}$ are $0$, $\pi$, and $\Phi_{\smalltriangleup/\smalltriangledown} = \pm \frac{\pi}{2}$.
\item For a $\mathrm{U(1)}$ {\it Ansatz}, $\chi_{ij}=0$, we can further express $\Phi_p$ in terms of $\phi_p$:
\begin{subequations}\label{Phiphi}
\begin{align}
e^{i \Phi_{\smalllozenge}}&\sim F_{\smalllozenge} \sim \cos(\phi_{\smalllozenge})\,,\\
e^{i \Phi_{\smalltriangleup/\smalltriangledown}}&\sim  F_{\smalltriangleup/\smalltriangledown} \sim i \sin(\phi_{\smalltriangleup/\smalltriangledown})\,,\\  e^{i \Phi_{\smallhexagon}}&\sim F_{\smallhexagon} \sim \cos(\phi_{\smallhexagon})\,.
\end{align}
\end{subequations}
\end{itemize}
We see that the $\mathrm{U(1)}$ flux $\phi_p$ provides a convenient description for a $\mathrm{U(1)}$ {\it Ansatz}. As we shall explain in the next section, the $\mathrm{U(1)}$ fluxes $\phi_{\smalllozenge,\smalltriangleup/\smalltriangledown,\smallhexagon}$ fully determine all non-chiral and chiral spin-singlet states at the NN level. The description of NN $\mathbb{Z}_2$ {\it Ans\"atze} in terms of the fluxes is more involved, which we will show in Table \ref{MFTparameters_z2_nn}. 

Before combining the flux information with the results of PSG, let us first look at the general symmetry properties of the SU(2) flux $\Phi_p$. By definition, $\Phi_p$ is odd under time-reversal operation; it is also a pseudoscalar that is odd under inversion. Further taking into account the transformation of the normal vectors of the plaquette, we obtain the transformation rule of $\Phi_p$ under $(I,S,\mathcal{T})$ as given in Table \ref{transunderIST} \footnote{By definition, a pseudoscalar is odd under parity transformation (i.e., inversion $I$), which seems to impose $\tilde{\Phi}_{I(\smalllozenge)}= -\Phi_{\smalllozenge}$ and $\tilde{\Phi}_{I(\smallhexagon)}= - \Phi_{\smallhexagon}$, at odds with Table \ref{transunderIST}. This naive claim is based on the fact that $\Phi \equiv \mathbf{B}\cdot \mathbf{n}$ where $\mathbf{B}$ is a pseudovector (even under parity: $\tilde{\mathbf{B}}(I(x))=\mathbf{B}(x)$) and $\mathbf{n}$ is a vector (odd under parity: $\tilde{\mathbf{n}}(I(x))=-\mathbf{n}(x)$). However, we adopt a uniform orientation $\mathbf{n}$ for all translation plaquettes (and similarly for all the hexagon plaquettes), as a consequence $\Phi_{I(\smalllozenge)} =- \tilde{\Phi}_{I(\smalllozenge)}$, $\Phi_{I(\smallhexagon)} = - \tilde{\Phi}_{\smallhexagon}$. This explains the parity rule in Table \ref{transunderIST}.}:

\begin{table}[!thb]
\centering
\begin{tabular}{c|ccc|c}
\hline
\hline
& $I$ & $S$ & $\mathcal{T}$ & Value of $(\Phi_p,\phi_p)$ \\
\hline
$\Phi_{\smalllozenge}$ 
& $\Phi_{\smalllozenge}$ 
& $-\Phi_{\smalllozenge}$ 
& $-\Phi_{\smalllozenge}$ 
& $(0,0),(\pi,\pi),(\forall,\frac{\pi}{2})$\\
$\Phi_{\smalltriangleup/\smalltriangledown}$ 
& $-\Phi_{\smalltriangledown/\smalltriangleup}$ 
& $\Phi_{\smalltriangledown/\smalltriangleup}$ 
& $-\Phi_{\smalltriangleup/\smalltriangledown}$ 
& $(\forall,0),(\forall,\pi),(\pm\frac{\pi}{2},\pm\frac{\pi}{2})$\\
$\Phi_{\smallhexagon}$ 
& $\Phi_{\smallhexagon}$ 
& $-\Phi_{\smallhexagon}$ 
& $-\Phi_{\smallhexagon}$ 
& $(0,0),(\pi,\pi),(\forall,\frac{\pi}{2})$\\
\hline
\hline
\end{tabular}
\caption{Transformation properties of the SU(2) gauge-invariant fluxes 
$\Phi_{\smalllozenge,\smalltriangleup / \smalltriangledown,\smallhexagon}$ 
under inversion $I$, screw $S$, and time reversal $\mathcal{T}$, together with the relation between 
$\Phi_p$ and the corresponding $\mathrm{U}(1)$ flux $\phi_p$, and their allowed values. 
For the triangular plaquettes, the signs in 
$(\pm\pi/2,\pm\pi/2)$ are correlated according to the chosen orientation convention.
\label{transunderIST}}
\end{table}

From Table \ref{transunderIST}, we see that the only well-defined values of $\Phi_{\smalltriangleup/\smalltriangledown}$, i.e., $\pm \frac{\pi}{2}$, break time-reversal symmetry, and can only appear in the $(I\mathcal{T},S,\xcancel{\mathcal{T}})$ and $(I,S\mathcal{T},\xcancel{\mathcal{T}})$ classes. In order to preserve all the three symmetries $I,S,\mathcal{T}$, we must have $\Phi_{\smalltriangleup/\smalltriangledown} = \forall$.

We will consider states preserving the following symmetries: $(I,S)$, $(I\mathcal{T},S)$, $(I,S\mathcal{T})$, and $(I\mathcal{T},S\mathcal{T})$. As we shall see, for spin singlet states at the NN bond level, both $(I,S)$ and $(I\mathcal{T},S\mathcal{T})$ imply $(I,S,\mathcal{T})$, i.e., time-reversal symmetry is restored. 

\section{Complete $\mathrm{U(1)}$ and $\mathbb{Z}_2$ chiral {\it Ans\"atze} from PSG}\label{sec:PSG}
\subsection{PSG: algebraic classes}
\label{classification} 

Here, we present the results of the PSG classifications of $\mathrm{U(1)}$ and $\mathbb{Z}_{2}$ chiral spin liquids for the different chiral symmetry classes, and discuss how they connect to the fully symmetric spin liquids that were previously classified in Ref.~\cite{PhysRevB.104.054401}. A complementary projective-symmetry-group classification of chiral $\mathbb{Z}_2$ spin liquids on the pyrochlore lattice within a Schwinger-boson framework was given in Ref.~\cite{Schneider-2022}. The consequences of the PT theorem for $\mathrm{U(1)}$ spin liquids and lack thereof for $\mathbb{Z}_{2}$ spin liquids are discussed.

The PSG classifications of $\mathrm{U(1)}$ and $\mathbb{Z}_2$ PSGs for the non-chiral classes were given in Ref.~\cite{PhysRevB.104.054401}. For each chiral type, $(I\mathcal{T},S)$, $(I,S\mathcal{T})$, or $(I\mathcal{T},S\mathcal{T})$, the number of PSG classes is exactly the same as their corresponding non-chiral types. Below we quote the PSG solution from Ref.~\cite{PhysRevB.104.054401}:
\begin{subequations}\label{www}
\begin{align}
W_{T_i}(\mathbf{r}_\mu) &= e^{i\sigma^3\phi_{T_i}(\mathbf{r}_\mu)},\quad i=1,2,3\,,\\
W_{\overline{C}_6}(\mathbf{r}_\mu) &= W_{\overline{C}_6,\mu}e^{i \sigma^3\phi_{\overline{C}_6}(\mathbf{r}_\mu)}\,,\label{eq:wc6}\\
W_S(\mathbf{r}_\mu) &= W_{S,\mu}e^{i \sigma^3\phi_S
(\mathbf{r}_\mu)}\,,\label{eq:ws}
\end{align}
\end{subequations}
with
\begin{subequations}
\begin{align}
\phi_{T_1}(\mathbf{r}_\mu) &= 0\,,\\
\phi_{T_2}(\mathbf{r}_\mu) &= -\chi_1 r_1\,,\\
\phi_{T_3}(\mathbf{r}_\mu) &= \chi_1(r_1-r_2)\,,\\
\phi_{\overline{C}_6}(\mathbf{r}_\mu) &= -\chi_1r_1(r_2-r_3)  \notag\\
&\quad-[{2\chi_{ST_1}+2 \chi_1}+(\delta_{\mu,2}-\delta_{\mu,3}) \chi_1]r_1\notag\\
&\quad+\delta_{\mu,2}\chi_1r_3\,,\\
\phi_S(\mathbf{r}_\mu) &=
\chi_1\left[\frac{(r_1+1)r_1}{2} - \frac{(r_2+1)r_2}{2}-r_1r_2\right]\notag\\
&\quad+
[(\delta_{\mu,1}-\delta_{\mu,2})\chi_1+(2\chi_1 - \chi_{ST_1})]r_1\notag\\
&\quad+
[(2\delta_{\mu,1}-\delta_{\mu,2})\chi_1 + 3\chi_{ST_1}]r_2\notag\\
&\quad+ [(\delta_{\mu,1}-\delta_{\mu,2}) + 2]\chi_1r_3\,,
\end{align}
\end{subequations}
the parameters $\chi_1$ and $\chi_{ST_1}$ label elements of the IGG. 

Recall that the PSG parameter for translation is $\chi_1 = 0$, $\pi$, and $\pi/2$, which defines the so-called $0$-flux, $\pi$-flux, and $\frac{\pi}{2}$-flux parton {\it Ans\"atze}. This label still applies to the chiral states and we will extensively study them below. The $\pi$-flux and $\frac{\pi}{2}$ states have an enlarged primitive cell that is 2 and 4 times that of the pyrochlore lattice, respectively. We show their parton Brillouin zones in Fig.~\ref{fig:BZs}.

\subsection{$\mathrm{U(1)}$ and $\mathbb{Z}_2$ mean-field {\it Ans\"atze} from PSG}
\label{mf}
In this section, we construct quadratic parton Hamiltonians realizing different $\mathrm{U(1)}$ and $\mathbb{Z}_{2}$ PSGs, whose ground states then serve as variational mean-field wave-functions for chiral spin liquids.

We give in Tables \ref{onsiteu1} and \ref{MFTparametersU1chiral} the $\mathrm{U(1)}$ classification and Tables
\ref{MFTparameters_z2_oo} and \ref{MFTparameters_z2_nn} the $\mathbb{Z}_2$ classification for the $(I,S)$, $(I\mathcal{T},S)$, $(I,S\mathcal{T})$, and $(I\mathcal{T},S\mathcal{T})$ PSGs including both the spin singlet and triplet parameters up to nearest neighbor (NN) level. We specify all the 12 singlet states of Table \ref{Tab1} in these PSGs. We also label the singlet states in terms of the fluxes $(\Phi_{\smalllozenge},\Phi_{\smalltriangleup},\Phi_{\smallhexagon})$ defined in Sec.~\ref{sub:fluxes}. 

The bond results for further neighbors are given in Appendix \ref{app:further_bonds}.

\begin{table}
\centering
\caption{
Independent onsite mean-field parameters allowed by the projective symmetry group (PSG) for $\mathrm{U(1)}$ chiral spin liquids on the pyrochlore lattice. 
Each row corresponds to a PSG class labeled by the parameters 
$\phi_1$-$(m_{C_6}m_S)$-$n_{ST_1}$-$n_{C_6S}$ introduced in Sec.~\ref{sec:PSG}. 
The columns indicate which onsite singlet parameters are permitted in the four chiral symmetry classes 
$(I,S)$, $(I\mathcal{T},S)$, $(I,S\mathcal{T})$, and $(I\mathcal{T},S\mathcal{T})$. 
Entries list the independent complex amplitudes allowed by symmetry, while ``--'' indicates that no independent onsite parameter is allowed in that class. These parameters serve as building blocks for constructing quadratic spinon Hamiltonians consistent with the corresponding PSG.}
\label{onsiteu1}
\begin{tabular}{c|cccc}
\hline
\hline
Class $\phi_1$-$(m_{\overline{C}_6}m_{S})$& \multicolumn{4}{l}{Nonzero onsite parameter in class}\\
\cline{2-5}
-${n}_{ST_1}$-$n_{\overline{C}_6S}$&$(I,S)$&$(I,S\mathcal{T})$&$(I\mathcal{T},S)$&$(I\mathcal{T},S\mathcal{T})$\\
\hline
$0$- or $\pi$-(00)-0-0 &$\mathrm{Im}\alpha$&--&--&--\\ 
$0$- or $\pi$-(00)-0-1 &$\mathrm{Im}\alpha$&--&--&--\\
$0$- or $\pi$-(01)-0-0 &--&$\mathrm{Im}\alpha$&--&--\\ 
$0$- or $\pi$-(01)-0-1 &--&$\mathrm{Im}\alpha$&--&--\\
$0$- or $\pi$-(10)-0-0 &--&--&$\mathrm{Im}\alpha$&--\\ 
$0$- or $\pi$-(10)-0-1 &--&--&$\mathrm{Im}\alpha$&--\\
$0$- or $\pi$-(11)-0-0 &--&--&--&$\mathrm{Im}\alpha$\\ 
$0$- or $\pi$-(11)-0-1 &--&--&--&$\mathrm{Im}\alpha$\\
\hline
$\frac{\pi}{2}$-(01)-{1}-0 & -- & $\mathrm{Im}\alpha$ &--&--\\
$\frac{\pi}{2}$-(01)-{1}-1 & -- & $\mathrm{Im}\alpha$ &--&--\\
\hline\hline
\end{tabular}
\end{table}
      
\begin{table*}
\centering
\caption{
Independent nearest-neighbor (NN) mean-field parameters and symmetry constraints for $\mathrm{U(1)}$ chiral spin liquids in the four PSG symmetry classes 
$(I,S)$, $(I\mathcal{T},S)$, $(I,S\mathcal{T})$, and $(I\mathcal{T},S\mathcal{T})$. 
For each PSG class (first column), the table lists the allowed NN hopping amplitudes together with the algebraic relations imposed by the PSG. 
The entries specify the independent parameters from which all other NN amplitudes can be generated by symmetry operations. These NN parameters determine the quadratic spinon Hamiltonians and the associated gauge-flux configurations discussed in Sec.~\ref{sec:PSG}. In the last column, the 12 nearest-neighbor spin-singlet {\it Ans\"atze} are highlighted (see Table \ref{Tab1} for further details on these 12 states).}
\label{MFTparametersU1chiral}\label{ITSandIST}
\begin{tabular}{c|c|c|cc}
\hline\hline
$(I,S)$ and $(I\mathcal{T},S\mathcal{T})$, &{Nonzero}&\multicolumn{1}{c|}{\multirow{ 2}{*}{Constraints}}&\multicolumn{2}{c}{Restrict to spin singlet: $(\phi_{\smalllozenge},\phi_{\smalltriangleup},\phi_{\smallhexagon})$ states}\\
\cline{4-5}
Class $\phi_1$-$(m_{\overline{C}_6}m_{S})$-${n}_{ST_1}$-$n_{\overline{C}_6S}$&NN  parameters
&
&$(I,S)$ type & $(I\mathcal{T},S\mathcal{T})$ type\\
\hline
0- or $\pi$-$(00)$-$0$-$0$   &$\mathrm{Im}a,\mathrm{Im}c$ 
& $\mathrm{Im}c=-\mathrm{Im}d$ 
&$(0,0,0)$ and $(\pi,0,\pi)$ & $(0,0,\pi)$ and $(\pi,0,0)$\\
0- or $\pi$-$(00)$-$0$-$1$ &$\mathrm{Im}b,\mathrm{Re}c$ 
& $\mathrm{Re}c=\mathrm{Re}d$   
&N/A&N/A\\
\hline
0- or $\pi$-$(01)$-$0$-$0$&$\mathrm{Im}c$ 
& $\mathrm{Im}c=\mathrm{Im}d$  
&N/A&N/A\\
0- or $\pi$-$(01)$-$0$-$1$&$\mathrm{Re}c$ 
& $\mathrm{Re}c=\mathrm{Re}d$   
&N/A&N/A\\
\hline
0- or $\pi$-$(10)$-$0$-$0$&$\mathrm{Im}c$ 
& $\mathrm{Im}c=\mathrm{Im}d$  
&N/A&N/A\\
0- or $\pi$-$(10)$-$0$-$1$&$\mathrm{Re}c$ 
& $\mathrm{Re}c=\mathrm{Re}d$  
&N/A&N/A\\
\hline
0- or $\pi$-$(11)$-$0$-$0$&$\mathrm{Im}a$, $\mathrm{Im}c$ 
& $\mathrm{Im}c=-\mathrm{Im}d$  
&$(0,0,\pi)$ and $(\pi,0,0)$ & $(0,0,0)$ and $(\pi,0,\pi)$ \\
0- or $\pi$-$(11)$-$0$-$1$&$\mathrm{Im}b,\mathrm{Re}c$  
& $\mathrm{Re}c=\mathrm{Re}d$ 
&N/A&N/A\\
\hline
$\frac{\pi}{2}$-$(01)$-${1}$-$0$&$\mathrm{Im}c$ 
& $\mathrm{Im}c=\mathrm{Im}d$  
&N/A&N/A\\
$\frac{\pi}{2}$-$(01)$-${1}$-$1$&$\mathrm{Re}c$ 
& $\mathrm{Re}c=\mathrm{Re}d$   
&N/A&N/A\\
\hline\hline
$(I,S\mathcal{T})$ and $(I\mathcal{T},S)$, &{Nonzero}&\multicolumn{1}{c|}{\multirow{ 2}{*}{Constraints}}&\multicolumn{2}{c}{Restrict to spin singlet: $(\phi_{\smalllozenge},\phi_{\smalltriangleup},\phi_{\smallhexagon})$ states}\\
\cline{4-5}
Class $\phi_1$-$(m_{\overline{C}_6}m_{S})$-${n}_{ST_1}$-$n_{\overline{C}_6S}$&NN parameters
&
& $(I\mathcal{T},S)$ type & $(I,S\mathcal{T})$ type\\
\hline
0- or $\pi$-$(00)$-$0$-$0$  &$\mathrm{Re}b$, $\mathrm{Im}c$ 
& $\mathrm{Im}c=\mathrm{Im}d$ 
&N/A&N/A\\
0- or $\pi$-$(00)$-$0$-$1$  &$\mathrm{Re}a$, $\mathrm{Re}c$ 
& $\mathrm{Re}c=-\mathrm{Re}d$  
& \textcolor{RoyalBlue}{$(0,\pi/2,0)$}~\cite{Burnell-2009}, \textcolor{Dandelion}{$(\pi,\pi/2,\pi)$} & \textcolor{ForestGreen}{$(0,\pi/2,\pi)$}, \textcolor{BrickRed}{$(\pi,\pi/2,0)$}~\cite{Kim-2008}\\
\hline
0- or $\pi$-$(01)$-$0$-$0$&$\mathrm{Im}a, \mathrm{Re}b, \mathrm{Im}c$ 
& $\mathrm{Im}c=-\mathrm{Im}d$  
&$(0,0,\pi)$ and $(\pi,0,0)$&$(0,0,0)$ and $(\pi,0,\pi)$\\
0- or $\pi$-$(01)$-$0$-$1$&$\mathrm{Re}a, \mathrm{Im}b,\mathrm{Re}c$ 
& $\mathrm{Re}c=-\mathrm{Re}d$  
& \textcolor{RoyalBlue}{$(0,\pi/2,0)$}~\cite{Burnell-2009}, \textcolor{Dandelion}{$(\pi,\pi/2,\pi)$} & \textcolor{ForestGreen}{$(0,\pi/2,\pi)$}, \textcolor{BrickRed}{$(\pi,\pi/2,0)$}~\cite{Kim-2008}\\
\hline
0- or $\pi$-$(10)$-$0$-$0$ &$\mathrm{Im}a, \mathrm{Re}b,\mathrm{Im}c$ 
& $\mathrm{Im}c=-\mathrm{Im}d$  
& $(0,0,0)$ and $(\pi,0,\pi)$$^{\ddag}$&$(0,0,\pi)$ and $(\pi,0,0)$\\
0- or $\pi$-$(10)$-$0$-$1$ &$\mathrm{Re}a,\mathrm{Im}b,\mathrm{Re}c$ 
& $\mathrm{Re}c=-\mathrm{Re}d$  
&  \textcolor{RoyalBlue}{$(0,\pi/2,0)$}~\cite{Burnell-2009}, \textcolor{Dandelion}{$(\pi,\pi/2,\pi)$} & \textcolor{ForestGreen}{$(0,\pi/2,\pi)$}, \textcolor{BrickRed}{$(\pi,\pi/2,0)$}~\cite{Kim-2008}\\
\hline
0- or $\pi$-$(11)$-$0$-$0$ &$\mathrm{Re}b$, $\mathrm{Im}c$ 
& $\mathrm{Im}c=\mathrm{Im}d$  
&N/A&N/A \\
0- or $\pi$-$(11)$-$0$-$1$ &$\mathrm{Re}a,\mathrm{Re}c$ 
& $\mathrm{Re}c=-\mathrm{Re}d$ 
& \textcolor{RoyalBlue}{$(0,\pi/2,0)$}~\cite{Burnell-2009}, \textcolor{Dandelion}{$(\pi,\pi/2,\pi)$} & \textcolor{ForestGreen}{$(0,\pi/2,\pi)$}, \textcolor{BrickRed}{$(\pi,\pi/2,0)$}~\cite{Kim-2008}\\
\hline
$\frac{\pi}{2}$-$(01)$-${1}$-$0$ &$\mathrm{Im}a, \mathrm{Re}b, \mathrm{Im}c$ 
& $\mathrm{Im}c=-\mathrm{Im}d$  
&\textcolor{Plum}{$(\frac{\pi}{2},\frac{\pi}{2},\frac{\pi}{2})$}& \textcolor{TealBlue}{$(\frac{\pi}{2},\frac{\pi}{2},-\frac{\pi}{2})$}\\
$\frac{\pi}{2}$-$(01)$-${1}$-$1$ &$\mathrm{Re}a, \mathrm{Im}b,\mathrm{Re}c$  
& $\mathrm{Re}c=-\mathrm{Re}d$  
& \textcolor{Plum}{$(\frac{\pi}{2},0,-\frac{\pi}{2})$} & \textcolor{TealBlue}{$(\frac{\pi}{2},0,\frac{\pi}{2})$}\\
\hline\hline
\end{tabular}
\end{table*}

\begin{table*}
\caption{
Independent onsite mean-field parameters for $\mathbb{Z}_2$ chiral spin liquids in the four symmetry classes 
$(I,S)$, $(I\mathcal{T},S)$, $(I,S\mathcal{T})$, and $(I\mathcal{T},S\mathcal{T})$ obtained from the PSG classification. 
Compared with the $\mathrm{U(1)}$ case, additional pairing terms are allowed once the invariant gauge group is reduced to $\mathbb{Z}_2$. 
All other onsite terms are fixed by the PSG constraints.}\label{MFTparameters_z2_oo}
\centering
\begin{tabular}{c|cccc}
\hline\hline
\multirow{2}{*}{Class ($\chi_1\chi_{ST_1}$)\---$(\chi_{\overline{C}_6S}\chi_{S\overline{C}_6}\chi_{\overline{C}_6})_j$}
&\multicolumn{4}{c}{Independent onsite parameters}\\
\cline{2-5}
&$(I,S)$&$(I,S\mathcal{T})$&$(I\mathcal{T},S)$&$(I\mathcal{T},S\mathcal{T})$\\
\hline
$(00)$\---  or $(\pi\pi)$\---$(000)$&$\mathrm{Im}\alpha_h,\alpha_p$&-&-&-\\
$(00)$\---  or $(\pi\pi)$\---$(\pi00)$&$\mathrm{Im}\alpha_h,\alpha_p$&-&-&-\\
$(00)$\---  or $(\pi\pi)$\---$(0\pi\pi)$&$\mathrm{Re}\alpha_p$&-&$\mathrm{Im}\alpha_h,\mathrm{Im}\alpha_p$&-\\
$(00)$\---  or $(\pi\pi)$\---$(\pi\pi\pi)$&$\mathrm{Re}\alpha_p$&-&$\mathrm{Im}\alpha_h,\mathrm{Im}\alpha_p$&-\\
\hline
$(0\pi)$\---  or $(\pi0)$\---$(0\pi0)_1$&{-}&$\mathrm{Im}\alpha_h$&-&-\\
$(0\pi)$\---  or $(\pi0)$\---$(0\pi0)_3$&$\mathrm{Re}\alpha_p$&$\mathrm{Im}\alpha_h,\mathrm{Im}\alpha_p$&-&-\\
$(0\pi)$\---  or $(\pi0)$\---$(\pi\pi0)_0$&$\mathrm{Re}\alpha_p$&$\mathrm{Im}\alpha_h,\mathrm{Im}\alpha_p$&-&-\\
$(0\pi)$\---  or $(\pi0)$\---$(\pi\pi0)_2$&{-}&$\mathrm{Im}\alpha_h$&-&-\\
\hline
$(0\pi)$\---  or $(\pi0)$\---$(00\pi)$&$\mathrm{Re}\alpha_p$&-&-&$\mathrm{Im}\alpha_h,\mathrm{Im}\alpha_p$\\
$(0\pi)$\---  or $(\pi0)$\---$(\pi0\pi)$&$\mathrm{Re}\alpha_p$&-&-&$\mathrm{Im}\alpha_h,\mathrm{Im}\alpha_p$\\
\hline
$(0\pi)$\---  or $(\pi0)$\---$(0\pi\pi)_0$&{-}&$\mathrm{Im}\alpha_h$&-&-\\
$(0\pi)$\---  or $(\pi0)$\---$(\pi\pi\pi)_0$&{- }&$\mathrm{Im}\alpha_h$&-&-\\
$(0\pi)$\---  or $(\pi0)$\---$(0\pi\pi)_1$&{- }&$\mathrm{Im}\alpha_h$&$\mathrm{Re}\alpha_p$&$\mathrm{Im}\alpha_p$\\
$(0\pi)$\---  or $(\pi0)$\---$(\pi\pi\pi)_1$&{- }&$\mathrm{Im}\alpha_h$&$\mathrm{Re}\alpha_p$&$\mathrm{Im}\alpha_p$\\
\hline
\hline
\end{tabular}
\end{table*}

\begin{table*}
\caption{
Independent nearest-neighbor (NN) mean-field parameters and symmetry constraints for $\mathbb{Z}_2$ chiral spin liquids in the four PSG symmetry classes 
$(I,S)$, $(I\mathcal{T},S)$, $(I,S\mathcal{T})$, and $(I\mathcal{T},S\mathcal{T})$. 
Both hopping and pairing amplitudes are included, as well as the algebraic relations required by symmetry. 
Spin-singlet parameters are underlined. 
All other NN matrix elements of the mean-field Hamiltonian follow from these parameters by applying the PSG operations.\\
$^{\text{a}}$ $\Theta=0,\pi$.\\
$^\S$ $\mathrm{Im}c_p = \sqrt{3} \mathrm{Re}c_p$, $c_p=d_p$.\\
$^{\S\S}$ $\mathrm{Im} c_p = \frac{1}{\sqrt{3}}\mathrm{Re} c_p$, $c_p=d_p$.\\
$^*$ $\mathrm{Im}c_p = \frac{1}{\sqrt{3}} \mathrm{Re}c_p$, $d_h = e^{\frac{\pi}{3}i} c_h^*$, $c_p = d_p$.\\
$^{**}$ $d_h = e^{\frac{\pi}{3}i}c^*_h$, $\mathrm{Im}c_p = -\sqrt{3}\mathrm{Re}c_p$, $c_p=d_p$.\\
$^\dag$ $\mathrm{Im}b_h = - \frac{1}{\sqrt{3}} \mathrm{Re}b_h$, $\mathrm{Im}b_p = -\frac{1}{\sqrt{3}}\mathrm{Re}b_p$,  $\mathrm{Im}c_p = \sqrt{3} \mathrm{Re}c_p$, $c_p=-d_p$.\\
$^\ddag$ $\mathrm{Im}b_p = -\sqrt{3}\mathrm{Re}b_p$, $\mathrm{Im}c_p = \frac{1}{\sqrt{3}} \mathrm{Re}c_p$, $d_h = -e^{\frac{\pi}{3}i} c_h^*$, $c_p = -d_p$.\\
$^\#$ $\mathrm{Im} b_h = \frac{1}{\sqrt{3}}\mathrm{Re} b_h$, $\mathrm{Im}b_p = -\sqrt{3}\mathrm{Re}b_p$, $\mathrm{Im} c_p = \frac{1}{\sqrt{3}}\mathrm{Re} c_p$, $c_p=-d_p$.\\
$^{\#\#}$ $\mathrm{Im}b_p = \frac{1}{\sqrt{3}} \mathrm{Re}b_p$, $\mathrm{Im}c_p = -\sqrt{3}\mathrm{Re}c_p$, $d_h = -e^{\frac{\pi}{3}i}c^*_h$, $c_p=-d_p$.}\label{MFTparameters_z2_nn}
\begin{tabular}{l|l|l|cc}
\hline\hline
$(I,S)$ and $(I\mathcal{T},S\mathcal{T})$ Classes  &Independent &\multirow{ 2}{*}{Constraints}&\multicolumn{2}{c}{Restrict to spin singlet: $(\Phi_{\smalllozenge},\Phi_{\smalltriangleup},\Phi_{\smallhexagon})$ states}\\
\cline{4-5}
($\chi_1\chi_{ST_1}$)\---$(\chi_{\overline{C}_6S}\chi_{S\overline{C}_6}\chi_{\overline{C}_6})_j$ 
&NN parameters&&$(I,S)$ type & $(I\mathcal{T},S\mathcal{T})$ type\\
\hline
$(00)$\---  or $(\pi\pi)$\---$(000)$& {$\underline{\mathrm{Im}a_h,a_p},\mathrm{Im}c_h,c_p$} & $\mathrm{Im}c_h=-\mathrm{Im}d_h,c_p=-d_p$&$(0,\forall,0)$ and $(\pi,\forall,\pi)$ & $(0,\forall,\pi)$ and $(\pi,\forall,0)$\\
$(00)$\---  or $(\pi\pi)$\---$(\pi00)$& {$\mathrm{Im}b_h,b_p,\mathrm{Re}c_h$} &  $\mathrm{Re}c_h=\mathrm{Re}d_h$ &NA&NA\\
$(00)$\---  or $(\pi\pi)$\---$(0\pi\pi)$& {$\underline{\mathrm{Re}a_h},c_h,\mathrm{Im}c_p$} &  $c_h = -d_h^*$, $\mathrm{Im}c_p = \mathrm{Im}d_p$&$(0,\forall,0)$ and $(\pi,\forall,\pi)$&$(0,\forall,\pi)$ and $(\pi,\forall,0)$\\
$(00)$\---  or $(\pi\pi)$\---$(\pi\pi\pi)$& {$\mathrm{Re}b_h,\mathrm{Re}c_p$} &  $\mathrm{Re}c_p=\mathrm{Re}d_p$&NA&NA\\
\hline
$(0\pi)$\---  or $(\pi0)$\---$(0\pi0)_1$& {$\mathrm{Re}b_h,\mathrm{Re}c_p$} &  $\mathrm{Im}b_h = \sqrt{3} \mathrm{Re}b_h$, +2$^{\S}$&NA&NA\\
$(0\pi)$\---  or $(\pi0)$\---$(0\pi0)_3$& {$\mathrm{Re}b_h,\mathrm{Re}c_p$} &  $\mathrm{Re}c_p=\mathrm{Re}d_p$&NA&NA\\
$(0\pi)$\---  or $(\pi0)$\---$(\pi\pi0)_0$& {$\underline{\mathrm{Re}a_h},c_h,\mathrm{Im}c_p$} & $c_h = -d^*_h$, $\mathrm{Im}c_p = \mathrm{Im}d_p$&$(0,\forall,0)$ and $(\pi,\forall,\pi)$& $(0,\forall,\pi)$ and $(\pi,\forall,0)$\\
$(0\pi)$\---  or $(\pi0)$\---$(\pi\pi0)_2$& {$\underline{\mathrm{Re}a_h},c_h,\mathrm{Re}c_p$} &  $\mathrm{Im}a_h = -\sqrt{3}\mathrm{Re}a_h$, +3$^{*}$&$(0,\forall,0)$ and $(\pi,\forall,\pi)$& $(0,\forall,\pi)$ and $(\pi,\forall,0)$\\
\hline
$(0\pi)$\---  or $(\pi0)$\---$(00\pi)$& {$\underline{\mathrm{Im}a_h,a_p},\mathrm{Im}c_h$, $c_p$} &  $\mathrm{Im}c_h =-\mathrm{Im}d_h,c_p=-d_p$&$(0,0,\Theta)^{\text{a}}$ and $(\pi,0,\Theta)^{\text{a}}$& $(0,0,\Theta)^{\text{a}}$ and $(\pi,0,\Theta)^{\text{a}}$\\
$(0\pi)$\---  or $(\pi0)$\---$(\pi0\pi)$& {$\mathrm{Im}b_h,b_p,\mathrm{Re}c_h$} &  $\mathrm{Re}c_h = \mathrm{Re}d_h$&NA&NA\\
\hline
$(0\pi)$\---  or $(\pi0)$\---$(0\pi\pi)_0$& {$\mathrm{Re}b_h,\mathrm{Re}c_p$} & $\mathrm{Im} b_h = -\sqrt{3}\mathrm{Re} b_h$, +2$^{\S\S}$&NA&NA\\
$(0\pi)$\---  or $(\pi0)$\---$(\pi\pi\pi)_0$& {$\underline{\mathrm{Re}a_h},c_h,\mathrm{Re}c_p$} & $\mathrm{Im}a_h = -\sqrt{3} \mathrm{Re}a_h$, +3$^{**}$&$(0,\forall,\pi)$ and $(\pi,\forall,0)$ & $(0,\forall,0)$ and $(\pi,\forall,\pi)$\\
$(0\pi)$\---  or $(\pi0)$\---$(0\pi\pi)_1$&{$\mathrm{Re}b_h,\mathrm{Im}c_p$} &  $\mathrm{Im}c_p=\mathrm{Im}d_p$&NA&NA\\
$(0\pi)$\---  or $(\pi0)$\---$(\pi\pi\pi)_1$& {$\underline{\mathrm{Re}a_h},c_h,\mathrm{Re}c_p$} &  $c_h= -d^*_h$, $\mathrm{Re}d_p = \mathrm{Re}c_p$&$(0,\forall,\pi)$ and $(\pi,\forall,0)$&$(0,\forall,0)$ and $(\pi,\forall,\pi)$\\
\hline
\hline
$(I,S\mathcal{T})$ and $(I\mathcal{T},S)$ Classes
&Independent &\multirow{ 2}{*}{Constraints}&\multicolumn{2}{c}{Restrict to spin singlet: $(\Phi_{\smalllozenge},\Phi_{\smalltriangleup},\Phi_{\smallhexagon})$ states}\\
\cline{4-5}
($\chi_1\chi_{ST_1}$)\---$(\chi_{\overline{C}_6S}\chi_{S\overline{C}_6}\chi_{\overline{C}_6})_j$ 
&NN parameters&&$(I\mathcal{T},S)$ type & $(I,S\mathcal{T})$ type\\
\hline
$(00)$\---  or $(\pi\pi)$\---$(000)$& {$\mathrm{Re}b_h,\mathrm{Im}c_h,c_p$} & $\mathrm{Im}c_h=\mathrm{Im}d_h$, $c_p=d_p$ &NA&NA\\
$(00)$\---  or $(\pi\pi)$\---$(\pi00)$& {$\underline{\mathrm{Re}a_h},\mathrm{Re}c_h$} &  $\mathrm{Re}c_h=-\mathrm{Re}d_h$ &$(0,\frac{\pi}{2},0)$ and $(\pi,\frac{\pi}{2},\pi)$&$(0,\frac{\pi}{2},\pi)$ and $(\pi,\frac{\pi}{2},0)$\\
\multirow{2}{*}{$(00)$\---  or $(\pi\pi)$\---$(0\pi\pi)$}& {$\underline{\mathrm{Im}a_h,\mathrm{Im}a_p},$} &  \multirow{2}{*}{$c_h = d^*_h$, $\mathrm{Im}c_p = -\mathrm{Im}d_p$} &\multirow{2}{*}{$(0,\forall,0)$ and $(\pi,\forall,\pi)$}&\multirow{2}{*}{$(0,\forall,\pi)$ and $(\pi,\forall,0)$}\\
&$\mathrm{Re}b_p, c_h, \mathrm{Im}c_p$ &&&\\
$(00)$\---  or $(\pi\pi)$\---$(\pi\pi\pi)$& {$\underline{\mathrm{Re}a_p}, \mathrm{Im}b_h,\mathrm{Im}b_p,\mathrm{Re}c_p$} &  $\mathrm{Re}c_p=-\mathrm{Re}d_p$ &$(0,\frac{\pi}{2},0)$ and $(\pi,\frac{\pi}{2},\pi)$&$(0,\frac{\pi}{2},\pi)$ and $(\pi,\frac{\pi}{2},0)$\\
\hline
$(0\pi)$\---  or $(\pi0)$\---$(0\pi0)_1$& {$\underline{\mathrm{Re}a_p},\mathrm{Re}b_h$, $\mathrm{Re}b_p,\mathrm{Re}c_p$} & $\mathrm{Im}a_p = \sqrt{3}\mathrm{Re}a_p$,+4$^{\dag}$ &$(0,\frac{\pi}{2},0)$ and $(\pi,\frac{\pi}{2},\pi)$&$(0,\frac{\pi}{2},\pi)$ and $(\pi,\frac{\pi}{2},0)$\\
$(0\pi)$\---  or $(\pi0)$\---$(0\pi0)_3$& {$\underline{\mathrm{Re}a_p},\mathrm{Im}b_h$, $\mathrm{Im}b_p,\mathrm{Re}c_p$} &  $\mathrm{Re}c_p=-\mathrm{Re}d_p$ &$(0,\frac{\pi}{2},0)$ and $(\pi,\frac{\pi}{2},\pi)$&$(0,\frac{\pi}{2},\pi)$ and $(\pi,\frac{\pi}{2},0)$\\
\multirow{2}{*}{$(0\pi)$\---  or $(\pi0)$\---$(\pi\pi0)_0$}& $\underline{\mathrm{Im}a_h,\mathrm{Im}a_p},$ & \multirow{2}{*}{$c_h = d^*_h$, $\mathrm{Im}c_p = -\mathrm{Im}d_p$} &\multirow{2}{*}{$(0,\forall,\pi)$ and $(\pi,\forall,0)$}&\multirow{2}{*}{$(0,\forall,0)$ and $(\pi,\forall,\pi)$}\\
&$\mathrm{Re}b_p,c_h,\mathrm{Im}c_p$&&&\\
\multirow{2}{*}{$(0\pi)$\---  or $(\pi0)$\---$(\pi\pi0)_2$}& $\underline{\mathrm{Re}a_h,\mathrm{Re}a_p},$ &  $\mathrm{Im}a_h = \frac{1}{\sqrt{3}}\mathrm{Re}a_h$,&\multirow{2}{*}{$(0,\frac{\pi}{2},\Theta)^{\text{a}}$  and $(\pi,\frac{\pi}{2},\Theta)^{\text{a}}$}& \multirow{2}{*}{$(0,\frac{\pi}{2},\Theta)^{\text{a}}$ and $(\pi,\frac{\pi}{2},\Theta)^{\text{a}}$}\\
&$\mathrm{Re}b_p,c_h,\mathrm{Re}c_p$& $\mathrm{Im}a_p = \frac{1}{\sqrt{3}}\mathrm{Re}a_p$,+4$^{\ddag}$&&\\
\hline
$(0\pi)$\---  or $(\pi0)$\---$(00\pi)$& {$\mathrm{Re}b_h,\mathrm{Im}c_h,c_p$} &  $\mathrm{Im}c_h =\mathrm{Im}d_h$, $c_p=d_p$ &NA&NA\\
$(0\pi)$\---  or $(\pi0)$\---$(\pi0\pi)$& {$\underline{\mathrm{Re}a_h},\mathrm{Re}c_h$} &  $\mathrm{Re}c_h = -\mathrm{Re}d_h$ &$(0,\frac{\pi}{2},0)$ and $(\pi,\frac{\pi}{2},\pi)$&$(0,\frac{\pi}{2},\pi)$  and $(\pi,\frac{\pi}{2},0)$\\
\hline
$(0\pi)$\---  or $(\pi0)$\---$(0\pi\pi)_0$& {$\underline{\mathrm{Re}a_p},\mathrm{Re}b_h,\mathrm{Re}c_p$} & $\mathrm{Im}a_p = \frac{1}{\sqrt{3}}\mathrm{Re}a_p$,+4$^{\#}$ &$(0,\frac{\pi}{2},0)$ and $(\pi,\frac{\pi}{2},\pi)$&$(0,\frac{\pi}{2},\pi)$ and $(\pi,\frac{\pi}{2},0)$\\
\multirow{2}{*}{$(0\pi)$\---  or $(\pi0)$\---$(\pi\pi\pi)_0$}& {$\underline{\mathrm{Re}a_h,\mathrm{Re}a_p},$} & $\mathrm{Im}a_h = \frac{1}{\sqrt{3}} \mathrm{Re}a_h$, &\multirow{2}{*}{$(0,\frac{\pi}{2},\Theta)^{\text{a}}$  and $(\pi,\frac{\pi}{2},\Theta)^{\text{a}}$}& \multirow{2}{*}{$(0,\frac{\pi}{2},\Theta)^{\text{a}}$ and $(\pi,\frac{\pi}{2},\Theta)^{\text{a}}$}\\
&$\mathrm{Re}b_p,c_h,\mathrm{Re}c_p$&$\mathrm{Im}a_p = -\sqrt{3}\mathrm{Re}a_p$,+4$^{\#\#}$ &&\\
$(0\pi)$\---  or $(\pi0)$\---$(0\pi\pi)_1$&{$\underline{\mathrm{Im}a_p},\mathrm{Im}b_h,\mathrm{Re}b_p,\mathrm{Im}c_p$} &  $\mathrm{Im}c_p=-\mathrm{Im}d_p$ &$(0,\frac{\pi}{2},0)$ and $(\pi,\frac{\pi}{2},\pi)$&$(0,\frac{\pi}{2},\pi)$ and $(\pi,\frac{\pi}{2},0)$\\
\multirow{2}{*}{$(0\pi)$\---  or $(\pi0)$\---$(\pi\pi\pi)_1$}& {$\underline{\mathrm{Im}a_h,\mathrm{Re}a_p},$} &  \multirow{2}{*}{$c_h= d^*_h$, $\mathrm{Re}d_p =- \mathrm{Re}c_p$} &\multirow{2}{*}{$(0,0,\Theta)^{\text{a}}$ and $(\pi,0,\Theta)^{\text{a}}$}&\multirow{2}{*}{$(0,0,\Theta)^{\text{a}}$ and $(\pi,0,\Theta)^{\text{a}}$}\\
&$\mathrm{Im}b_p,c_h,\mathrm{Re}c_p$&&&\\
\hline
\hline
\end{tabular}
\end{table*}

\subsection{Flux classification of NN singlet {\it Ans\"atze}}\label{sec:fluxes}

Here we restrict our consideration to nearest-neighbor (NN), spin-singlet {\it Ans\"atze}. While these states are subject to the classification of PSG,
we will show that several PSG classes can give rise to the same state, as a consequence of restricting to NN hopping
and pairing parameters. A separate argument in terms of the symmetry properties of the fluxes can instead effectively classify these states, thereby providing a more physically transparent interpretation of the PSG parameters.

As explained in detail in Sec.~\ref{sub:fluxes}, for $\mathrm{U(1)}$ {\it Ans\"atze} the gauge invariant SU(2) flux $\Phi_p$ are determined by the $\mathrm{U(1)}$ flux $\phi_p$. A $\mathrm{U(1)}$ flux $\phi_p$ is abelian and its values on the diamond $\smalllozenge$, triangle $\smalltriangleup/ \smalltriangledown$ and hexagon $\smallhexagon$ plaquettes completely determine the spin singlet states at the NN level. The results are summarized in Table \ref{Tab1}. We arrive at these results as follows:
\begin{enumerate}
\item Due to three-fold rotation symmetries\footnote{Note that all the three-fold symmetries can be obtained from $C_3$ composed with translations and even numbers of screw $S$.}, all four triangle faces of a tetrahedron must have the same flux ($\phi_{\smalltriangleup}$ for up tetrahedron and $\phi_{\smalltriangledown}$ for down tetrahedron). Given that the sum of four triangle fluxes must be multiples of $2\pi$, the triangle flux can only be $\phi_{\smalltriangleup,\smalltriangledown} = 0,\pi$ or $\phi_{\smalltriangleup,\smalltriangledown} = \pm\pi/2$ ($\phi_{\smalltriangleup}$ and $\phi_{\smalltriangledown}$ are independent at this point), corresponding to $\Phi_{\smalltriangleup,\smalltriangledown}= \forall$ in the former case and $\Phi_{\smalltriangleup,\smalltriangledown} = \pm\pi/2$ in the latter. 
\item We see from Table \ref{transunderIST} that $\Phi_{\smalltriangleup} = \pi/2$ breaks time-reversal symmetry, and can only appear in the classes $(I\mathcal{T},S,\xcancel{\mathcal{T}})$ and $(I,S\mathcal{T},\xcancel{\mathcal{T}})$. Specifically, $\Phi_{\smalltriangleup} = \Phi_{\smalltriangledown} = \pm\pi/2$ can only appear in class $(I\mathcal{T},S,\xcancel{\mathcal{T}})$, while $\Phi_{\smalltriangleup} = - \Phi_{\smalltriangledown} = \pm\pi/2$ can only appear in class $(I,S\mathcal{T},\xcancel{\mathcal{T}})$.
\item $\phi_{\smalllozenge}$, $\phi_{\smalltriangleup/\smalltriangledown}$, and $\phi_{\smallhexagon}$ are not all independent; they are constrained by \begin{equation}\label{eq:u1_flux_constr}
\phi_{\smalllozenge} = \phi_{\smallhexagon} + \phi_{\smalltriangleup} - \phi_{\smalltriangledown}\,,
\end{equation}
here the minus sign in front of $\phi_{\smalltriangledown}$ is due to the orientation of the normal vector for the ``down'' triangle faces, which is opposite to that for the ``up'' triangle faces.
\item We can then enumerate all the possible values for $\phi_{\smalllozenge,\smalltriangleup,\smalltriangledown}$ to obtain the complete set of NN spin-singlet states: $(\phi_{\smalllozenge},\phi_{\smalltriangleup},\phi_{\smalltriangledown}) = (\phi_1,0,0)$, $(\phi_1,0,\pi)$, $(\phi_1,\frac{\pi}{2},\frac{\pi}{2})$, $(\phi_1,\frac{\pi}{2},-\frac{\pi}{2})$ with $\phi_1 = 0$, $\pi$, or $\frac{\pi}{2}$.
\end{enumerate}

The NN spin-singlet states resulting from these general rules are given in Table \ref{Tab1}. The $\mathrm{U}(1)$ PSG classification correctly produces these states, as shown in Table \ref{MFTparametersU1chiral}.

\begin{table*}[!thb]
\centering
\begin{tabular}{cccc}
\hline
\hline
Type & Class labeled by $(\phi_{\smalllozenge}, \phi_{\smalltriangleup}, \phi_{\smallhexagon})$ & Relation between $\phi_{\smalltriangleup}$ and $\phi_{\smalltriangledown}$&Corresponding $(\Phi_{\smalllozenge},\Phi_{\smalltriangleup},\Phi_{\smalltriangledown},\Phi_{\smallhexagon})$\\
\hline
$(I,S,\mathcal{T})$ & $(0,0,0)${$^a$}, $(\pi,0,\pi)${$^b$} & $\phi_{\smalltriangleup}=\phi_{\smalltriangledown}=0$ &$(0,\forall,\forall,0)$, $(\pi,\forall,\forall,\pi)$ \\
$(I,S,\mathcal{T})$ & $(0,0,\pi)$, $(\pi,0,0)$ & $\phi_{\smalltriangleup}=0, \phi_{\smalltriangledown}=\pi$&$(0,\forall,\forall,\pi)$, $(\pi,\forall,\forall,0)$ \\
$(I\mathcal{T},S,\xcancel{\mathcal{T}})$ & \textcolor{RoyalBlue}{$(0,\frac{\pi}{2},0)$}{$^c$}, \textcolor{Dandelion}{$(\pi,\frac{\pi}{2},\pi)$}{$^d$}, \textcolor{Plum}{$(\frac{\pi}{2},\frac{\pi}{2},\frac{\pi}{2})$}& $\phi_{\smalltriangleup}=\phi_{\smalltriangledown} = \pi/2$& $(0,\frac{\pi}{2},\frac{\pi}{2},0)$, $(\pi,\frac{\pi}{2},\frac{\pi}{2},\pi)$, $(\forall,\frac{\pi}{2},\frac{\pi}{2},\forall)$\\
$(I\mathcal{T},S,\xcancel{\mathcal{T}})$ & \textcolor{Plum}{$(\frac{\pi}{2},0,-\frac{\pi}{2})$} & $\phi_{\smalltriangleup}=0$, $\phi_{\smalltriangledown} = \pi$ & $(\forall,\forall,\forall,\forall)$\\
$(I,\mathcal{T} S,\xcancel{\mathcal{T}})$ & \textcolor{ForestGreen}{$(0,\frac{\pi}{2},\pi)$}, \textcolor{BrickRed}{$(\pi,\frac{\pi}{2},0)$}{$^e$}, \textcolor{TealBlue}{$(\frac{\pi}{2},\frac{\pi}{2},-\frac{\pi}{2})$} & $\phi_{\smalltriangleup}=-\phi_{\smalltriangledown} = \pi/2$&$(0,\frac{\pi}{2},-\frac{\pi}{2},\pi)$, $(\pi,\frac{\pi}{2},-\frac{\pi}{2},0)$, $(\forall,\frac{\pi}{2},-\frac{\pi}{2},\forall)$\\
$(I,\mathcal{T} S,\xcancel{\mathcal{T}})$ & \textcolor{TealBlue}{$(\frac{\pi}{2},0,\frac{\pi}{2})$} & $\phi_{\smalltriangleup}=\phi_{\smalltriangledown} = 0$& $(\forall,\forall,\forall,\forall)$\\
\hline
\hline
\end{tabular}
\caption{The 12 $\mathrm{U(1)}$ spin singlet states at the NN bond level (non-chiral/chiral), classified by the $\mathrm{U(1)}$ fluxes on the diamond plaquettes $\phi_{\smalllozenge}$, triangle plaquettes $\phi_{\smalltriangleup/\smalltriangledown}$, and hexagon plaquettes $\phi_{\smallhexagon}$.\\
$^{a}$ $(0,0,0)$: This {\it Ansatz} is labelled as $[0,0,0]$ in Ref.~\cite{Kim-2008}, and referred to as the uniform state in Ref.~\cite{Burnell-2009}.\\
$^{b}$ $(\pi,0,\pi)$: This {\it Ansatz} is labelled as $[0,0,\pi]$ in Ref.~\cite{Kim-2008}, and as $(\pi,\pi)$ in Table I of Ref.~\cite{Burnell-2009}.\\
$^{c}$ \textcolor{RoyalBlue}{$(0,\pi/2,0)$}: This {\it Ansatz} is referred to as the monopole flux state in Ref.~\cite{Burnell-2009}. In Ref.~\cite{Kim-2008}, this state is labelled as $[\frac{\pi}{2},\frac{\pi}{2},0]$ and is referred to as the uniform flux state.\\
$^{d}$ \textcolor{Dandelion}{$(\pi,\pi/2,\pi)$}: This {\it Ansatz} is labelled as $(\pi/2,\pi)$ in Table I of Ref.~\cite{Burnell-2009}.\\
$^{e}$ \textcolor{BrickRed}{$(\pi,\pi/2,0)$}: This {\it Ansatz} is referred to as the staggered flux state and labelled $[\frac{\pi}{2},-\frac{\pi}{2},0]$ in Ref.~\cite{Kim-2008}.\\
}\label{Tab1}
\end{table*}

A $\mathbb{Z}_2$ {\it Ansatz} generally contains both hopping and pairing terms. The projective symmetry transformation generally contains particle-hole transformation, therefore the $\mathrm{U(1)}$ flux $\phi_p$ is no longer a meaningful quantity and we should only study the SU(2) flux $\Phi_p$. Unlike $\phi_p$ in the $\mathrm{U(1)}$ {\it Ans\"atze} which obeys Eq.~\eqref{eq:u1_flux_constr}, the SU(2) flux $\Phi_p$ instead obeys a slightly more loosened constraint
\begin{equation}
\label{eq:z2_flux_constr}
\Phi_{\smalllozenge} \equiv \Phi_{\smallhexagon}+\Phi_{\smalltriangleup}-\Phi_{\smalltriangledown}\,\,(\text{mod}\,\,\pi)\,.
\end{equation}
As mentioned before, $\Phi_{\smalllozenge}$ and $\Phi_{\smallhexagon}$, when well-defined, can only be $0$ or $\pi$, and $\Phi_{\smalltriangleup/\smalltriangledown}$ when well-defined can only be $\pm \frac{\pi}{2}$. Therefore the only possible values for $(\Phi_{\smalllozenge},\Phi_{\smalltriangleup},\Phi_{\smallhexagon})$ are $(\phi_1,\forall,0)$, $(\phi_1,\forall,\pi)$, $(\phi_1,\frac{\pi}{2},0)$, $(\phi_1,\frac{\pi}{2},\pi)$ for $\phi_1 = 0$ or $\pi$. Again $\Phi_{\smalltriangleup}=\Phi_{\smalltriangledown} = \frac{\pi}{2}$ appears only in an $(I\mathcal{T},S,\xcancel{\mathcal{T}})$ state and $\Phi_{\smalltriangleup} = - \Phi_{\smalltriangledown} = \frac{\pi}{2}$ appears only in an $(I,S\mathcal{T},\xcancel{\mathcal{T}})$ state.

Now it is crucial to realize that, within one $\mathbb{Z}_2$ PSG class, the flux $\Phi_{\smallhexagon}$ can still change: for example, in the class $(I,S)$\---$(0\pi)$\---$(00\pi)$, $\Phi_{\smallhexagon}$ can be  either 0 or $\pi$, depending on the relative strength of $\mathrm{Im}a_h$, $\mathrm{Re}a_p$ and $\mathrm{Im}a_p$. We use the notation $\Phi_{\smallhexagon} = \Theta$ to label such classes, where $\Theta=0$ or $\pi$. 

We note that $\Phi_{\smallhexagon}$ can change between $0$ and $\pi$ within one $\mathbb{Z}_2$ PSG class. This is a consequence of the absence of time-reversal symmetry: recall that time-reversal operation maps $\Phi_{\smallhexagon}$ to $-\Phi_{\smallhexagon}$, and therefore pins $\Phi_{\smallhexagon}$ to a definite value being either $0$ or $\pi$. The lack of time-reversal symmetry allows the value of $\Phi_{\smallhexagon}$ to change in certain PSG classes. The change of $\Phi_{\smallhexagon}$ in such classes is exactly due to the fact that we have several independent spin-singlet parameters. In fact, if only the hopping $\mathrm{Im}a_h$ is present, we will have $\Phi_{\smallhexagon}=0$, whereas if we only have pairing $a_p$, then $\Phi_{\smallhexagon} = \pi$. Remember that the $\mathbb{Z}_2$ PSG classes further split upon adding time-reversal symmetry. The parameters $a_p$ and $\mathrm{Im}a_h$ precisely represent two ways of adding time-reversal symmetry, which leads to two distinct time-reversal invariant PSG classes, with $\Phi_{\smallhexagon}=0$ and $\Phi_{\smallhexagon}=\pi$, respectively. 

Furthermore, remember that the $\mathrm{U(1)}$ PSG classes do not split upon adding time-reversal symmetry. We can then infer the following: upon switching some mean-field parameters, a $\mathbb{Z}_2$ PSG (without imposing time-reversal symmetry) may reduce to a $\mathrm{U(1)}$ PSG class; exactly which $\mathrm{U(1)}$ PSG class depends on the way the mean-field parameters are switched off. In the current gauge choice, quite often a hopping parameter is mapped under projective symmetries to a pairing parameter, making it hard to track the dependence relation between $\mathrm{U(1)}$ and $\mathbb{Z}_2$ {\it Ans\"atze}. In general, it is possible for a $\mathbb{Z}_2$ PSG class to have \emph{several} parent $\mathrm{U(1)}$ PSG classes. 

\subsection{PT theorem and its (in)applicability to $\mathbb{Z}_2$ states}
\label{sec:PT_Z2}

An important constraining principle for singlet parton mean-field theories is the ``PT theorem''~\cite{Bieri-2016}. In many $\mathrm{U}(1)$ spin liquids, time-reversal symmetry $\mathcal{T}$ and a spatial operation $\mathcal{P}$ that reverses loop orientation (e.g., inversion or a mirror) impose equivalent constraints on gauge-invariant fluxes. As a result, breaking $\mathcal{T}$ typically forces the breaking of $\mathcal{P}$, while the combined antiunitary $\mathcal{PT}$ may remain a symmetry.

Within the PSG framework this is naturally phrased in terms of SU(2) Wilson loops. Following Ref.~\cite{Bieri-2016}, the gauge-invariant flux through a loop $p$ is characterized by an SU(2) matrix $W_p$, whose trace defines a flux angle $\Phi_p$ via $\mathrm{Tr}\,W_p = 2\cos\Phi_p$. Under time reversal one has $\Phi_p\to -\Phi_p$, and any spatial operation $\mathcal{P}$ that reverses the loop orientation produces the same sign change. A genuine PT theorem corresponds to the situation in which $\mathcal{T}$ and $\mathcal{P}$ therefore enforce identical constraints on $\Phi_p$ for all relevant loops.

Ref.~\cite{Bieri-2016} shows that this equivalence holds quite generally for $\mathrm{U}(1)$ spin liquids. In that case, all Wilson loops can be gauge-rotated into a common Abelian $\mathrm{U(1)}$ subgroup: the loop matrices commute and the associated flux angles are strictly additive under loop composition. For the pyrochlore lattice this additivity is reflected, for example, in the Abelian loop-flux relation
$\phi_{\smalllozenge} = \phi_{\smallhexagon}+\phi_{\smalltriangleup}-\phi_{\smalltriangledown}$ [Eq.~\eqref{eq:u1_flux_constr}], while for $\mathbb{Z}_2$ IGG the corresponding SU(2) fluxes obey only the weaker congruence
$\Phi_{\smalllozenge} \equiv \Phi_{\smallhexagon}+\Phi_{\smalltriangleup}-\Phi_{\smalltriangledown}\,\,(\text{mod}\,\,\pi)$ [Eq.~\eqref{eq:z2_flux_constr}].

This additivity enforces a one-to-one correspondence between $\mathcal{T}$ and $\mathcal{P}$ constraints, implying that a state which breaks $\mathcal{T}$ necessarily breaks $\mathcal{P}$ as well. Consequently, chiral $\mathrm{U}(1)$ spin liquids must preserve $\mathcal{PT}$, and purely $\mathcal{T}$-breaking states are excluded at the singlet mean-field level.

The situation changes qualitatively for $\mathbb{Z}_2$ spin liquids. Pairing terms render Wilson loops genuinely non-Abelian: flux ``directors'' on different loops need not be collinear, loop matrices need not commute, and flux angles need not be strictly additive~\cite{Bieri-2016}. Therefore, the equivalence between $\mathcal{T}$ and orientation-reversing $\mathcal{P}$ constraints that underlies the PT theorem need not hold for $\mathbb{Z}_2$ IGG. In contrast to the $\mathrm{U}(1)$ case, breaking $\mathcal{T}$ does \emph{not} generically force the breaking of parity, and genuinely $\mathcal{T}$-breaking $\mathbb{Z}_2$ spin liquids are therefore allowed.

In particular, starting from a fully symmetric, gapped singlet $\mathbb{Z}_2$ state with real pairing, one may generate a chiral descendant by introducing an imaginary singlet pairing component (proportional to $\sigma^2$ in our convention), e.g., on-site and on third-neighbor bonds in our construction. This deformation changes the SU(2) flux pattern and breaks $\mathcal{T}$ while remaining compatible with lattice symmetries, realizing a legitimate chiral $\mathbb{Z}_2$ state with no counterpart among $\mathrm{U}(1)$ {\it Ans\"atze} constrained by the PT theorem~\cite{Bieri-2016}.

\section{Nearest-neighbor singlet $\mathrm{U(1)}$ chiral {\it Ans\"atze}}\label{sec:NNU(1)}

\begin{figure*}
  \centering
  \def\twidth{0.245}
\subfloat[  $(0,0,0)$ state\label{bond_subfig:a}]{    \includegraphics[width=\twidth\columnwidth]{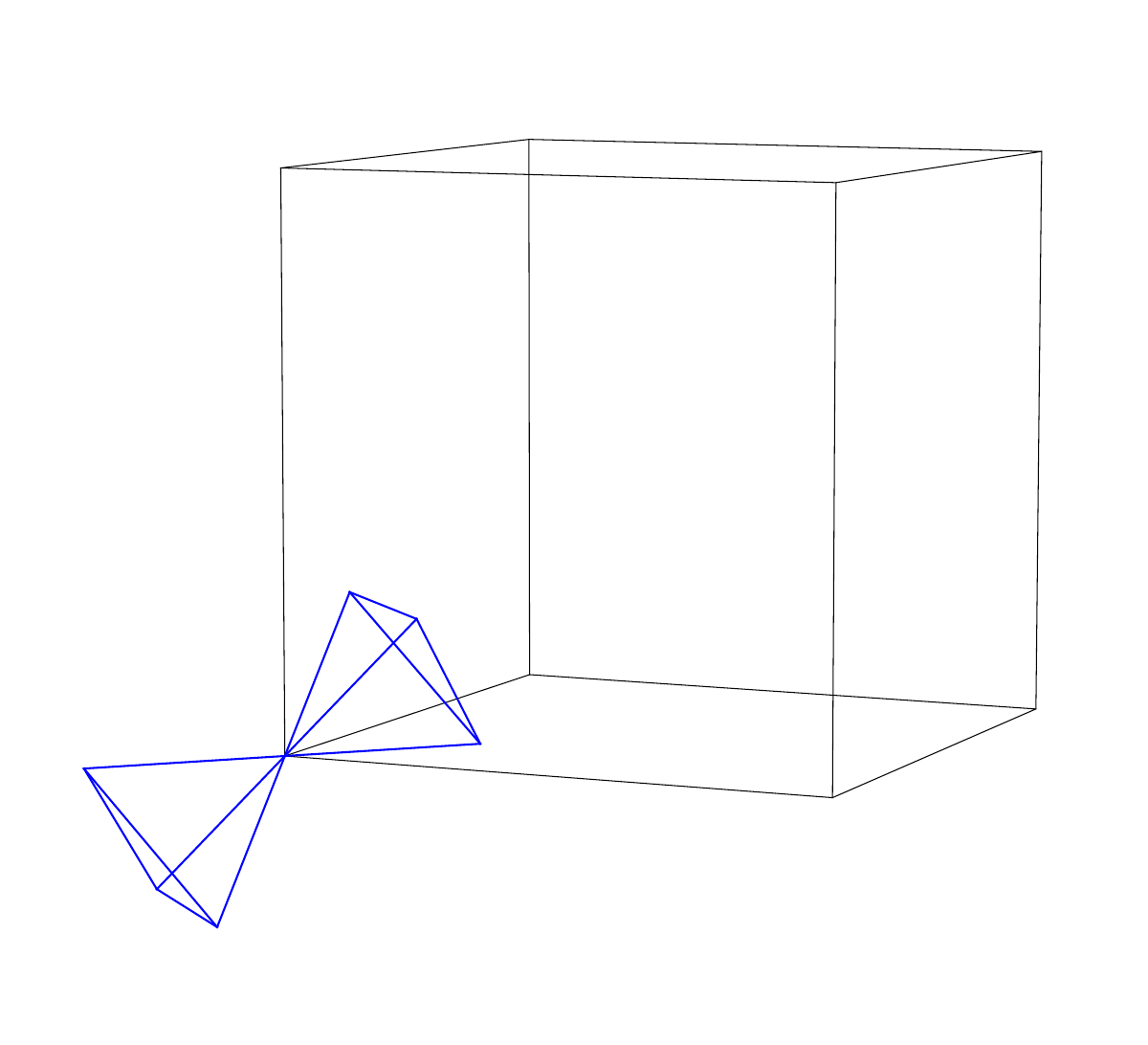}}
  \hspace{-0.2cm}
\subfloat[  $(0,0,\pi)$ state\label{bond_subfig:b}]{
    \includegraphics[width=\twidth\columnwidth]{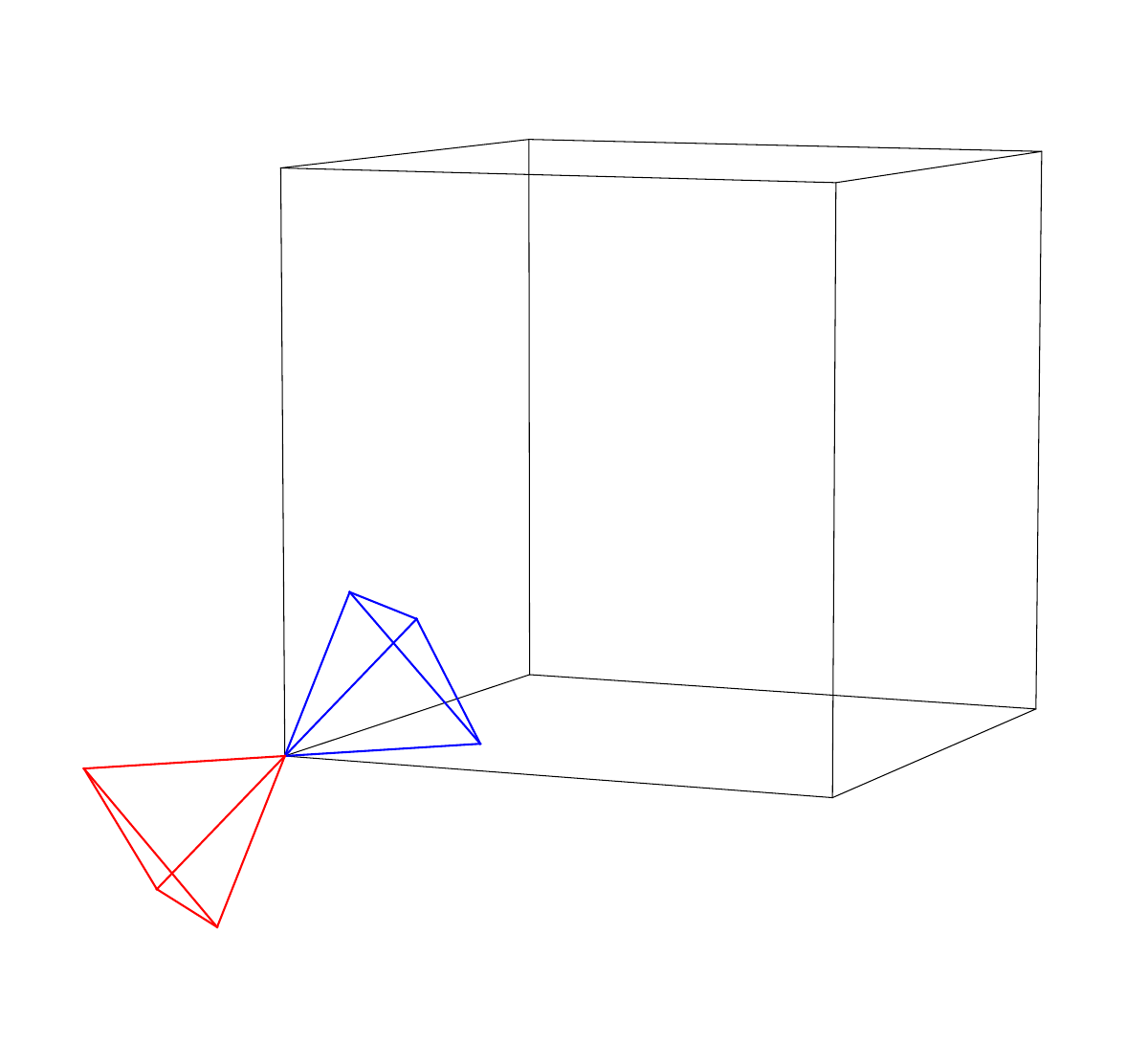}}
 \hspace{-0.2cm}
\subfloat[\textcolor{RoyalBlue}{$(0,\frac{\pi}{2},0)$ state}\label{bond_subfig:c}]{    \includegraphics[width=\twidth\columnwidth]{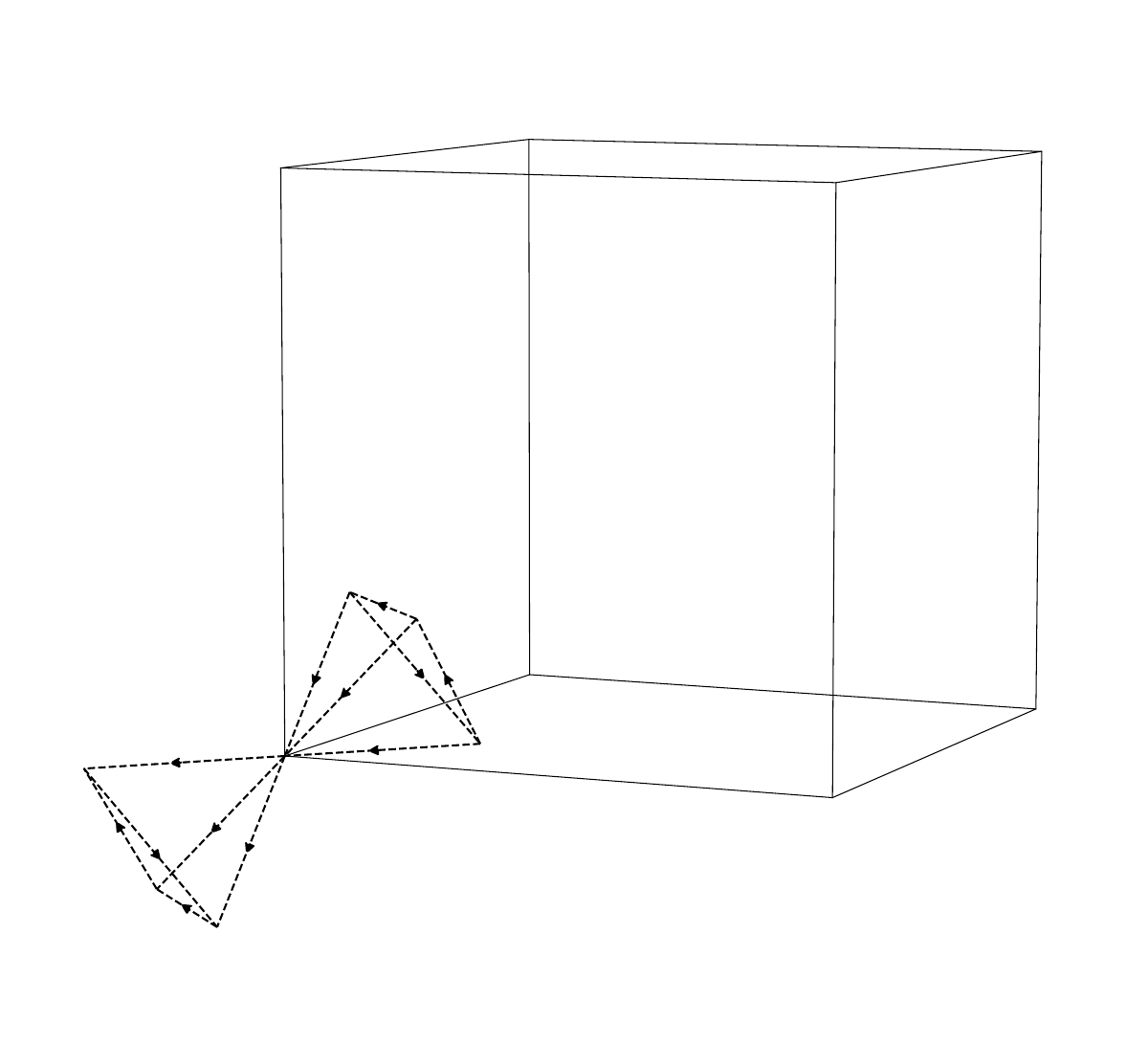}}
  \hspace{-0.2cm}
\subfloat[\textcolor{ForestGreen}{$(0,\frac{\pi}{2},\pi)$ state}\label{bond_subfig:d}]{    \includegraphics[width=\twidth\columnwidth]{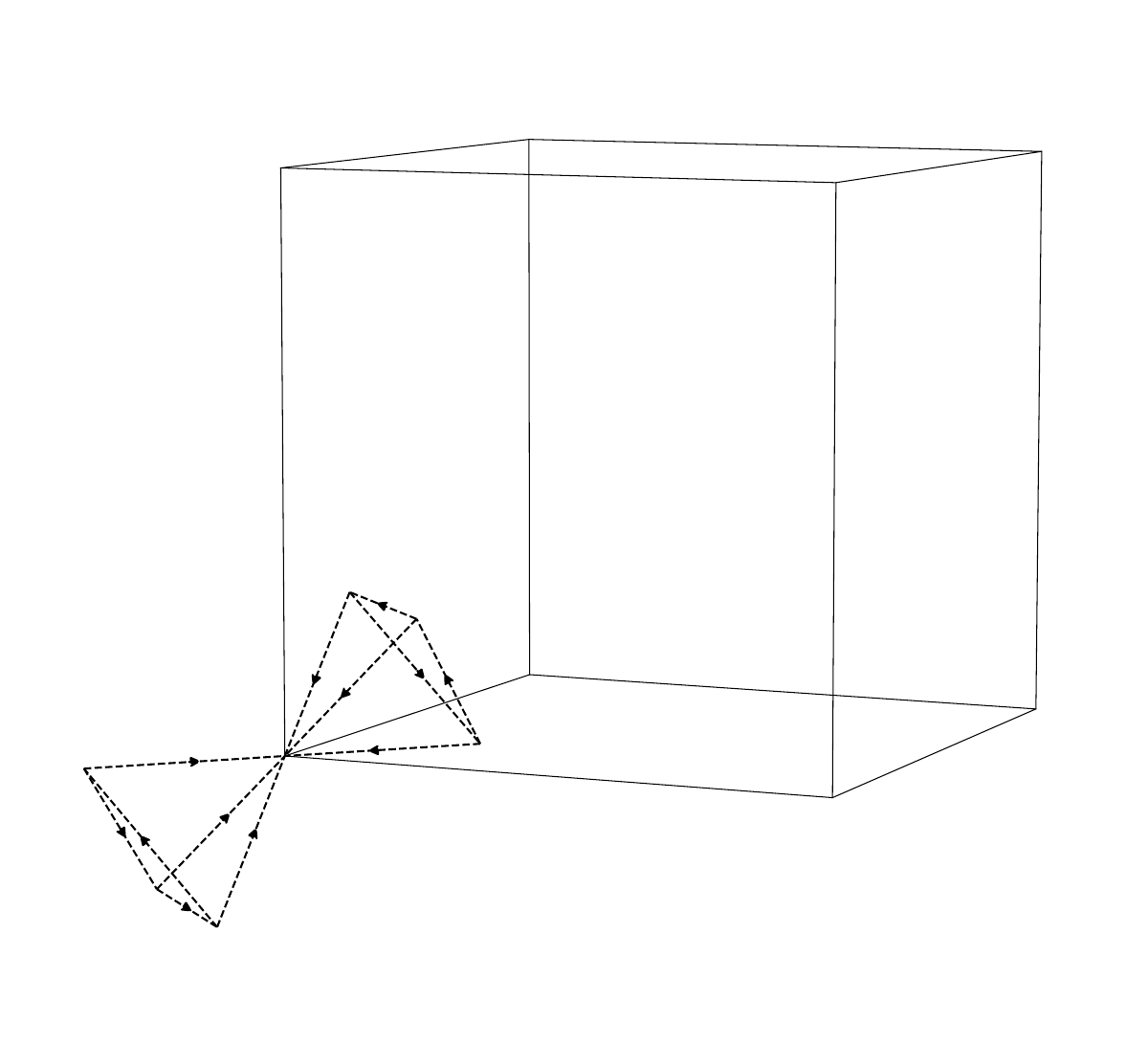}}

\subfloat[$(\pi,0,\pi)$ state \label{bond_subfig:e}]{
    \includegraphics[width=\twidth\columnwidth]{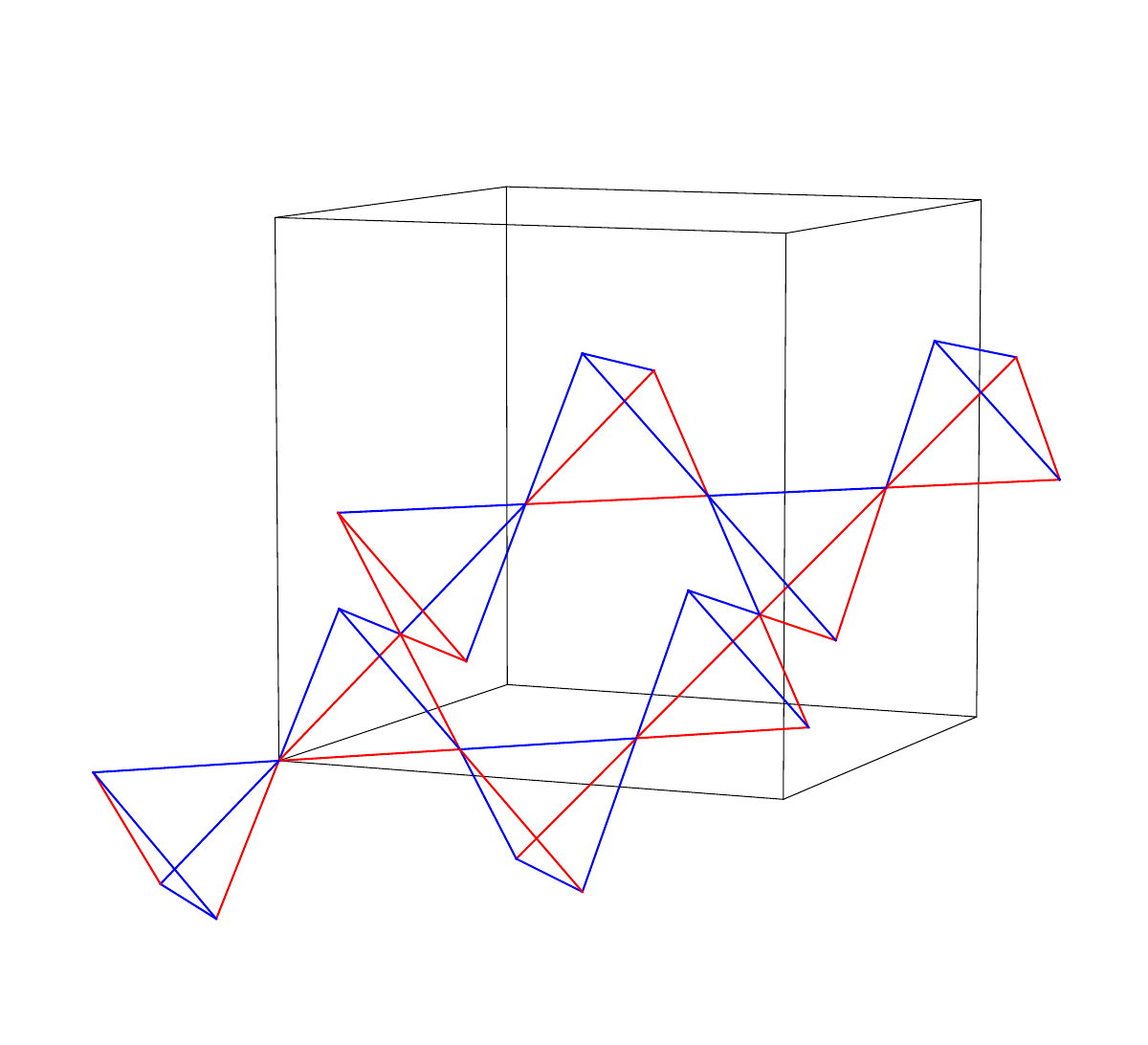}}
 \hspace{-0.2cm}
\subfloat[$(\pi,0,0)$ state\label{bond_subfig:f}]{
    \includegraphics[width=\twidth\columnwidth]{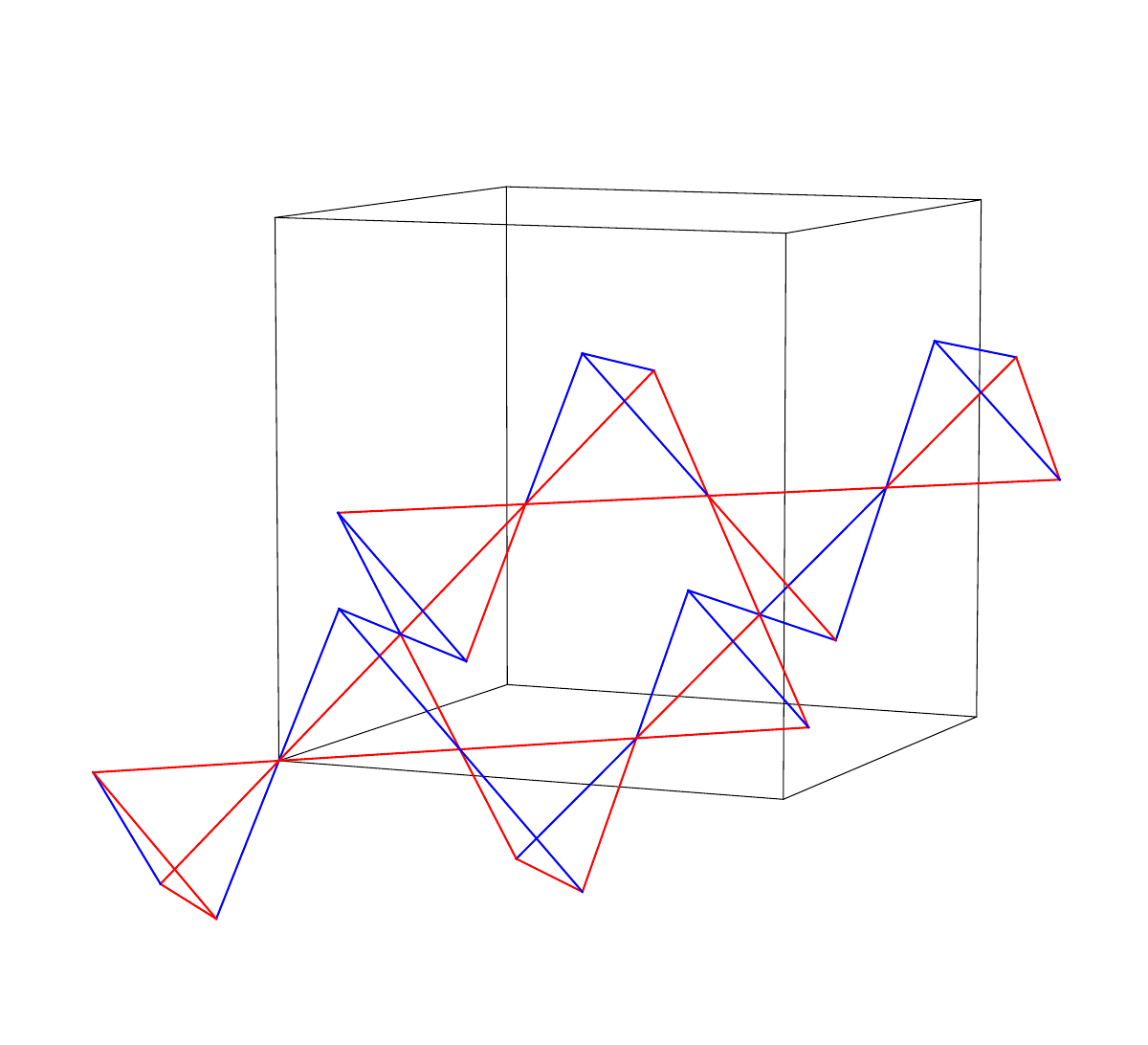}}
  \hspace{-0.2cm}
\subfloat[\textcolor{Dandelion}{$(\pi,\frac{\pi}{2},\pi)$ state}\label{bond_subfig:g}]{    \includegraphics[width=\twidth\columnwidth]{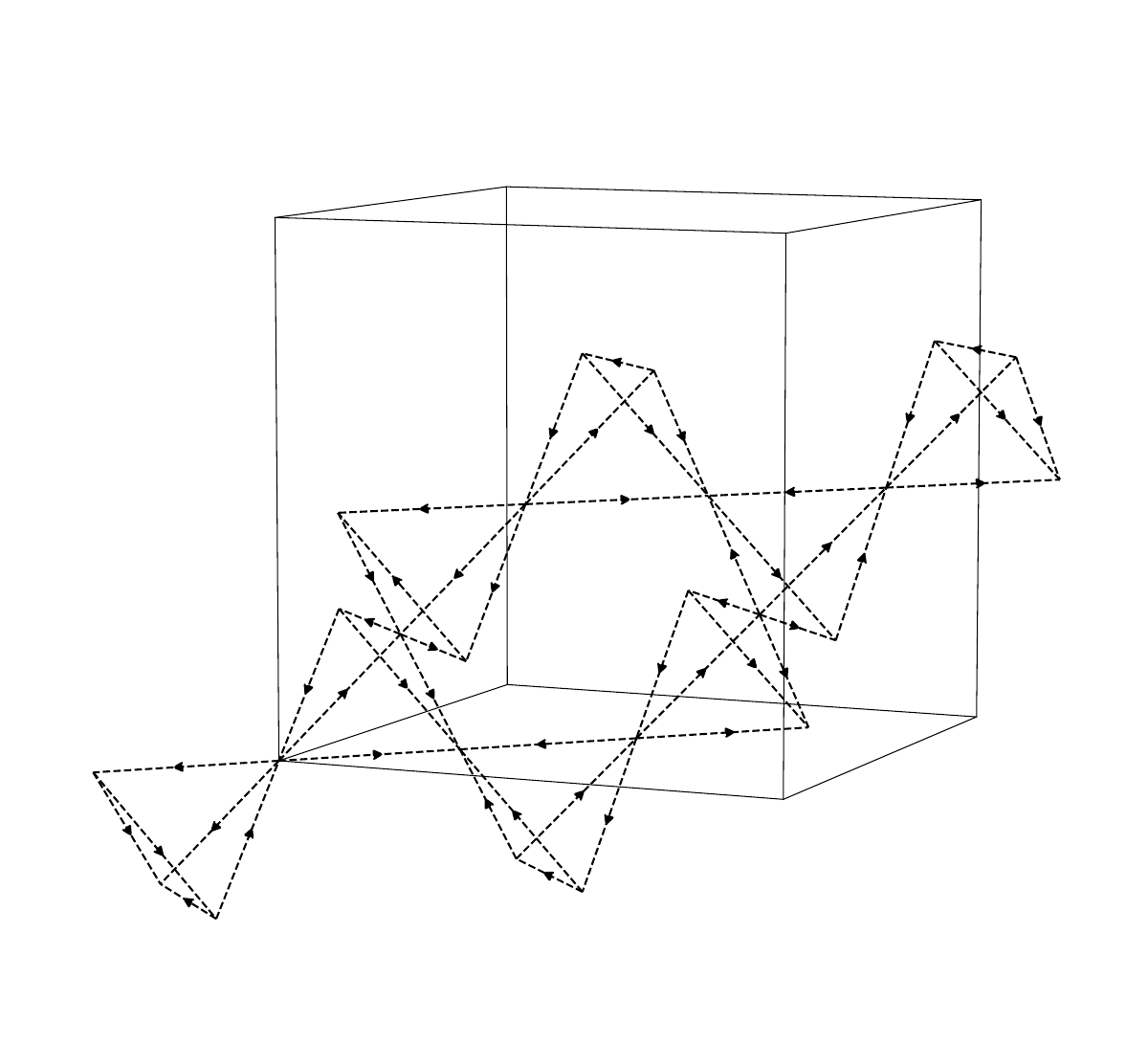}}
  \hspace{-0.2cm}
\subfloat[\textcolor{BrickRed}{$(\pi,\frac{\pi}{2},0)$ state}\label{bond_subfig:h}]{    \includegraphics[width=\twidth\columnwidth]{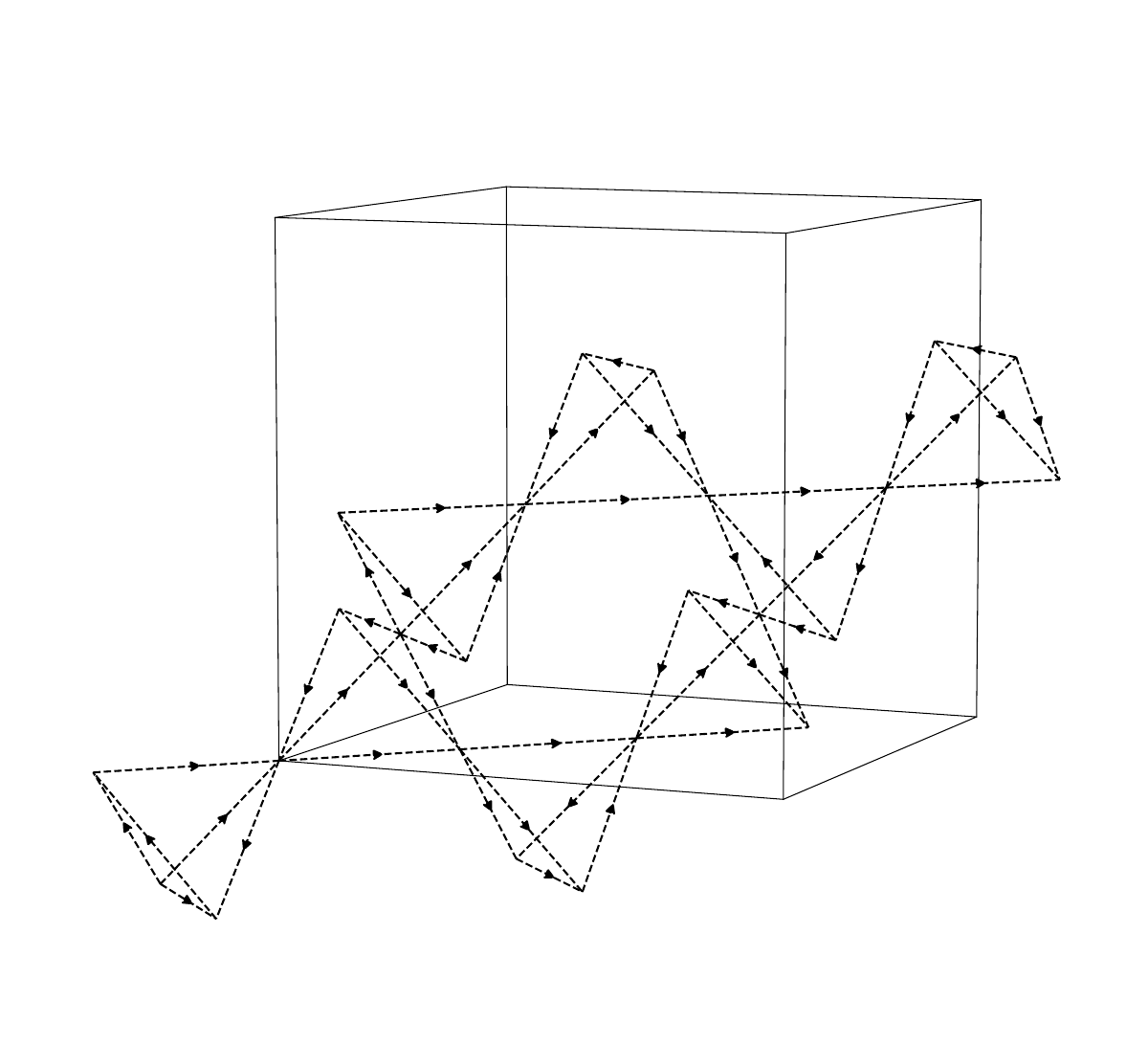}}

\subfloat[\textcolor{Plum}{$(\frac{\pi}{2},\frac{\pi}{2},\frac{\pi}{2})$ state}\label{bond_subfig:i}]{
    \includegraphics[width=\twidth\columnwidth,trim=0cm 0cm 0cm 6cm,clip]{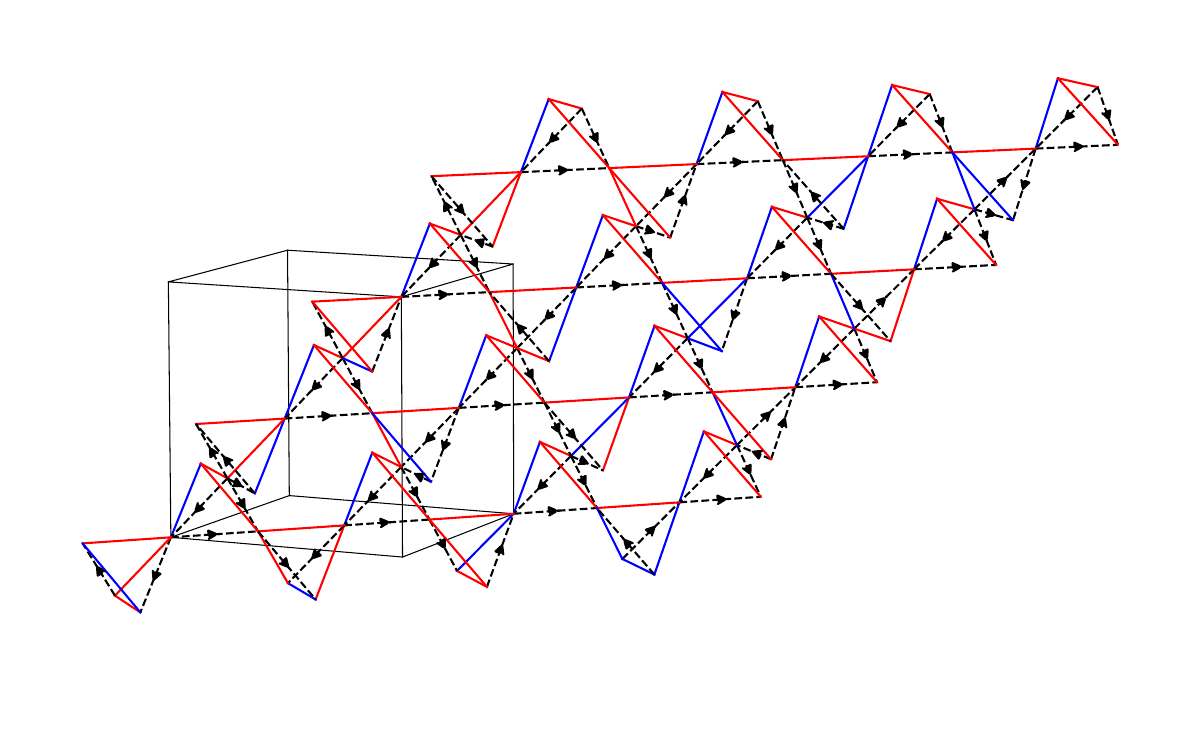}}
 \hspace{-0.2cm}
\subfloat[\textcolor{Plum}{$(\frac{\pi}{2},0,-\frac{\pi}{2})$ state}\label{bond_subfig:j}]{
    \includegraphics[width=\twidth\columnwidth,trim=0cm 0cm 0cm 6cm,clip]{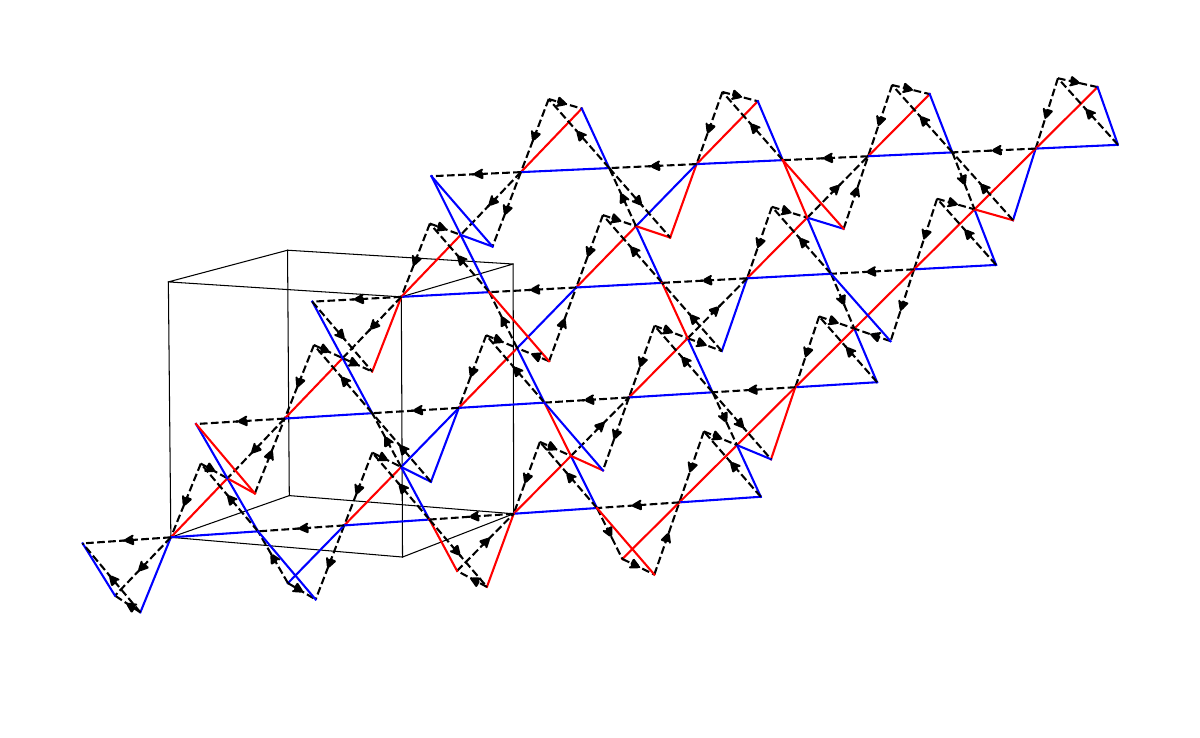}}
  \hspace{-0.2cm}
\subfloat[\textcolor{TealBlue}{$(\frac{\pi}{2},\frac{\pi}{2},-\frac{\pi}{2})$ state}\label{bond_subfig:k}]{    \includegraphics[width=\twidth\columnwidth,trim=0cm 0cm 0cm 6cm,clip]{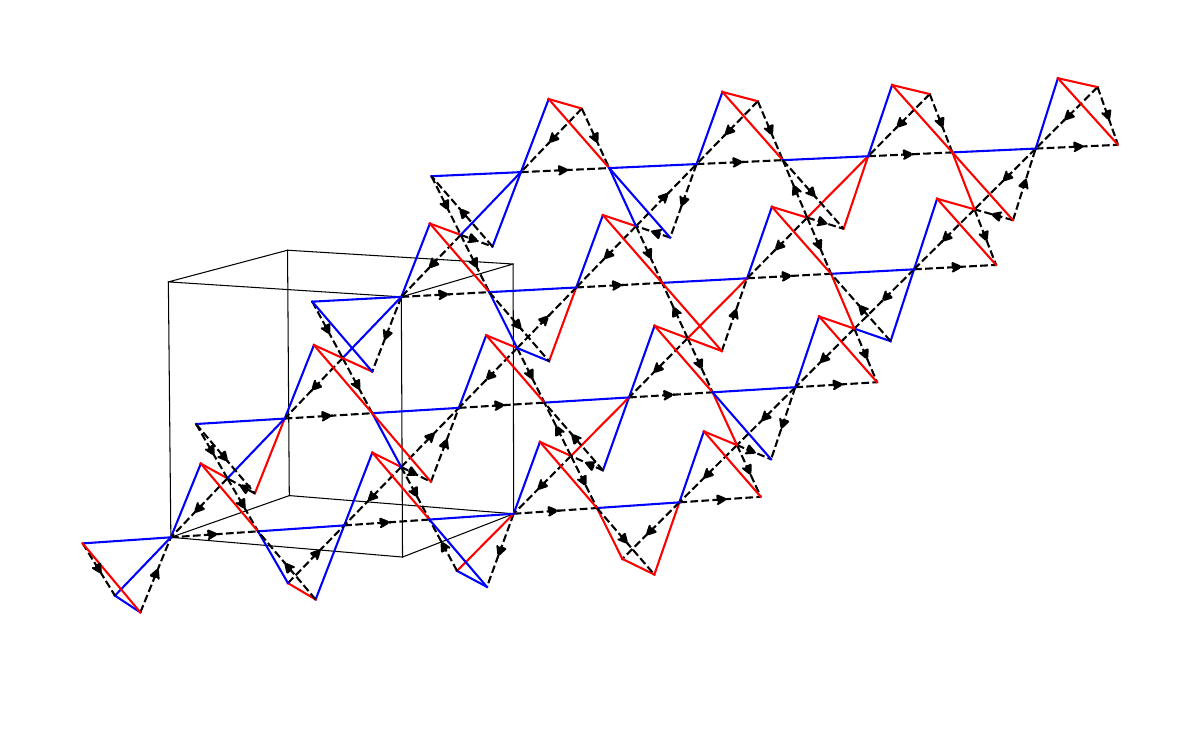}}
  \hspace{-0.2cm}
\subfloat[\textcolor{TealBlue}{$(\frac{\pi}{2},0,\frac{\pi}{2})$ state}\label{bond_subfig:l}]{    \includegraphics[width=\twidth\columnwidth,trim=0cm 0cm 0cm 6cm,clip]{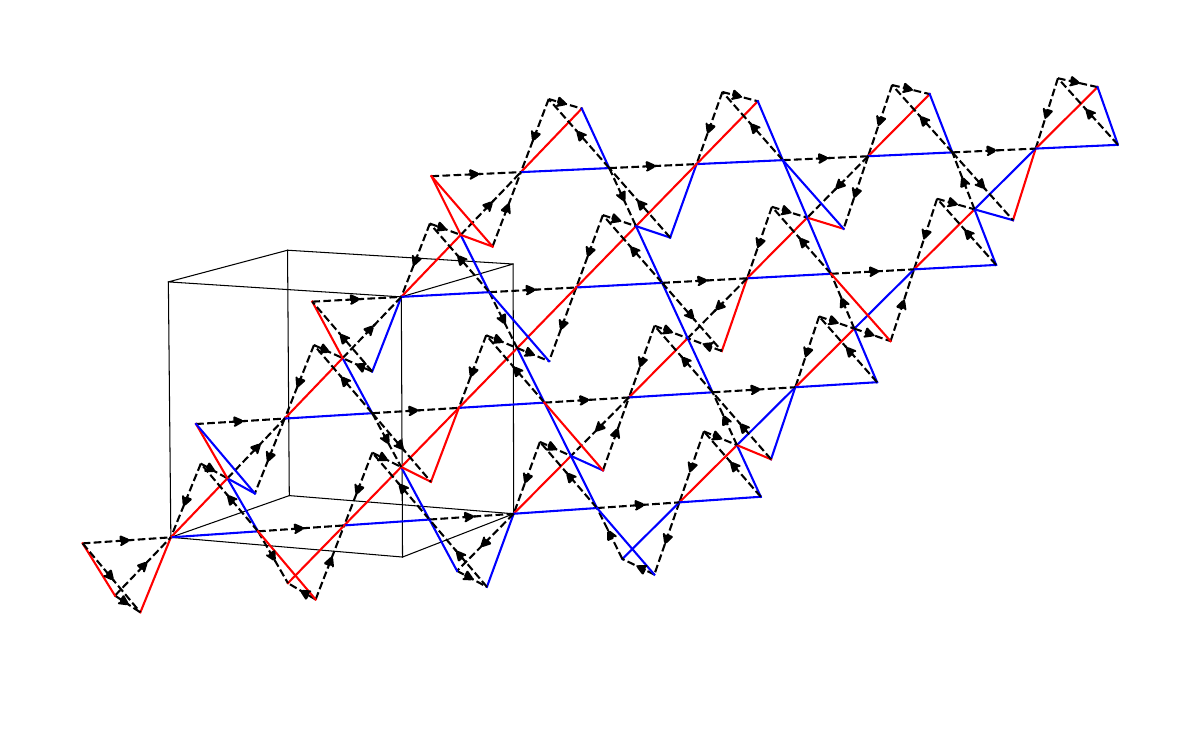}}
  \caption{Nearest-neighbor hopping patterns for the 12 $\mathrm{U}(1)$ spin-singlet {\it Ans\"atze} listed in Table~\ref{Tab1}. Solid bonds represent real hopping amplitudes and dashed bonds purely imaginary ones. Blue (red) bonds denote positive (negative) real hoppings, while a dashed black arrow denotes a purely imaginary hopping with phase $+i$ along the arrow direction.}\label{bond_configurations_12_states}
\end{figure*}

The PSG classification in principle allows us to construct symmetric mean-field {\it Ans\"atze} for chiral spin liquids up to arbitrary range, while physically local {\it Ans\"atze} are expected to be dominated by short-range hopping and pairing amplitudes. In this section, we restrict ourselves to the 12 nearest-neighbor $\mathrm{U(1)}$ spin-singlet {\it Ans\"atze}, including the eight chiral states, and conduct a detailed
study of their spectral signatures.
We emphasize that these are spin-rotation-invariant spin-liquid
{\it Ans\"atze}. They therefore do not describe antiferromagnetic states
with static dipolar order: after Gutzwiller projection the corresponding
trial wave functions remain nonmagnetic, with
\begin{equation}
    \langle \mathbf{S}_i\rangle = 0 .
\end{equation}
Their chirality resides instead in time-reversal-odd flux and scalar-chirality
patterns. A finite magnetic moment would require enlarging the variational
space to include spin-rotation-breaking mean fields, or an instability toward
a magnetically ordered phase~\cite{Iqbal-2016a}, neither of which is part of the spin-singlet PSG classification considered here. The choice of $\mathrm{U(1)}$ states is motivated by quantum spin ice, where the low-energy theory contains an emergent $\mathrm{U(1)}$ gauge field. As we will show, the 12 NN singlet states already exhibit rich physics, with unique band structures and gauge field dependence. They lead to distinct features in their equal-time spin structure factors which serve as an initial diagnostic in distinguishing spin liquid states.

\subsection{Spectrum of the 12 $\mathrm{U(1)}$ singlet states}
\label{sec:spec_12}

We first show in Fig.~\ref{bond_configurations_12_states} the phase of the NN singlet bands of the 12 states. The bond phases are chosen with a particular gauge fixing; the invariant data are the SU(2) phases on the triangle, hexagon, and diamond plaquettes as labeled in Table~\ref{Tab1}. 

\begin{figure*}[t]
  \centering
  \def\twidth{0.245}
\subfloat[  $(0,0,0)$ state\label{spectrum_subfig:a}]{    \includegraphics[width=\twidth\columnwidth]{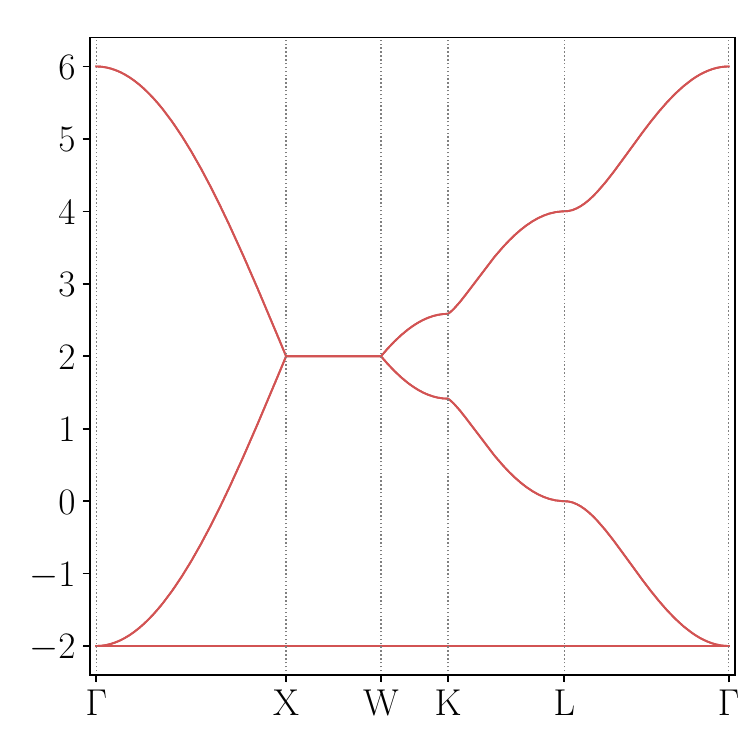}}
  \hspace{-0.2cm}
\subfloat[  $(0,0,\pi)$ state\label{spectrum_subfig:b}]{
    \includegraphics[width=\twidth\columnwidth]{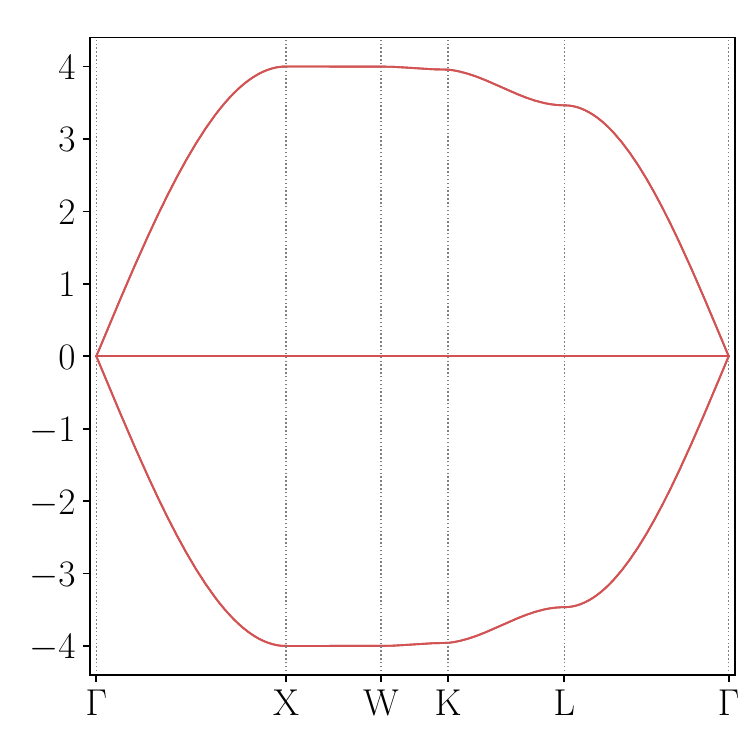}}
 \hspace{-0.2cm}
\subfloat[\textcolor{RoyalBlue}{$(0,\frac{\pi}{2},0)$ state}\label{spectrum_subfig:c}]{    \includegraphics[width=\twidth\columnwidth]{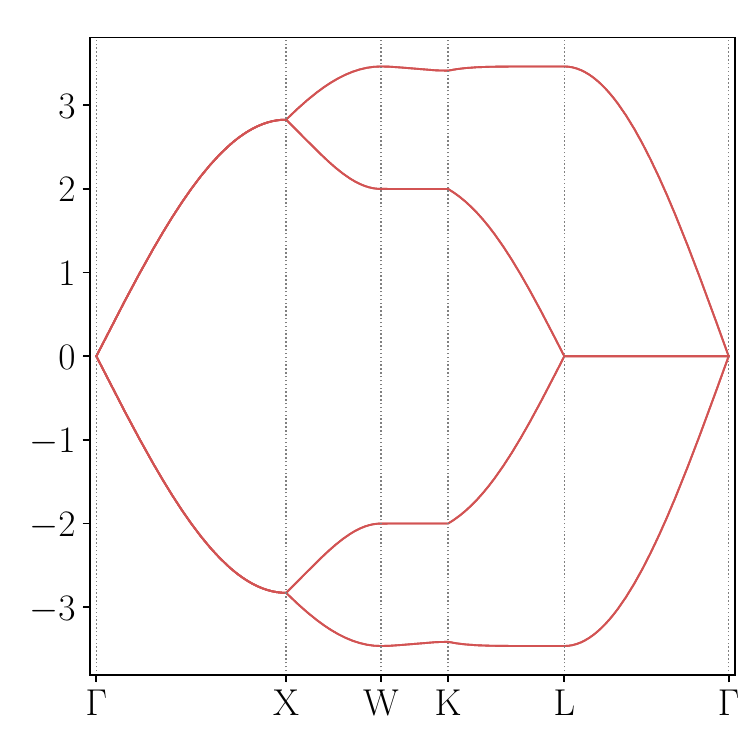}}
  \hspace{-0.2cm}
\subfloat[\textcolor{ForestGreen}{$(0,\frac{\pi}{2},\pi)$ state}\label{spectrum_subfig:d}]{    \includegraphics[width=\twidth\columnwidth]{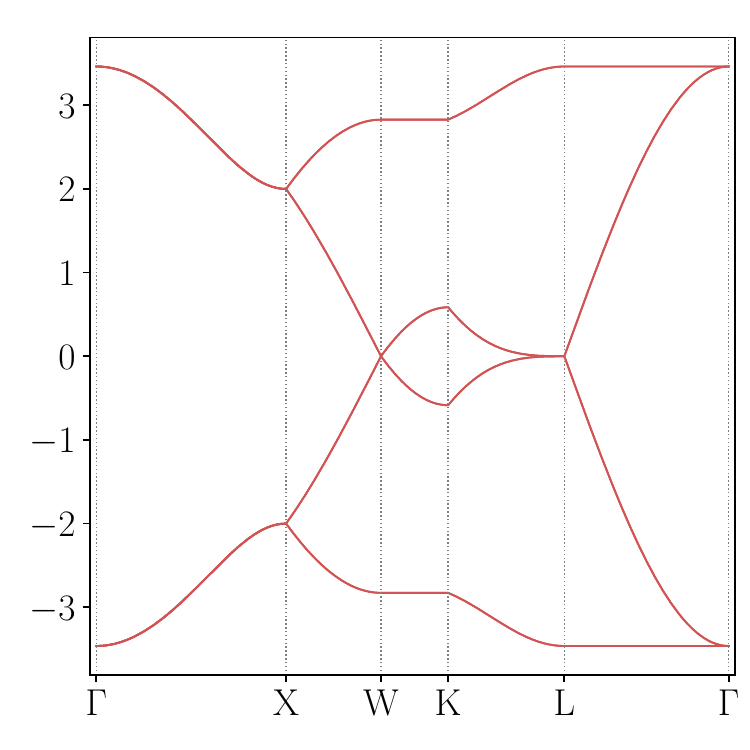}}

\subfloat[$(\pi,0,\pi)$ state \label{spectrum_subfig:e}]{
    \includegraphics[width=\twidth\columnwidth]{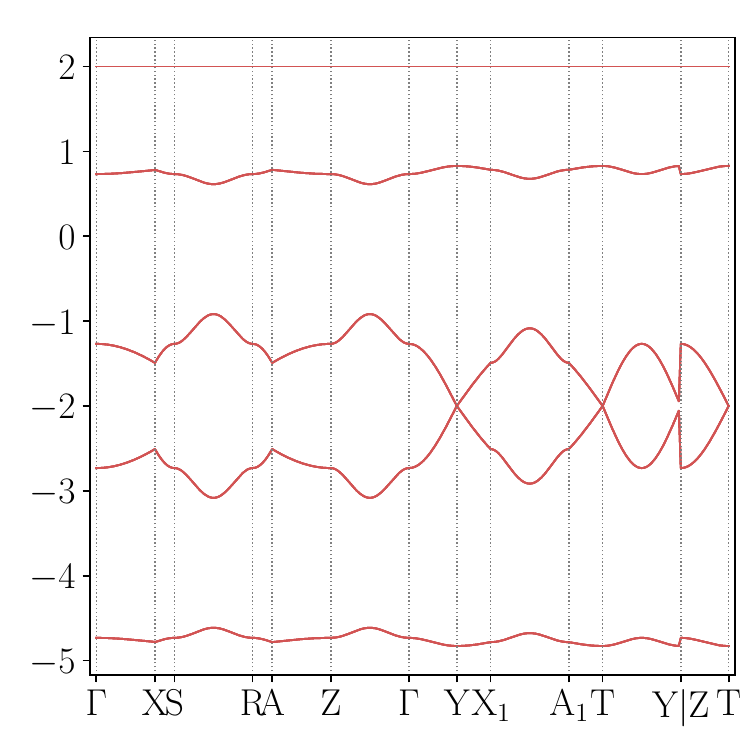}}
 \hspace{-0.2cm}
\subfloat[$(\pi,0,0)$ state\label{spectrum_subfig:f}]{
    \includegraphics[width=\twidth\columnwidth]{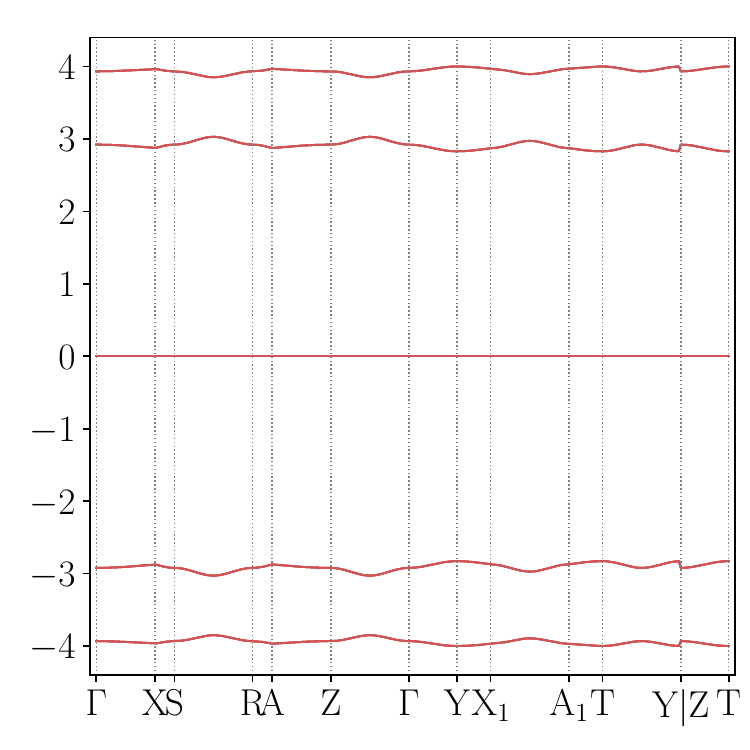}}
  \hspace{-0.2cm}
\subfloat[\textcolor{Dandelion}{$(\pi,\frac{\pi}{2},\pi)$ state}\label{spectrum_subfig:g}]{    \includegraphics[width=\twidth\columnwidth]{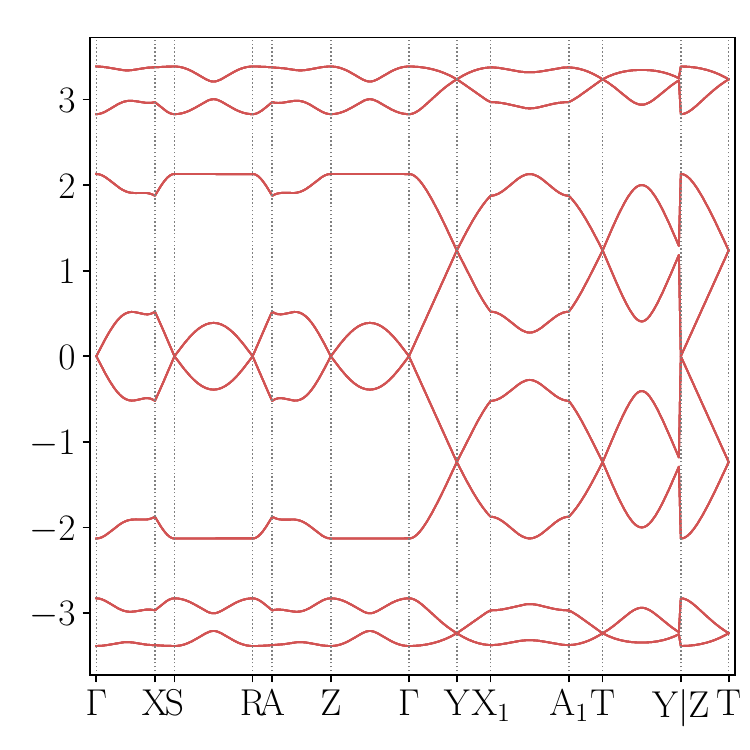}}
  \hspace{-0.2cm}
\subfloat[\textcolor{BrickRed}{$(\pi,\frac{\pi}{2},0)$ state}\label{spectrum_subfig:h}]{    \includegraphics[width=\twidth\columnwidth]{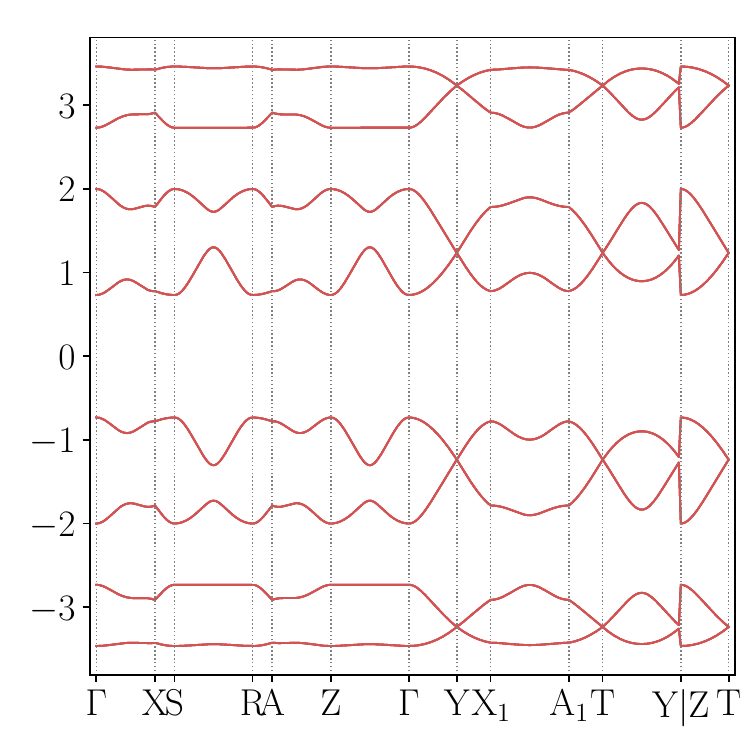}}

\subfloat[\textcolor{Plum}{$(\frac{\pi}{2},\frac{\pi}{2},\frac{\pi}{2})$ state}\label{spectrum_subfig:i}]{
    \includegraphics[width=\twidth\columnwidth]{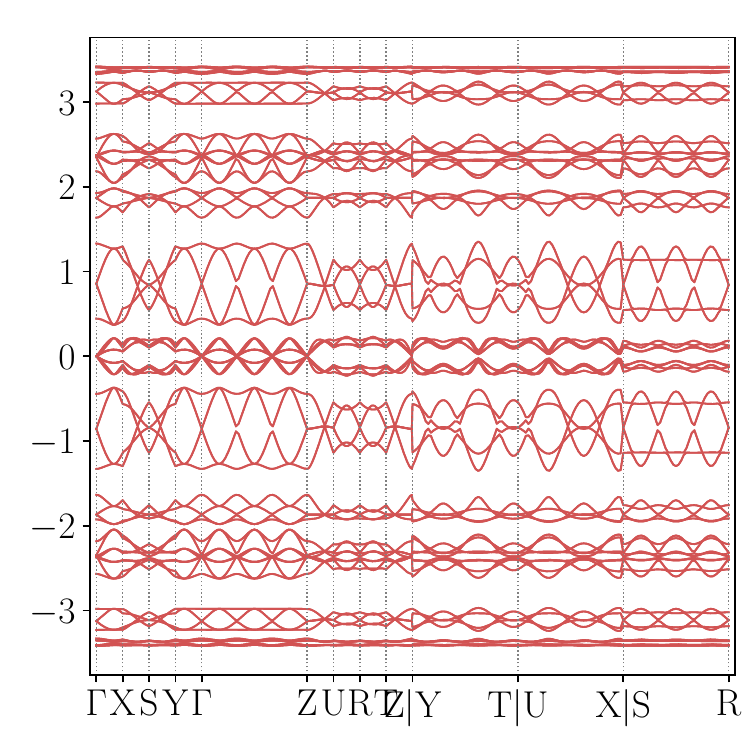}}
 \hspace{-0.2cm}
\subfloat[\textcolor{Plum}{$(\frac{\pi}{2},0,-\frac{\pi}{2})$ state}\label{spectrum_subfig:j}]{
    \includegraphics[width=\twidth\columnwidth]{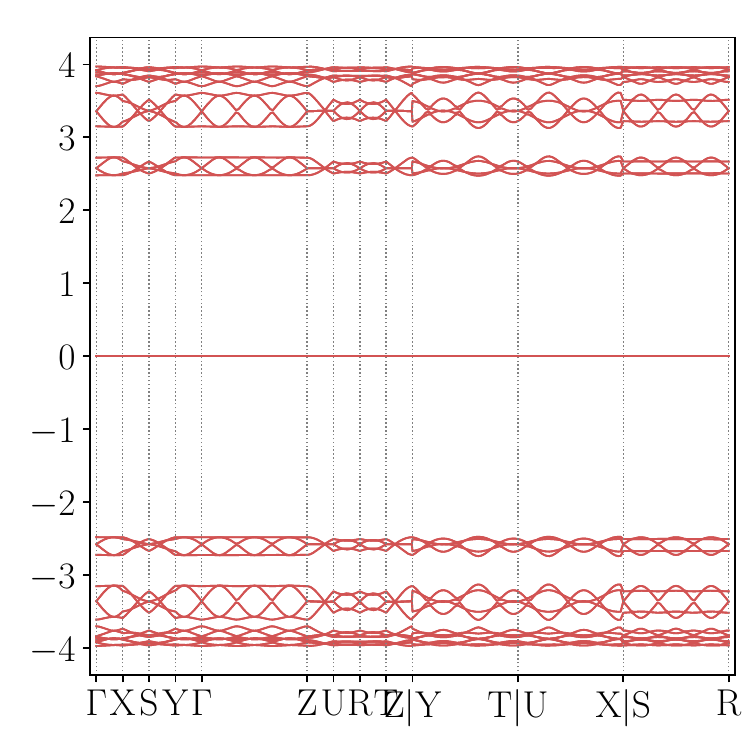}}
  \hspace{-0.2cm}
\subfloat[\textcolor{TealBlue}{$(\frac{\pi}{2},\frac{\pi}{2},-\frac{\pi}{2})$ state}\label{spectrum_subfig:k}]{    \includegraphics[width=\twidth\columnwidth]{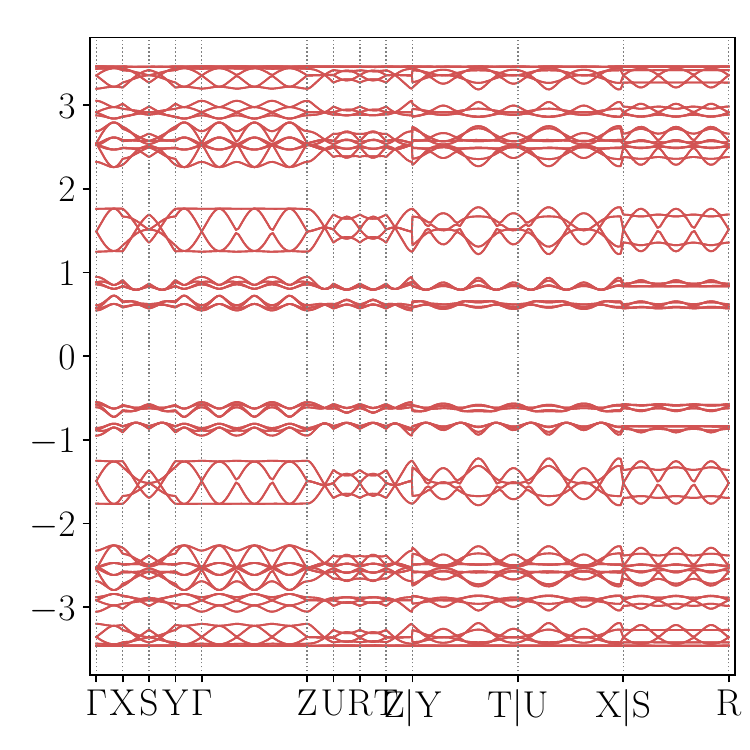}}
  \hspace{-0.2cm}
\subfloat[\textcolor{TealBlue}{$(\frac{\pi}{2},0,\frac{\pi}{2})$ state}\label{spectrum_subfig:l}]{    \includegraphics[width=\twidth\columnwidth]{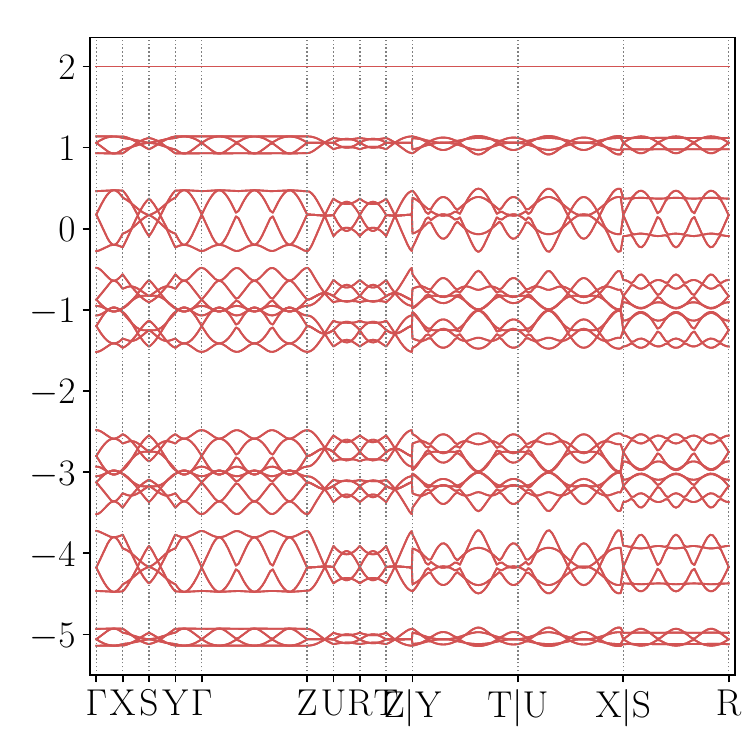}}
  \caption{Spinon band structures for the 12 nearest-neighbor spin-singlet $\mathrm{U}(1)$ {\it Ans\"atze} on the pyrochlore lattice listed in Table~\ref{Tab1}. The spectra are plotted along the high-symmetry momentum paths shown in Fig.~\ref{fig:BZs} for the corresponding parton Brillouin zones. Eight of these states are chiral, while the remaining four are fully symmetric.} \label{spectraof12states}
\end{figure*}

We plot the band structure for the 12  NN singlet hopping states with chosen bond phase of Table \ref{Tab1}, assuming that the hopping amplitude is unity. Their spectra along high symmetry paths are shown in Fig.~\ref{spectraof12states}: the high symmetry points for the $0$-flux, $\pi$-flux, and $\pi/2$-flux {\it Ans\"atze} are given in Fig.~\ref{fig:BZs}. First we note that several spectra exhibit spectral symmetry upon reverting the sign of energy $E\leftrightarrow -E$. This can happen whenever the inversion $I$ exists as a (projective) symmetry that conjugates the spinon $\mathrm{U(1)}$ charge $f\rightarrow f^\dag$, or the chiral inversion $I\mathcal{T}$ exists as a symmetry that does not conjugate the spinon charge $f\rightarrow e^{i\theta} f$ (here a possible nonzero phase $\theta$ is allowed by $I\mathcal{T}$).

On the other hand, the spectral symmetry $E(\mathbf{k})\leftrightarrow -E(\mathbf{k})$ is manifestly absent for the $(\frac{\pi}{2},\frac{\pi}{2},\frac{\pi}{2})$ and the $(\frac{\pi}{2},0,-\frac{\pi}{2})$ states. In fact, these two states are in the $\frac{\pi}{2}\text{-}(01)\text{-}1\text{-}*$ classes with $m_{\overline{C}_6} = 0$ and symmetry $I\mathcal{T}$. This means that they have the property $W_{I\mathcal{T}} \mathcal{H}(\mathbf{k}) W_{I\mathcal{T}}^\dag = - \mathcal{H}(-\mathbf{k})$, here the minus sign in $-\mathbf{k}$ comes from the effect of inversion, whereas the minus sign in front of $- \mathcal{H}(-\mathbf{k})$ comes from the effect of $\mathcal{T}$ (after factoring out this minus sign, $\mathcal{T}$ operates as a unitary symmetry in PSG). This means that the energy has the usual particle-hole symmetry $E(\mathbf{k}) = - E(-\mathbf{k})$, which is not necessarily manifest in the plots.

A second notable feature is the existence of flat bands for the $(0,0,0)$, $(0,0,\pi)$, $(\pi,0,0)$, and the $(\frac{\pi}{2},0,-\frac{\pi}{2})$. We have checked that these flat bands acquire dispersion when further hoppings are included. The state $(0,\frac{\pi}{2},0)$ has a gapless nodal star band structure, which is robust against adding further hopping parameters as long as the projective symmetry holds.

\subsection{Monopole fluxes and two notable states}\label{sec:mfs_and_sfs}

Among the 12 NN $\mathrm{U(1)}$ singlet {\it Ans\"atze} classified in Table~\ref{TB:monopole_and_staggered}, chirality can arise only through a nontrivial triangle flux. As discussed in Sec.~\ref{sec:fluxes}, the SU(2) flux on diamond and hexagon plaquettes is restricted to $0$ or $\pi$, while the only time-reversal-odd possibility at the NN level is
\begin{equation}
\Phi_{\smalltriangleup/\smalltriangledown} = \pm \frac{\pi}{2}\,.
\end{equation}
Because the four triangular faces of a tetrahedron form a closed surface, their $\mathrm{U(1)}$ fluxes obey the lattice constraint \eqref{eq:u1_flux_constr}, so assigning $|\phi_{\smalltriangleup/\smalltriangledown}| = \pi/2$ corresponds to inserting a quantized $2\pi$ gauge flux through each tetrahedron. In this sense, the $\pi/2$ triangle flux may be interpreted as threading an effective monopole of the emergent $\mathrm{U(1)}$ gauge field through every tetrahedron.

Once this local monopole flux structure is imposed, there are two maximally symmetric global realizations compatible with the space group and the chiral symmetry classes identified in Sec.~\ref{symmetries}. First, one may assign the same chirality to all tetrahedra,
\begin{equation}
\phi_{\smalltriangleup} = \phi_{\smalltriangledown} = \frac{\pi}{2}\,,
\end{equation}
yielding the \emph{monopole flux state}~\cite{Burnell-2009}. Alternatively, one may assign opposite chiralities to up- and down-tetrahedra,
\begin{equation}
\phi_{\smalltriangleup} = -\phi_{\smalltriangledown} = \frac{\pi}{2}\,,
\end{equation}
yielding the \emph{staggered flux state}~\cite{Kim-2008}.  

Although both states share the same local monopole magnitude $|\Phi_{\smalltriangleup/\smalltriangledown}|=\pi/2$, they differ in the global arrangement of monopole charge and therefore belong to distinct chiral symmetry families. The monopole flux state preserves chiral inversion $I\mathcal{T}$ and nonchiral screw $S$, while the staggered flux state preserves inversion $I$ and chiral screw $S\mathcal{T}$. Their symmetry properties are summarized in Table~\ref{TB:monopole_and_staggered}. 

Importantly, at the NN singlet level these states appear across several PSG classes (see Table~\ref{MFTparametersU1chiral}) as an artifact of restricting to NN hoppings. All such realizations share $n_{ST_1}=0$ and $n_{\overline{C}_6S}=1$, irrespective of the values of $m_{\overline{C}_6}, m_S = 0,1$. The physical distinction between the two states is therefore not encoded in short-range energetics but in the structure of their emergent gauge flux and its consequences for the spinon band topology. As we now show, the two realizations lead to qualitatively different low-energy spinon and gauge-field regimes.

\subsubsection{Monopole flux state and nodal line spin liquids}\label{sec:monopole-flux}

The $(0,\pi/2,0)$ state corresponds to the monopole flux state introduced in Ref.~\cite{Burnell-2009} and further analyzed in Ref.~\cite{PhysRevB.104.054401}. In this configuration all triangular plaquettes carry the same chiral flux $\pi/2$, while the hexagon flux vanishes. As a result, each tetrahedron encloses a net $2\pi$ gauge flux, which can be interpreted as an effective monopole of the emergent $\mathrm{U(1)}$ gauge field located at the center of the tetrahedron. The flux pattern is therefore uniform across the lattice.

At the mean-field level this flux pattern produces a gapless spinon spectrum characterized by nodal lines in momentum space. The origin of these nodal structures can be traced to the combination of lattice symmetry and the projective implementation of inversion and time-reversal operations. In particular, the preserved symmetry $I\mathcal{T}$ constrains the spinon Hamiltonian such that band crossings along one-dimensional manifolds are protected, leading to nodal loops rather than isolated Dirac points. The resulting nodal structure forms characteristic star-shaped patterns in the Brillouin zone, 
and was referred to as a nodal star $\mathrm{U(1)}$ spin liquid in Ref.~\cite{PhysRevB.104.054401}. Here we claim that the nodal star structure exists for all $\mathrm{U(1)}$ {\it Ans\"atze} in the $(I\mathcal{T},S)$ chiral class $0$-$(11)$-$0$-$1$: it is protected by the projective symmetry group, and is robust against the addition of arbitrarily long-range neighbor hoppings. The proof exactly parallels the one in Ref.~\cite{PhysRevB.104.054401}.

The low-energy theory therefore consists of gapless fermionic spinons residing on these nodal lines coupled to a dynamical $\mathrm{U(1)}$ gauge field. This combination realizes a three-dimensional gapless $\mathrm{U(1)}$ spin liquid. Compared with conventional Coulomb phases with gapped matter, the presence of nodal spinons modifies the low-energy density of states and produces additional contributions to thermodynamic and correlation functions. Notably, the nodal-line density of states vanishes linearly, $\rho(E)\propto |E|$, which implies parametrically weak Landau damping of the emergent photon and supports the stability of the deconfined $\mathrm{U(1)}$ gauge field.
The low-energy field theory is minimally described by a two-band nodal star dispersion coupled to a dynamical $\mathrm{U(1)}$ gauge field. Such a field theory was studied in detail in Ref.~\cite{PhysRevB.104.054401} and was predicted to have a $T^{3/2}/\ln T$ subleading term in the specific heat, in addition to the specific heat due to the emergent photon modes.

\subsubsection{Staggered flux state and gapped $\mathrm{U(1)}$ spin liquids}\label{sec:staggered-flux}

The $(\pi,\pi/2,0)$ state defined here can be identified with the staggered flux (SF) state introduced in Ref.~\cite{Kim-2008}, up to unitary transformations (see Appendix~\ref{app:equiv_staggered_flux}). The feature of this state is that the triangular plaquettes on up and down tetrahedra carry opposite chiral fluxes $\pm\pi/2$, while the hexagon flux vanishes. This staggered arrangement distinguishes it from the monopole-flux state, where the triangle flux has the same sign on all tetrahedra. Correspondingly, the two states belong to different chiral symmetry classes, summarized in Table~\ref{TB:monopole_and_staggered}.

At the mean-field level the staggered flux pattern produces a fully gapped spinon spectrum. In contrast to the monopole-flux state, where the flux configuration enforces nodal structures, the alternating chirality removes the band degeneracies and opens a gap throughout the Brillouin zone. The resulting phase therefore realizes a gapped fermionic parton sector coupled to a dynamical $\mathrm{U(1)}$ gauge field.

In three dimensions such a theory naturally gives rise to a stable Coulomb phase: the spinons are massive, while the low-energy sector is governed by the emergent photon of the $\mathrm{U(1)}$ gauge field. Consequently the universal long-wavelength physics is dominated by gauge fluctuations, with the spinon sector contributing only through activated processes.

Just as the monopole flux state, the staggered flux state also exhibits a spectral symmetry under $E(\mathbf{k})\leftrightarrow E(-\mathbf{k})$, which can be understood by restricting to PSG classes with $m_{\overline{C}_6}=1$.

\begin{table}
\centering
\caption{Symmetry properties of the monopole flux state \cite{Burnell-2009, Kim-2008} and the staggered flux state \cite{Kim-2008}.}\label{TB:monopole_and_staggered}
\begin{tabular}{c|c|c|c}
\hline\hline
State & $(\phi_{\smalllozenge}, \phi_{\smalltriangleup}, \phi_{\smallhexagon})$ & Preserves & Breaks\\
\hline
Monopole flux state & \textcolor{RoyalBlue}{$(0,\frac{\pi}{2},0)$} & $T_{1,2,3}$, $C_3$, $I \mathcal{T}$, $S$ & $I$, $\mathcal{T}$\\
Staggered flux state & \textcolor{BrickRed}{$(\pi,\frac{\pi}{2},0)$} & $T_{1,2,3}$, $C_3$, $I$, $S \mathcal{T}$ & $S$, $\mathcal{T}$\\
\hline\hline
\end{tabular}
\end{table}

\subsection{Variational Monte Carlo framework}
\label{sec:vmc}
Within the VMC approach, the variational state is obtained by projecting (to the spin Hilbert space) a Slater determinant, $|\Phi_{\rm 0}\rangle$, which is the ground state of an auxiliary quadratic Hamiltonian
\begin{equation}\label{eq:h0_vmc}
	\mathcal{H}_{\rm 0}
	=\sum_{i,j} t_{ij}(f_{i,\uparrow}^{\dagger}f_{j,\uparrow}^{\phantom\dagger}
	+ f_{i,\downarrow}^{\dagger}f_{j,\downarrow}^{\phantom\dagger})\,.
\end{equation}
The projection, which enforces the single fermionic occupation of each lattice site, is achieved by means of a Gutzwiller-projector $\hat{\mathcal{P}}^\infty_{\rm G}$
\begin{equation}\label{eq:gutzwiller}
  \hat{\mathcal{P}}^{\infty}_{\mbox{\footnotesize G}}=\prod_i (f^\dagger_{i,\uparrow}f_{i,\uparrow}-f^\dagger_{i,\downarrow}f_{i,\downarrow})^2\,,
\end{equation}
and can be performed exactly by an appropriate Monte Carlo sampling~\cite{BeccaBook}. Here, we are interested in obtaining the spin structure factors of the variational {\it Ans\"atze} for $\mathrm{U(1)}$ chiral spin liquids and their energetics for the spin-$1/2$ Heisenberg antiferromagnet. Thus, we restrict ourselves to NN singlet hopping terms as in Eq.~\eqref{eq:h0_vmc}.

The parameters $t_{ij}$ (hoppings) are complex with the phases determined by the PSGs, and thus the resulting wave functions have no free variational parameters (with $|t_{ij}|=1$ as reference). The Gutzwiller projection is then performed
on these systematically enumerated fermionic {\it Ans\"atze}.

We use VMC to obtain the energy of the variational states. The result is listed in Table~\ref{tab:variational_energy}.

\begin{table*}[t]
\begin{tabular}{lllll}
\hline\hline
 Symmetry & ($\Phi_{\smalllozenge},\Phi_{\smalltriangleup},\Phi_{\smallhexagon}$) & ($\Phi_{\smalltriangledown},\Phi_{\smalltriangleup}$) & $E_{0}/NJ$ & Proper/improper symmetries \\ \hline
 
  \multirow{4}{*}{$(I,S,\mathcal{T})$} & $(0,0,0)^{a}$~\cite{Kim-2008,Burnell-2009} &
  $\Phi_{\smalltriangleup}=\Phi_{\smalltriangledown}=0$  & $-0.37502(6)$ & \multirow{4}{*}{All symmetries nonchiral}\\ 
  & $(\pi,0,\pi)^{b}$~\cite{Kim-2008,Burnell-2009} & $\Phi_{\smalltriangleup}=\Phi_{\smalltriangledown}=0$ & $-0.37511(8)$ \\  
  & $(0,0,\pi)$ & $\Phi_{\smalltriangleup}=0, \Phi_{\smalltriangledown}=\pi$  & $-0.37457(5)$ \\ 
  & $(\pi,0,0)$ & $\Phi_{\smalltriangleup}=0, \Phi_{\smalltriangledown}=\pi$  & $-0.37499(8)$ \\ \\
  
  \multirow{4}{*}{$(I\mathcal{T},S,\xcancel{\mathcal{T}})$} & \textcolor{RoyalBlue}{$(0,\frac{\pi}{2},0)$}$^{c}$~\cite{Kim-2008,Burnell-2009} & $\Phi_{\smalltriangleup}=\Phi_{\smalltriangledown}=\frac{\pi}{2}$ & $-0.45756(1)$ & \multirow{ 2}{*}{nonchiral: $8C_3,3C_2,6\tilde{C}_4,6\tilde{C}_2$} \\
 & \textcolor{Dandelion}{$(\pi,\frac{\pi}{2},\pi)$}$^{d}$~\cite{Burnell-2009} & $\Phi_{\smalltriangleup}=\Phi_{\smalltriangledown}=\frac{\pi}{2}$ & $-0.43457(1)$  \\
 & \textcolor{Plum}{$(\frac{\pi}{2},\frac{\pi}{2},\frac{\pi}{2})$} & $\Phi_{\smalltriangleup}=\Phi_{\smalltriangledown}=\frac{\pi}{2}$ & $-0.43457(1)$  &\multirow{ 2}{*}{chiral: $6\sigma_d$,$6S_4$,$3\tilde{\sigma}_h$,$8\tilde{S}_6$} \\
 & \textcolor{Plum}{$(\frac{\pi}{2},0,-\frac{\pi}{2})$} & $\Phi_{\smalltriangleup}=0,\Phi_{\smalltriangledown}=\pi$ & $-0.29753(1)$ \\ \\
  
  \multirow{4}{*}{$(I,S\mathcal{T},\xcancel{\mathcal{T}})$} & \textcolor{ForestGreen}{$(0,\frac{\pi}{2},\pi)$} & $\Phi_{\smalltriangleup}=-\Phi_{\smalltriangledown}=\frac{\pi}{2}$ & $-0.43735(1)$ & \multirow{ 2}{*}{nonchiral: $8C_3,3C_2,3\tilde{\sigma}_h,8\tilde{S}_6$}\\
 & \textcolor{BrickRed}{$(\pi,\frac{\pi}{2},0)$}{$^{e}$}~\cite{Kim-2008} & $\Phi_{\smalltriangleup}=-\Phi_{\smalltriangledown}=\frac{\pi}{2}$ & $-0.45930(1)$  \\
 & \textcolor{TealBlue}{$(\frac{\pi}{2},\frac{\pi}{2},-\frac{\pi}{2})$} & $\Phi_{\smalltriangleup}=-\Phi_{\smalltriangledown}=\frac{\pi}{2}$ & $-0.44930(1)$  &\multirow{ 2}{*}{chiral: $6\sigma_d$,$6S_4$,$6\tilde{C}_4,6\tilde{C}_2$}\\
 & \textcolor{TealBlue}{$(\frac{\pi}{2},0,\frac{\pi}{2})$} & $\Phi_{\smalltriangleup}=\Phi_{\smalltriangledown}=0$ & $-0.37462(1)$ 

  \\ \hline \hline
\end{tabular}
\caption{For the nearest-neighbor spin-$1/2$ Heisenberg antiferromagnet, the energies per site $E_{0}/NJ$ of the twelve distinct $\mathrm{U(1)}$ {\it Ans\"atze} which feature non-vanishing nearest-neighbor singlet amplitudes. The proper and improper symmetries which become chiral or are nonchiral are listed in the last column using the notation of Ref.~\cite{Burnell-2009}. The VMC calculations are done on a $2048$-site $(=4\times8\times8\times8)$ cluster which preserves the full symmetry of the pyrochlore lattice.\\
$^{a}$ $(0,0,0)$: This {\it Ansatz} is labelled as $[0,0,0]$ in Ref.~\cite{Kim-2008}, and referred to as the uniform state in Ref.~\cite{Burnell-2009}.\\
$^{b}$ $(\pi,0,\pi)$: This {\it Ansatz} is labelled as $[0,0,\pi]$ in Ref.~\cite{Kim-2008}, and as $(\pi,\pi)$ in Table I of Ref.~\cite{Burnell-2009}.\\
$^{c}$ \textcolor{RoyalBlue}{$(0,\pi/2,0)$}: This {\it Ansatz} is referred to as the monopole flux state in Ref.~\cite{Burnell-2009}. In Ref.~\cite{Kim-2008}, this state is labelled as $[\frac{\pi}{2},\frac{\pi}{2},0]$ and is referred to as the uniform flux state.\\
$^{d}$ \textcolor{Dandelion}{$(\pi,\pi/2,\pi)$}: This {\it Ansatz} is labelled as $(\pi/2,\pi)$ in Table I of Ref.~\cite{Burnell-2009}.\\
$^{e}$ \textcolor{BrickRed}{$(\pi,\pi/2,0)$}: This {\it Ansatz} is referred to as the staggered flux state and labelled $[\frac{\pi}{2},-\frac{\pi}{2},0]$ in Ref.~\cite{Kim-2008}.\label{tab:variational_energy}}
\end{table*}

\begin{table*}[t]
\begin{footnotesize}
\centering
\newcommand{\energycomment}[1]{\parbox[t]{0.36\textwidth}{\raggedright #1}}
\begin{tabular}{lccl}
\hline\hline
\multicolumn{1}{c}{Method}
&
\multicolumn{1}{c}{System size / limit}
&
\multicolumn{1}{c}{$E_0/(NJ)$}
&
\multicolumn{1}{c}{Comment}
\\
\hline\hline

DMRG, Hagym\'asi {\it et al.}~\cite{Hagymasi-2021}
&
Thermodynamic
&
$-0.490(6)$
&
\energycomment{Best thermodynamic estimate from finite-cluster SU(2) DMRG scaling; consistent with the NLCE zero-temperature extrapolation $-0.495(15)$, while converged finite-temperature NLCE gives the bound $E_0/N\lesssim -0.471J$.}
\\

mVMC, Astrakhantsev {\it et al.}~\cite{Astrakhantsev-2021}
&
$4\times4^{3}$, $N=256$
&
$-0.4831(1)$
&
\energycomment{Largest finite equilateral-cluster mVMC energy reported in Ref.~\cite{Astrakhantsev-2021}.}
\\

mVMC, Astrakhantsev {\it et al.}~\cite{Astrakhantsev-2021}
&
Thermodynamic
&
$-0.477(3)$
&
\energycomment{Infinite-volume extrapolation of their mVMC energies.}
\\

Hard-hexagon variational state, Sch\"afer {\it et al.}~\cite{Schafer-2023}
&
Thermodynamic
&
$-0.489472(8)$
&
\energycomment{Optimized hard-hexagon variational energy $E_{\alpha_0}$; a strict variational upper bound for the nearest-neighbor model.}
\\

Hexagon NLCE, Sch\"afer {\it et al.}~\cite{Schafer-2023}
&
Thermodynamic
&
$-0.4917(5)$
&
\energycomment{Second-order hexagon-based NLCE estimate; not a variational upper bound, but consistent with the hard-hexagon variational state and earlier DMRG estimates.}
\\

mVMC-RBM/Lanczos, Pohle {\it et al.}~\cite{Pohle-2023}
&
$L=2$, $N_s=128$
&
$-0.49229(7)$
&
\energycomment{Best strictly variational finite-size energy reported in Ref.~\cite{Pohle-2023}, obtained for the spin-parity-even mVMC-RBM wave function with one Lanczos step.}
\\

mVMC, Pohle {\it et al.}~\cite{Pohle-2023}
&
Thermodynamic
&
$-0.4853(1)$
&
\energycomment{Thermodynamic extrapolation of strictly variational mVMC energies from clusters with $N=128,432,1024$.}
\\

mVMC/Lanczos, Pohle {\it et al.}~\cite{Pohle-2023}
&
Thermodynamic
&
$-0.4881(3)$
&
\energycomment{Thermodynamic extrapolation of strictly variational mVMC energies after one Lanczos step.}
\\

Variance extrapolation, Pohle {\it et al.}~\cite{Pohle-2023}
&
Thermodynamic
&
$-0.4921(4)$
&
\energycomment{Zero-variance extrapolated thermodynamic estimate from the mVMC-based sequence; not a strict variational upper bound.}
\\

Present chiral $\mathrm{U}(1)$ PSG {\it Ansatz}, Table~\ref{tab:variational_energy}
&
$L=8$
&
$\simeq -0.45930$
&
\energycomment{Lowest fixed-flux-sector energy in Table~\ref{tab:variational_energy}, obtained for the staggered-flux state $(\pi,\pi/2,0)$.}
\\

\hline\hline
\end{tabular}
\end{footnotesize}
\caption{Representative energy estimates for the nearest-neighbor spin-$1/2$ pyrochlore Heisenberg antiferromagnet,
$H=J\sum_{\langle ij\rangle}\mathbf{S}_i\cdot\mathbf{S}_j$, with $J=1$. Here, $N$ denotes the number of spins. The entries include both finite-size values and thermodynamic estimates, as well as both strict variational upper bounds and extrapolated estimates; they are therefore meant to provide an energetic scale rather than a direct one-to-one benchmark of identical variational classes. The final row gives the lowest energy among the fixed PSG/flux-sector chiral $\mathrm{U}(1)$ states in Table~\ref{tab:variational_energy}. These states are used here to classify and diagnose chiral spin-liquid Ans\"atze, and are not intended as unrestricted variational searches for the ground state of the nearest-neighbor Heisenberg model.}
\label{tab:nn-pyrochlore-energy-comparison}
\end{table*}

\subsubsection{Energetics}
\label{subsec:vmc_energy}
We now turn to the variational Monte Carlo (VMC) evaluation of the ground-state energies
of the eight $\mathrm{U(1)}$ chiral spin-liquid {\it Ans\"atze} introduced above.
The resulting energies are summarized in Table~\ref{tab:variational_energy}. 
For reference, and to place these values on the scale of the nearest-neighbor
spin-$1/2$ pyrochlore Heisenberg antiferromagnet, Table~\ref{tab:nn-pyrochlore-energy-comparison} collects representative energy estimates from recent numerical and variational studies.

The comparison shows that the fixed-flux chiral $\mathrm{U(1)}$ states considered here lie above the best available estimates for the pure nearest-neighbor model. For example,
recent DMRG and NLCE-based studies, hard-hexagon variational states, and large-scale
mVMC calculations give energies in the approximate range
$E_0/(NJ)\simeq -0.477$ to $-0.492$, depending on the method and on whether the quoted
number is a strict variational upper bound, a finite-size value, or an extrapolated estimate. By contrast, the lowest energy among the present chiral $\mathrm{U(1)}$ PSG states is approximately $-0.4593J$, obtained for the staggered-flux state
$(\pi,\pi/2,0)$ in Table~\ref{tab:variational_energy}. This observation is neither
unexpected nor problematic in the present context. Our primary goal is not to establish the lowest variational energy for the NN model, but rather to construct and characterize a controlled family of chiral $\mathrm{U(1)}$ spin
liquids and to elucidate their emergent gauge structure, correlation signatures, and
systematic dependence on flux patterns.

Indeed, Table~\ref{tab:variational_energy} reveals that the different chiral {\it Ans\"atze} form a relatively narrow
energy manifold, despite exhibiting markedly distinct behavior in their equal-time
spin structure factors and gauge-sector dominance (Sec.~\ref{sec:monopole-flux} and Table~\ref{tab:gauge_dominance}).
This separation between energetic ranking and correlation structure highlights an important
point: variational states that are not energetically optimal for the pure NN
model may nevertheless provide particularly clean realizations of specific emergent gauge
phenomena, such as strongly anisotropic pinch points associated with a Coulomb phase.

From a broader perspective, the chiral $\mathrm{U(1)}$ {\it Ans\"atze} studied here should be viewed as natural ground-state candidates for a wider class of pyrochlore Hamiltonians.
It is well established that further-neighbor exchange interactions, ring-exchange terms,
or symmetry-allowed Dzyaloshinskii--Moriya interactions can substantially modify the
energetic landscape on the pyrochlore lattice.
Such perturbations are expected to selectively stabilize different flux configurations
and may favor chiral spin-liquid phases even when they are not energetically competitive
in the pure NN limit.
Within this extended parameter space, the relative ordering of energies in Table~\ref{tab:variational_energy}
provides a useful guide to which chiral $\mathrm{U(1)}$ states are most proximate and therefore most
likely to be realized.

Finally, we emphasize that the present VMC results should be interpreted in conjunction
with the detailed analysis of correlation functions and structure factors presented below. Taken together, these results establish the chiral $\mathrm{U(1)}$ spin liquids as well-defined quantum phases with robust gauge-field signatures, independent of their precise energetic placement within the NN Heisenberg model. In this sense, the present study complements energy-focused numerical approaches by providing a systematic variational characterization of chiral $\mathrm{U(1)}$ spin liquids and their distinct correlation fingerprints.

\subsubsection{Equal-time spin structure factors}
\label{sec:equal_time_sq}

The equal-time spin structure factor
\begin{equation}\label{etssf}
S(\mathbf{q})=\frac{1}{N}\sum_{i,j}
e^{-i\mathbf{q}\cdot(\mathbf{r}_{i}-\mathbf{r}_{j})}
\langle \hat{\mathbf S}_{i}\cdot \hat{\mathbf S}_{j}\rangle
\end{equation}
provides a principal probe of magnetic correlations in frustrated quantum magnets and plays a particularly important role in diagnosing Coulomb phases on the pyrochlore lattice. In contrast to magnetically ordered states, which exhibit Bragg peaks signaling spontaneous symmetry breaking, quantum spin liquids display diffuse yet highly structured scattering patterns encoding emergent constraints and gauge-field physics.

A defining feature of $\mathrm{U(1)}$ Coulomb spin liquids is the appearance of pinch-point (bow-tie) singularities in the equal-time structure factor. These features originate from a local constraint enforcing a divergence-free condition on coarse-grained spin or flux variables, which leads to dipolar correlations in real space and a transverse projector structure in momentum space. This mechanism was first elucidated in the context of classical spin ice and Coulomb phases on the pyrochlore lattice~\cite{Isakov-2004,Henley-2010,Castelnovo-2008}, and subsequently extended to quantum spin liquids and three-dimensional frustrated magnets~\cite{Savary-2017}. Neutron-scattering signatures of such pinch points, including their characteristic angular anisotropy, have been extensively discussed in both classical and quantum pyrochlore systems.

From the gauge-theory perspective, pinch points reflect the presence of an emergent $\mathrm{U(1)}$ gauge field obeying Gauss’ law. At long wavelengths, the equal-time correlations take the universal form
\begin{equation}
\langle B_a(\mathbf{q}) B_b(-\mathbf{q}) \rangle
\propto \delta_{ab}-\frac{q_a q_b}{q^2}\,,
\end{equation}
which directly produces the bow-tie geometry in momentum space~\cite{Henley-2010,Savary-2017}. Crucially, this structure is a property of the \emph{gauge sector} and does not arise from the dynamics of the fractionalized matter fields alone.

Within fermionic parton constructions, this distinction becomes particularly transparent. At the mean-field level, only the spinon (matter-field) contribution is retained, and the resulting structure factor is governed by particle-hole processes set by the spinon band structure. As illustrated by the unprojected results in Appendix \ref{app:unprojected_SSF} and Fig.~\ref{fig:sq_unprojected} therein, this contribution provides a comparatively smooth background on the momentum scales of interest, although weak nonanalyticities may arise when the spinon spectrum is gapless. The enforcement of the local single-occupancy constraint (implemented here via Gutzwiller projection) is therefore essential. Projection restores gauge-field fluctuations, including the temporal component enforcing Gauss' law, and thereby generates the Coulombic correlations responsible for pinch points. This separation between comparatively smooth matter-field backgrounds (possibly with subleading weak nonanalytic corrections) and the dominant pinch-point singularity of the gauge sector is a hallmark of $\mathrm{U(1)}$ spin liquids in three dimensions~\cite{Kim-2008,Savary-2017,BeccaBook}.

\begin{figure*}
  \centering
  \includegraphics[width=0.95\linewidth]{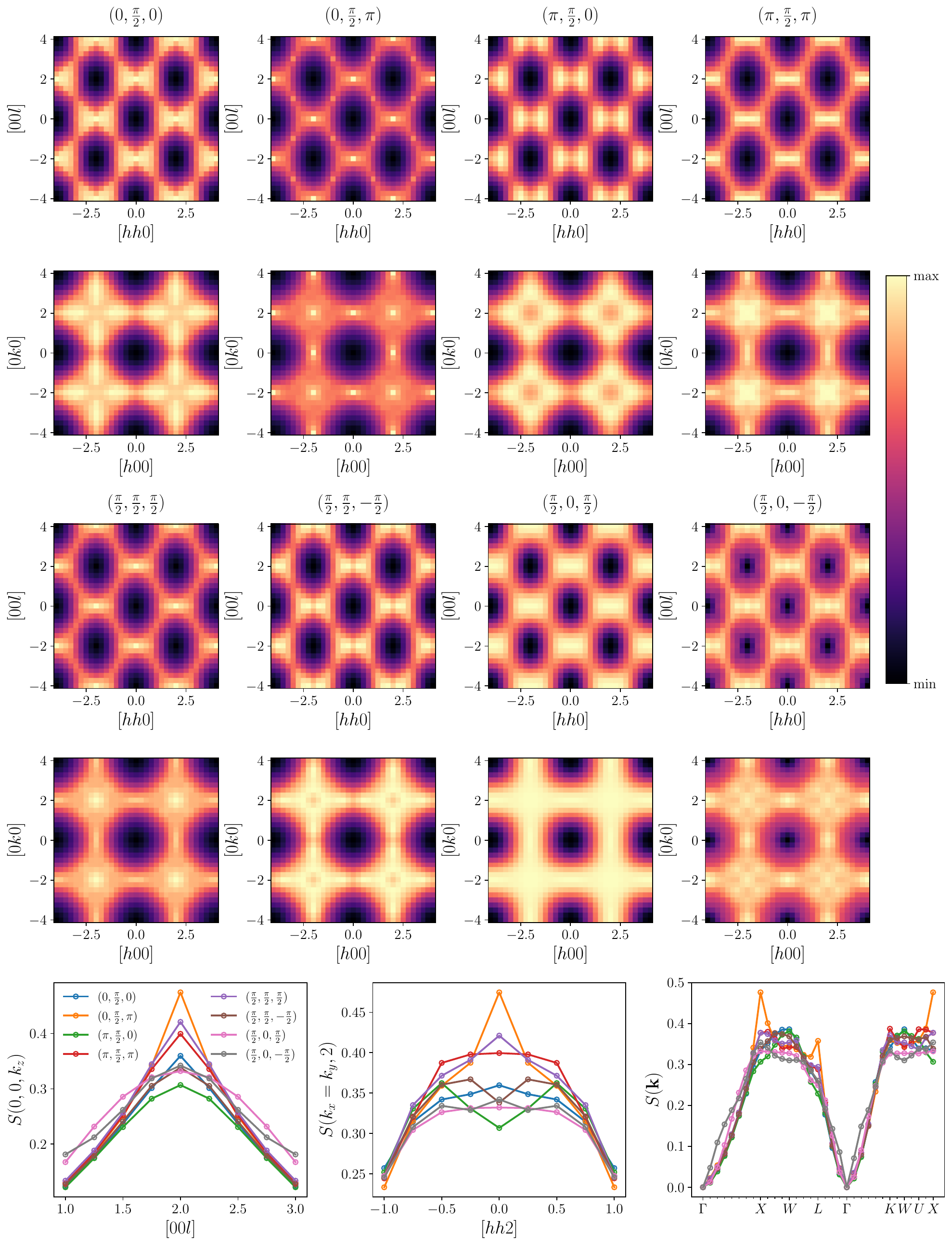}
  \caption{Equal-time spin structure factors of the Gutzwiller-projected nearest-neighbor $\mathrm{U}(1)$ chiral spin-liquid {\it Ans\"atze}. The top and middle rows show two-dimensional momentum cuts in the $[hhl]$ and $[hk0]$ planes, respectively. The bottom row shows one-dimensional cuts through the pinch point at $\mathbf G=(0,0,2)$ along the vertical $(00k_z)$ and transverse $(kk2)$ directions. While all states exhibit the characteristic pinch-point geometry of a three-dimensional $\mathrm{U}(1)$ gauge field, the contrast and apparent sharpness vary substantially across flux sectors because of differing background contributions from matter-field and short-range correlations.}
  \label{fig:sq}
\end{figure*}

In Fig.~\ref{fig:sq} we present the equal-time spin structure factors for the projected NN $\mathrm{U(1)}$ chiral spin-liquid {\it Ans\"atze} discussed in Sec.~\ref{sec:spec_12}. The first two rows show two-dimensional cuts in the $[hhl]$ and $[hk0]$ planes, while the third row displays one-dimensional cuts through the pinch point at $(0,0,2)$ along two orthogonal directions, together with a cut along a representative high-symmetry path.

Several robust features emerge. First, all projected $\mathrm{U(1)}$ chiral states exhibit well-defined pinch points at $(0,0,2)$ and symmetry-related reciprocal lattice vectors. The appearance of these pinch points at $(0,0,2)$, rather than at the zone center, is a well-understood consequence of the pyrochlore basis and neutron-scattering form factors: interference between the four sublattices suppresses intensity at certain reciprocal vectors while enhancing it at others, making $(0,0,2)$-type positions the most visible locations for Coulomb-phase signatures in the conventional cubic Brillouin zone~\cite{Fennell-2007,Benton-2012,Savary-2012a}. The angular anisotropy of the bow-tie patterns, clearly visible in both planar cuts and one-dimensional slices, is consistent with the transverse projector expected for a $\mathrm{U(1)}$ gauge field.

Second, while the \emph{geometry} of the pinch points is universal, their \emph{spectral weight and contrast} vary between different chiral {\it Ans\"atze}. This reflects the coexistence of gauge-field correlations with {\it Ansatz}-dependent matter-field contributions. Schematically, the structure factor can be decomposed as
\begin{equation}
S(\mathbf{q}) \simeq S_{\mathrm{gauge}}(\mathbf{q})
+ S_{\mathrm{matter}}(\mathbf{q})
+ S_{\mathrm{short\text{-}range}}(\mathbf{q})\,,
\end{equation}
where the first term encodes the Coulombic pinch-point singularity, while the latter two provide a comparatively smooth background {(on the momentum scales resolved in Fig.~\ref{fig:sq})} arising from spinon particle-hole excitations and short-range RVB correlations~\cite{Kim-2008,Savary-2017}. Depending on the underlying spinon dispersion and flux pattern, this background may either enhance the apparent intensity near the pinch point or partially fill in the pinched directions, reducing contrast without altering the underlying angular structure. In particular, gapless spinon manifolds (e.g., nodal lines) {may, in principle, contain weak nonanalytic corrections in $S_{\mathrm{matter}}(\mathbf{q})$, but in the present numerics with the given momentum resolution} they do not produce any obvious divergent or pinch-point-like singularity; the dominant nonanalytic cusp and angular ``pinch'' structure are controlled by the gauge sector. This effect is particularly evident in the one-dimensional cuts of Fig.~\ref{fig:sq}, where different {\it Ans\"atze} show distinct curvatures and slopes approaching $(0,0,2)$.

A useful way to disentangle these contributions is to analyze the local line shape of the one-dimensional cuts in the immediate vicinity of the pinch point. Writing $\mathbf{q}=\mathbf{G}+\mathbf{k}$ with $\mathbf{G}=(0,0,2)$ and small $\mathbf{k}$ along the cut direction, the gauge-sector contribution produces the characteristic nonanalytic, strongly anisotropic approach governed by the transverse projector. In particular, the intensity is suppressed along the ``pinched'' direction, while along the transverse direction it remains finite and is therefore larger relative to the pinched direction~\cite{Henley-2010,Savary-2017,Fennell-2007}. This transverse response should not be interpreted as a divergent enhancement; in the ideal Coulomb form it is diffuse, and over a small momentum window near the reciprocal-lattice vector it may appear rather flat. By contrast, the matter and short-range sectors provide a comparatively smooth background on the scales shown in Fig.~\ref{fig:sq}, but need not be strictly analytic to all orders when the spinon spectrum is gapless. More generally, one may write
$
S_{\mathrm{matter}}(\mathbf{G}+\mathbf{k})+S_{\mathrm{short\text{-}range}}(\mathbf{G}+\mathbf{k})
=
A_0 + A_1\!\cdot\!\mathbf{k} + \mathbf{k}^T A_2 \mathbf{k} + \cdots + \delta S_{\mathrm{weak}}(\mathbf{k})
$,
where the first terms are analytic and $\delta S_{\mathrm{weak}}(\mathbf{k})$ denotes possible weak nonanalytic corrections associated with gapless spinons. The Appendix \ref{app:unprojected_SSF} numerics show that, on the momentum cuts considered here, these matter-field contributions do not produce any obvious divergent or pinch-point-like singularity, and are therefore subleading compared with the dominant gauge-sector pinch-point structure. It is also useful to note that the component symmetric about the pinch point predominantly reflects the gauge-induced anisotropy, while asymmetric deviations are largely associated with nonuniversal background contributions. We use this observation only heuristically here, as the matter sector may itself contain weak nonanalytic corrections when the spinon spectrum is gapless.

The separation of the universal pinch-point form fixed by Gauss’ law from nonuniversal background contributions with possible weak nonanalytic correlations is implicit in the Coulomb-phase description; here we make this separation operational for projected parton wave functions and use it comparatively to quantify ‘gauge dominance’ across distinct chiral flux sectors. We now apply this diagnostic explicitly to the one-dimensional cuts shown in the bottom row of
Fig.~\ref{fig:sq} for all eight $\mathrm{U(1)}$ chiral {\it Ans\"atze}.
For each state, two orthogonal cuts through the pinch point $\mathbf{G}=(0,0,2)$ are shown: a vertical cut $(00k_z)$ and a horizontal transverse cut $(kk2)$. Within the three-sector decomposition, the gauge contribution is identified by a strong,
direction-dependent nonanalytic approach to $\mathbf{G}$, while matter-field and short-range
contributions enter as backgrounds with possible weak nonanalyticities
but cannot generate the pinch-point anisotropy by themselves.

A first robust observation is that \emph{all} eight {\it Ans\"atze} exhibit a pronounced enhancement
upon approaching $\mathbf{G}$ along the vertical $(00k_z)$ cut, demonstrating that a Coulombic
gauge sector is present in every case.
However, the sharpness of this enhancement varies significantly.
The $(0,\pi/2,\pi)$ and $(\pi,\pi/2,\pi)$ states show the largest cusp-like increase in the $(00k_z)$
direction, indicating a strong gauge-sector signal, whereas the $(\pi/2,0,\pi/2)$ state displays
a noticeably more rounded approach, consistent with a larger contribution from the matter and short-range sectors partially masking
the singular contribution.
The remaining states, $(0,\pi/2,0)$, $(\pi,\pi/2,0)$, $(\pi/2,0,-\pi/2)$, $(\pi/2,\pi/2,\pi/2)$,
and $(\pi/2,\pi/2,-\pi/2)$, exhibit intermediate behavior, with clearly visible cusps but varying
degrees of rounding near $\mathbf{G}$.

The horizontal $(kk2)$ cut provides a particularly sensitive probe of
nongauge-sector contributions because, for the pinch point and
spin-structure-factor component shown here, the ideal gauge-sector line shape
has only weak local contrast and is nearly flat along this transverse cut.
Pronounced additional intensity or peak-like filling in this direction therefore
signals smooth matter-field or short-range backgrounds rather than an enhanced
gauge singularity.

Here, the eight {\it Ans\"atze} separate into three qualitatively distinct groups. The $(\pi,\pi/2,0)$ and $(\pi/2,\pi/2,-\pi/2)$ states exhibit a suppression or near-flat behavior at $\mathbf{G}$ in the $(kk2)$ cut, i.e., a local minimum or at most a very weak maximum. This indicates that matter-field and short-range contributions are relatively small, allowing the gauge-sector anisotropy to
dominate the local line shape. At the opposite extreme, the $(0,\pi/2,\pi)$
state shows a pronounced peak even along the $(kk2)$ direction, while the
$(\pi/2,\pi/2,\pi/2)$ state also retains a clear enhancement. In these cases,
the nongauge background is sufficiently large to fill in the ideal pinch-point
suppression and substantially reduce the apparent directional contrast, despite
the continued presence of a strong gauge-sector cusp in the vertical cut. The
remaining {\it Ans\"atze}, $(0,\pi/2,0)$, $(\pi/2,0,\pi/2)$,
$(\pi/2,0,-\pi/2)$, and $(\pi,\pi/2,\pi)$, display nearly flat or weakly peaked
behavior along $(kk2)$, corresponding to an intermediate regime in which
backgrounds are present but do not overwhelm the directional anisotropy.

Taken together, the bottom-row cuts of Fig.~\ref{fig:sq} reveal that while the pinch-point geometry
is universal across all $\mathrm{U(1)}$ chiral {\it Ans\"atze}, the relative weights of the gauge, matter, and
short-range sectors vary substantially with the flux pattern.
States such as $(\pi,\pi/2,0)$ and $(\pi/2,\pi/2,-\pi/2)$ represent the most gauge-dominated cases,
exhibiting strong anisotropy with minimal peak-like filling along the transverse cut, whereas
$(0,\pi/2,\pi)$ and $(\pi/2,\pi/2,\pi/2)$ show the strongest backgrounds from the matter field and short-range sectors, leading to more
rounded pinch points.
The remaining {\it Ans\"atze} interpolate continuously between these limits, highlighting how changes
in the underlying flux configuration primarily redistribute  spectral weight without
destroying the emergent Coulomb-phase correlations.

\begin{table*}
\centering
\caption{Qualitative ranking of the eight $\mathrm{U(1)}$ chiral spin-liquid {\it Ans\"atze} by the relative
dominance of the gauge-sector contribution near the pinch point at $\mathbf{G}=(0,0,2)$.
The classification is based on the one-dimensional cuts shown in the bottom row of
Fig.~\ref{fig:sq}, comparing the sharpness of the cusp in the vertical $(00k_z)$ direction and the
degree to which the \textcolor{purple}{[``pinched'' $\rightarrow$] transverse} $(kk2)$ direction is filled in by matter-field and
short-range contributions.}
\label{tab:gauge_dominance}
\begin{ruledtabular}
\begin{tabular}{ccc}
Flux tuple $(\phi_{\smalllozenge},\phi_{\smalltriangleup},\phi_{\smallhexagon})$
& Gauge dominance
& Characteristic features in $1d$ cuts \\ \hline
\textcolor{BrickRed}{$(\pi,\pi/2,0)$}
& Strong
& Sharp cusp in $00k_z$; suppression / near-flatness in $kk2$ \\

\textcolor{TealBlue}{$(\pi/2,\pi/2,-\pi/2)$}
& Strong
& Pronounced anisotropy; minimal \textcolor{purple}{[``filling of pinched direction'' $\rightarrow$] transverse filling} \\

\textcolor{RoyalBlue}{$(0,\pi/2,0)$}
& Intermediate
& Clear cusp in $00k_z$; weak peak in $kk2$ \\

 \textcolor{Plum}{$(\pi/2,0,-\pi/2)$}
& Intermediate
& Visible anisotropy with moderate background \\

\textcolor{Dandelion}{$(\pi,\pi/2,\pi)$}
& Intermediate
& Cusp present; partial filling in $kk2$ \\

\textcolor{TealBlue}{$(\pi/2,0,\pi/2)$}
& Intermediate--weak
& Rounded approach to pinch point \\

\textcolor{Plum}{$(\pi/2,\pi/2,\pi/2)$}
& Weak
& Strong  background; peak persists in $kk2$ \\

\textcolor{ForestGreen}{$(0,\pi/2,\pi)$}
& Weak
& Largest background; pinch-point contrast strongly reduced \\
\end{tabular}
\end{ruledtabular}
\end{table*}

For clarity, we summarize in Table~\ref{tab:gauge_dominance} a qualitative ranking of the
$\mathrm{U(1)}$ chiral {\it Ans\"atze} according to the relative visibility of the gauge-sector contribution
near the pinch point, as inferred from the one-dimensional cuts through $(0,0,2)$.

We now comment on the relationship between the spinon band structures shown in Fig.~\ref{spectraof12states} and the hierarchy of gauge dominance summarized in Table~\ref{tab:gauge_dominance}. While the emergent $\mathrm{U}(1)$ gauge structure is fixed at the level of the projective symmetry group, the detailed distribution of low-energy spinon states differs substantially between distinct chiral flux sectors. These differences affect how strongly matter-field fluctuations compete with long-wavelength gauge-field correlations once the Gutzwiller projection is imposed.

In particular, {\it Ans\"atze} whose mean-field band structures feature extended nodal manifolds, multiple symmetry-related band touchings, or an enhanced low-energy phase space tend to generate sizable  background contributions to the equal-time spin structure factor. In such cases, gauge-field correlations, though present, are partially obscured by matter-induced short-range correlations, leading to reduced pinch-point contrast. Conversely, flux sectors whose band structures are more dispersive or possess fewer low-energy spinon channels allow gauge-field fluctuations to dominate the equal-time correlations more clearly, resulting in sharper and higher-contrast pinch-point features.

We emphasize that this connection is not a simple one-to-one correspondence between band topology and gauge dominance, nor is it captured by variational energetics alone. Rather, it reflects a competition between matter and gauge fluctuations that becomes manifest only at the level of projected wave functions and equal-time correlations. From this perspective, Fig.~\ref{spectraof12states} and Table~\ref{tab:gauge_dominance} provide complementary information: the former characterizes the low-energy spinon content of a given flux sector, while the latter quantifies how effectively emergent gauge-field correlations survive projection in equal-time observables.

The behavior observed here contrasts sharply with that expected for gapped $\mathbb{Z}_2$ spin liquids on the pyrochlore lattice. In $\mathbb{Z}_2$ phases, the gauge field is gapped and equal-time correlations are short-ranged; as a result, true pinch-point singularities are absent in the equal-time structure factor, even though diffuse maxima reflecting short-range correlations may still appear~\cite{Wen-2002,Sachdev-1991}. The persistence of pinch points in all projected states studied here therefore provides strong evidence that these chiral spin liquids retain a gapless $\mathrm{U(1)}$ gauge sector.

Finally, we note that the equal-time spin structure factor is even under time-reversal symmetry and therefore does not directly probe chirality. The similarity of pinch-point features across different chiral {\it Ans\"atze} is thus expected; signatures of time-reversal symmetry breaking must instead be sought in time-reversal-odd observables such as scalar spin chirality correlations or thermal transport responses~\cite{Wen-1990,Lee-2002}. Moreover, the pinch points observed here are point-like singularities characteristic of rank-1 $\mathrm{U(1)}$ gauge theories and should be clearly distinguished from pinch-line features associated with higher-rank gauge structures~\cite{Benton-2016,Niggemann-2023,Gresista-2025}.

The equal-time spin structure factors presented above reveal that chiral $\mathrm{U}(1)$ spin liquids belonging to different flux sectors can display markedly different correlation profiles, despite sharing the same emergent gauge group and closely related mean-field energetics. In particular, some states exhibit sharp, high-contrast pinch-point singularities characteristic of a gauge-dominated Coulomb phase, while others show substantially reduced pinch-point visibility accompanied by a large, relatively smooth background contribution. This observation motivates a distinction between gauge-dominated and matter-dominated chiral $\mathrm{U}(1)$ spin liquids at the level of equal-time correlations.

Operationally, this distinction may be viewed as reflecting the relative weight of long-wavelength transverse gauge-field fluctuations versus short-range spinon and constraint-induced correlations in the projected wave function. In gauge-dominated states, the pinch-point geometry is sharply resolved and largely controls the momentum-space structure of $S(\mathbf{q})$, closely resembling the correlations of a three-dimensional Coulomb phase. In contrast, in matter-dominated states the same gauge structure is present but masked by sizable contributions from the matter and short-range sectors, leading to reduced pinch-point contrast and a less distinctive Coulombic signature.

Importantly, this distinction is not encoded at the level of projective symmetry group classification or $\mathrm{U}(1)$ gauge structure alone, nor is it reliably captured by variational energies. Rather, it emerges only upon analyzing equal-time correlations of the fully projected wave functions. In this sense, equal-time correlation functions provide an experimentally relevant and theoretically robust diagnostic that distinguishes proximate chiral spin liquids which are otherwise difficult to separate. The results presented here demonstrate that different chiral flux sectors on the pyrochlore lattice realize distinct realizations of emergent gauge physics, even within the same $\mathrm{U}(1)$ spin-liquid framework.

It is useful to compare this diagnostic role of the equal-time structure factor with existing pyrochlore quantum spin-ice candidates. Ce-based
dipolar-octupolar pyrochlores such as Ce$_2$Zr$_2$O$_7$~\cite{Gaudet-2019,Bhardwaj-2022,Gao-2025}, Ce$_2$Hf$_2$O$_7$~\cite{Poree-2025} and Ce$_2$Sn$_2$O$_7$~\cite{Sibille-2020} provide important experimental reference points for three-dimensional $\mathrm{U}(1)$ gauge physics on the pyrochlore lattice.
Their neutron spectra and thermodynamic responses have been discussed in
terms of fractionalized excitations, emergent photons, or quantum liquids of
multipolar degrees of freedom. These observations, however, are not by
themselves signatures of a chiral spin liquid. The chiral states studied here
require, in addition to $\mathrm{U}(1)$ gauge correlations, a time-reversal-odd
and orientation-sensitive flux or scalar-chirality sector. Thus, comparison
with experiment should examine not only whether diffuse scattering,
pinch-point-like features, or continua are present, but also whether the
detailed momentum-space anisotropy, polarization dependence, energy
dependence, and independent probes of time-reversal/parity breaking are
consistent with a chiral flux pattern. At present, the reported signatures in
these compounds can be interpreted without invoking such a chiral sector; by
themselves they do not demonstrate zero-field time-reversal breaking, a
scalar-chirality-sensitive response, or a domain-trainable chiral signal.

\section{Discussion and Outlook}
\label{outlook}

The present work establishes a systematic classification of chiral $\mathrm{U(1)}$ and chiral $\mathbb{Z}_2$
spin liquids on the pyrochlore lattice and introduces correlation-based diagnostics that sharply
distinguish gauge-dominated Coulomb phases from states with substantial  backgrounds that are relatively smooth.
Several open directions naturally follow from these results, many of which touch on fundamental
questions concerning the infrared structure of chiral quantum phases in three dimensions.

The perspective developed in this work suggests several broader implications. First, it highlights that chiral quantum spin liquids in three dimensions should be viewed not as isolated phases but as manifolds of proximate states whose physical character can vary substantially depending on flux sector, even when symmetry and gauge structure are identical. Second, it shows that equal-time correlation functions---particularly the geometry and contrast of pinch-point features---provide a powerful way to distinguish such phases, complementing symmetry-based classifications and variational energetics. This is especially relevant for experimental probes such as neutron scattering, where static structure factors are directly accessible and may distinguish chiral spin liquids with otherwise similar thermodynamic properties.

We also emphasize the symmetry meaning of ``chiral'' in the present context. A chiral spin liquid need not rely on explicit inversion breaking of the crystal structure. The microscopic Hamiltonian may be inversion symmetric, while the 
many-body ground state spontaneously selects one of two chiral sectors related 
by inversion and/or time reversal. This spontaneous domain selection is the reason explicit structural inversion breaking is not a prerequisite for the chiral phases considered here. In the PSG classes studied in this work, the selected chiral state may nevertheless preserve combined operations such as 
$I\mathcal{T}$ or $S\mathcal{T}$ may remain symmetries even though $I$ or 
$\mathcal{T}$ separately is broken. External perturbations, including magnetic 
field, strain, pressure, or weak symmetry-lowering interactions, can nevertheless 
be useful experimentally because they may train chiral domains or tune the 
system closer to a chiral instability. Neutron scattering primarily probes 
two-spin correlations, and is therefore most directly sensitive to the 
gauge-sector fingerprints discussed in this work: absence of magnetic Bragg 
peaks, structured diffuse scattering, pinch-point anisotropy, and continuum 
response. Establishing scalar chirality itself is more demanding because 
$\chi_{ijk}$ is a three-spin pseudoscalar. A compelling experimental case would 
therefore require combining neutron scattering with probes of 
time-reversal/parity breaking and chirality-sensitive responses, such as 
zero-field $\mu$SR, NMR, Kerr rotation, polarized neutron measurements of 
antisymmetric correlations where symmetry permits, field-training protocols for  opposite chiral domains, and thermal Hall or nonreciprocal responses.

The present framework opens the door to systematic studies of chiral $\mathrm{U}(1)$ and $\mathbb{Z}_2$ spin liquids in extended pyrochlore Hamiltonians beyond the nearest-neighbor Heisenberg model. Further-neighbor exchange~\cite{Lapa-2012,Okubo-2011,Iqbal-2017,Iqbal-2019}, ring-exchange processes~\cite{Kranitz-2026,Motrunich-2005,Balents-2010}, spin-anisotropic~\cite{Noculak-2023,Gresista-2025,Gresista-2026} or chiral interactions~\cite{Gomez-2024} provide concrete routes to favor particular flux patterns and noncoplanar tendencies identified here. Since such interactions already stabilize noncoplanar ground states in corresponding classical models, it is natural to ask whether quantum fluctuations for $S=1/2$ can restore spin-rotation symmetry while preserving time-reversal breaking, thereby yielding chiral spin liquids. In this broader setting, the correlation-based diagnostics introduced here may be especially useful for distinguishing regimes in which emergent gauge-field physics remains dominant from those in which it is partially masked by short-range contributions. We therefore expect the classification and diagnostics developed in this work to be applicable well beyond the nearest-neighbor model, and to provide a useful bridge between microscopic realization and the broader theoretical description of three-dimensional chiral quantum spin liquids.

A promising avenue concerns the construction of three-dimensional chiral spin liquids from lower-dimensional building blocks. The pyrochlore lattice may be viewed as an arrangement of interpenetrating kagome and triangular planes, each of which is known to host chiral spin liquids in two dimensions~\cite{Kalmeyer-1987,Wen-1991,Gong-2015}. This suggests the possibility of controlled \emph{coupled-plane} constructions, analogous in spirit to coupled-wire approaches~\cite{Meng-2015,Pereira-2018}, also developed for other strongly correlated and topological settings~\cite{Kane-2002,Teo-2014}, but now with genuinely three-dimensional gauge dynamics. More broadly, one may also envision complementary coupled-chain or coupled-wire formulations built from quasi-one-dimensional chiral ingredients. In either language, the essential question is whether the inter-unit coupling stabilizes a true three-dimensional $\mathrm{U(1)}$ Coulomb phase, or instead drives confinement or Higgsing into a $\mathbb{Z}_2$ phase~\cite{Fradkin-1979,Senthil-2000,Wu-2023}. The correlation diagnostics developed here---most notably the geometry and contrast of pinch-point singularities---provide a concrete way to address this question.

A second set of questions concerns the relationship between energetic proximity and gauge-sector
purity. Our variational results demonstrate that {\it Ans\"atze} which are energetically close can exhibit
markedly different degrees of gauge dominance in their equal-time correlations. This separation
suggests that phase transitions between chiral spin liquids may be driven primarily by changes in
the gauge sector rather than by conventional energetic considerations alone, consistent with
earlier insights from gauge-theoretic studies of quantum spin liquids~\cite{Senthil-2004,Savary-2017,Feuerpfeil-2026a}.
Understanding whether such transitions admit universal descriptions~\cite{Maity-2026,Feuerpfeil-2026b}, and how characteristic
features such as pinch-point singularities evolve or collapse across them, constitutes an
important open problem.

The coexistence of chiral $\mathrm{U(1)}$ and chiral $\mathbb{Z}_2$ spin liquids within a single classification
framework further motivates a systematic study of Higgs transitions between these phases. In
particular, it would be valuable to elucidate how time-reversal symmetry breaking is encoded in
Higgsed phases, how chirality survives the loss of a gapless gauge field, and how these changes are
reflected in both static and dynamical correlation functions~\cite{Fradkin-1979,Senthil-2000}. The
present PSG classification provides a natural starting point for formulating continuum field
theories that capture these transitions.

More broadly, a fundamental open problem concerns the infrared structure of genuinely
three-dimensional chiral spin liquids. In contrast to two dimensions, where chiral phases admit a
well-defined Chern--Simons description~\cite{Wen-1991}, there is currently no established low-energy
field theory that universally characterizes time-reversal–breaking spin liquids in 3+1D. This issue
is particularly acute for gapless chiral $\mathrm{U(1)}$ phases, where fermionic matter fields coexist with a
dynamical three-dimensional gauge field~\cite{Hermele-2004,Savary-2017}. Determining whether such
phases flow to stable infrared fixed points, and if so how they should be characterized in terms of
continuum gauge theories, remains an open theoretical challenge. Closely related questions concern
the existence and nature of topological response terms~\cite{Qi-2008}, the role of boundary
anomalies~\cite{Vishwanath-2013}, and the structure of surface states in three-dimensional chiral
spin liquids.

An additional feature emerging from our PSG analysis is the presence of symmetry-protected nodal
stars in several chiral $\mathrm{U(1)}$ spin-liquid {\it Ans\"atze}. These consist of multiplets of symmetry-related
band touchings enforced by the projective realization of the pyrochlore space group on the fermionic
spinons. Unlike isolated Dirac points familiar from two-dimensional spin liquids, nodal stars form
extended low-energy structures in momentum space whose multiplicity and geometry are fixed by
symmetry rather than fine tuning, paralleling recent developments in symmetry-enforced band
topology~\cite{Bradlyn-2016}.

The existence of nodal stars in three dimensions raises a number of important open questions. In
particular, their stability in the presence of a gapless $\mathrm{U(1)}$ gauge field is far from obvious, as
gauge fluctuations may couple more strongly to an extended nodal manifold than to isolated point
nodes~\cite{Hermele-2004}. Understanding the resulting low-energy field theories—and whether they
define universality classes distinct from Dirac or spinon–Fermi–surface spin liquids—represents a
promising direction for future work. It would also be interesting to determine how
symmetry-breaking perturbations, either explicit or spontaneous, can gap, split, or transmute
nodal stars into other types of gapless or gapped chiral phases. Finally, identifying experimental
signatures that distinguish nodal-star spin liquids from other gapless quantum spin liquids
remains an open challenge.

This work shows how the interplay of emergent gauge structure, time-reversal--breaking chirality, and three-dimensional lattice geometry gives rise to quantum spin liquids whose properties, diagnostics, and low-energy physics extend well beyond two-dimensional paradigms.

From a materials perspective, the most promising route is not necessarily the nearest-neighbor Heisenberg limit, but pyrochlores in which further-neighbor exchange, ring exchange, Dzyaloshinskii--Moriya interactions, or multipolar
anisotropic exchange place the system near a noncoplanar or chiral instability.
A practical search for a chiral $\mathrm{U}(1)$ spin liquid should combine:
(i) thermodynamic, neutron, and local-probe evidence for the absence of
conventional magnetic order; (ii) diffuse and inelastic neutron signatures of
a $\mathrm{U}(1)$ gauge sector; (iii) local or optical probes of time-reversal
breaking in zero field; and (iv) polarization-, field-training-, or
strain-training protocols capable of distinguishing opposite chiral domains.
Since scalar chirality is a three-spin pseudoscalar, no ordinary two-spin
measurement alone is expected to prove it unambiguously. The theoretical structure factors presented here should therefore be used as part of a broader diagnostic package rather than as a standalone criterion.

\section*{Acknowledgments}
We thank Francesco Ferrari, Federico Becca, Hee Seung Kim, SungBin Lee, Joel Moore, and Johannes Reuther for helpful discussions. L.B. is supported by the NSF CMMT program under Grant No. DMR-2419871, and the Simons Collaboration on Ultra-Quantum Matter, which is a grant from the Simons Foundation (Grant No. 651440). The work of Y.I. was performed in part at the Aspen Center for Physics, which is supported by a grant from the Simons Foundation (1161654, Troyer). This research was also supported in part by grant NSF PHY-2309135 to the Kavli Institute for Theoretical Physics. Y.I. acknowledges support from the Abdus Salam International Centre for Theoretical Physics through the Associates Programme, from the Simons Foundation through Grant No.~284558FY19, from IIT Madras through the Institute of Eminence program for establishing QuCenDiEM (Project No. SP22231244CPETWOQCDHOC), and the International Centre for Theoretical Sciences for participation in the Discussion Meeting --- Fractionalized Quantum Matter (code: ICTS/DMFQM2025/07).

\section*{Data Availability Statement} The data generated during the current study are available from the corresponding author upon reasonable request.

\appendix

\section{Staggered flux state is the $(\pi,\frac{\pi}{2},0)$ state: gauge transformation }\label{app:equiv_staggered_flux}

The \textcolor{BrickRed}{$(\pi,\frac{\pi}{2},0)$} chiral state is the same as the staggered flux (SF) state defined in Ref.~\cite{Kim-2008} up to unitary transformations. Below we provide the exact mapping of their momentum space Hamiltonian:  
\begin{equation}
\mathcal{H}_{\text{SF state}}(\mathbf{k})=U^\dag_3(\mathbf{k}) \mathcal{H}_{(\pi,\frac{\pi}{2},0)}(\mathbf{k}')
U_3(\mathbf{k})\,,
\end{equation}
here $\mathcal{H}_{\text{SF state}}(\mathbf{k})$ is the Hamiltonian of the SF state defined according to its bond configurations shown in Fig.~1 of Ref.~\cite{Kim-2008}, and $\mathcal{H}_{(\pi,\frac{\pi}{2},0)}$ is the Hamltonian of the \textcolor{BrickRed}{$(\pi,\frac{\pi}{2},0)$} state defined according to the bond configurations shown in Fig.~\ref{bond_subfig:h}.
We defined $\mathbf{k} = (k_x,k_y,k_z)$ and $\mathbf{k}' = (k'_x,k'_y,k'_z)$, with
\begin{equation}
\begin{aligned}
k'_x &= -\frac{2}{3}(3k_x-\sqrt{3}k_y + \sqrt{6} k_z) +\frac{\pi}{2}\,,\\
k'_y &= -\frac{2}{3}(3k_x+\sqrt{3}k_y- \sqrt{6}k_z) - \frac{\pi}{2}\,,\\ 
k'_z &= - \frac{2}{3}(2\sqrt{3}k_y+\sqrt{6}k_z)+\frac{\pi}{2}\,.
\end{aligned}
\end{equation}
The unitary matrix $U_3(\mathbf{k})$ has the form $U_3(\mathbf{k})= U_p(\mathbf{k}) U_b U_g$, where $U_p(\mathbf{k})$ is the ``form factor'' matrix $U_p(\mathbf{k}) = 1_{4\times 4}\otimes \mathrm{Diag}(1, e^{\frac{i}{4}(k_y+k_z)}, e^{\frac{i}{4}(k_z+k_x)},e^{\frac{i}{4}(k_x+k_y)})$, $U_b$ is the permutation matrix $U_b = 1_{4\times 4} \otimes \left(\begin{smallmatrix} 1 &&&\\ &&&1\\&1 &&\\&&1&\end{smallmatrix}\right)$, and $U_g$ is a gauge transformation matrix 
\begin{equation}
\begin{aligned}
U_g = \mathrm{diag}(&1,-1,-1, e^{\frac{3\pi i}{4}},1,-1,-1,e^{\frac{3\pi i}{4}}\,,\\
&i,i,i,e^{\frac{\pi i}{4}},-i,-i,-i,e^{\frac{5 \pi i}{4}})\,.
\end{aligned}
\end{equation}

\section{Chiral {\it Ans\"atze}  from PSG: 2nd and 3rd neighbor amplitudes}\label{app:further_bonds}

Above sets the convention for the NN bond $\mathbf{0}_0\leftarrow \mathbf{0}_1$. All other NN bonds can be obtained by performing certain PSG operation from this bond. The convention carries over to other bonds in a natural way. More generally, for other types of bonds, we define
\begin{itemize}
\item Onsite bond $(0,0,0)_0\leftarrow(0,0,0)_0$:$$(\alpha_h,0,0,0,0,\beta_p,\gamma_p,\delta_p),$$
\item NN bond $(0,0,0)_0\leftarrow(0,0,0)_1$:$$(a_h,b_h,c_h,d_h,a_p,b_p,c_p,d_p),$$
\item 2nd NN bond $(0,0,0)_1\leftarrow(0,-1,0)_2$:
$$(A_h,B_h,C_h,D_h,A_p,B_p,C_p,D_p),$$
\item 3rd NN bond Type I
$(0,0,0)_0\leftarrow(1,0,0)_0$:$$(A_{3h},B_{3h},C_{3h},D_{3h},A_{3p},B_{3p},C_{3p},D_{3p}),$$
\item 3rd NN bond Type II $(0,0,0)_0\leftarrow(1,-1,0)_0$:$$(A'_{3h},B'_{3h},C'_{3h},D'_{3h},A'_{3p},B'_{3p},C'_{3p},D'_{3p})$$
\end{itemize}

Note that for onsite bond, we have $\beta_h=\gamma_h=\delta_h=\alpha_p=0$ due to fermion anticommutativity and hermiticity of the Hamiltonian.

The independent 2nd NN and 3rd NN bonds for the $\mathrm{U(1)}$ states are given in Table \ref{NNNparams}. The independent 2nd NN bonds for the $\mathbb{Z}_2$ states are given in Table \ref{MFTparameters_z2_nnn}.

\begin{table*}
\centering
\caption{Independent 2nd NN and 3rd NN mean-field parameters and constraints for the $(I,S)$, $(I\mathcal{T},S)$, $(I,S\mathcal{T})$, and $(I\mathcal{T},S\mathcal{T})$ $\mathrm{U(1)}$ spin liquids. The spin-singlet parameters are underlined. The parameters labeled with a ``$^\dag$" is for the {staggered flux state}, i.e., the \textcolor{BrickRed}{$(\pi,\pi/2,0)$ state}. The parameters in color are the further neighbor singlet bonds that can be added to the nearest-neighbor states in Table \ref{Tab1} labeled by the \emph{same} color.} \label{NNNparams}
\begin{tabular}{c|c|c|c|c}
\hline\hline
Chiral class $(I,S)$ &{Nonzero 2nd NN}&\multirow{2}{*}{Constraint}&\multirow{2}{*}{Nonzero 3rd NN parameters}&\multirow{2}{*}{Constraints}\\
 $\phi_1$-$(m_{\overline{C}_6}m_{S})$-${n}_{ST_1}$-$n_{\overline{C}_6S}$ &parameters&&&\\
\hline
0-$(00)$-$0$-$0$ or 0-$(00)$-$0$-$1$&$\underline{\mathrm{Im}A},B,\mathrm{Im}D$& $B=-C^*$& $\underline{\mathrm{Im}A_3},\mathrm{Re}C_3,\underline{\mathrm{Im}A'_3},\mathrm{Re}B'_3$&$C_3=-D_3,B'_3=-C'_3$\\
$\pi$-$(00)$-$0$-$0$ or $\pi$-$(00)$-$0$-$1$&$\underline{\mathrm{Im}A},B,\mathrm{Im}D$& $B=-C^*$&$\underline{\mathrm{Im}A_3},\mathrm{Re}C_3,\mathrm{Im}B'_3,\mathrm{Im}D'_3$&$C_3=-D_3,B'_3=C'_3$\\
\hline
0-$(01)$-$0$-$0$ or 0-$(01)$-$0$-$1$&$B$ & $B=-C$&$\mathrm{Re}C_3,\mathrm{Re}B'_3$&$C_3=-D_3,B'_3=-C'_3$ \\
$\pi$-$(01)$-$0$-$0$ or $\pi$-$(01)$-$0$-$1$&$B$ & $B=-C$& $\mathrm{Re}C_3,\mathrm{Im}B'_3$&$C_3=-D_3,B'_3=-C'_3$\\
\hline
0-$(10)$-$0$-$0$ or 0-$(10)$-$0$-$1$&$\underline{\mathrm{Im}A},B,\mathrm{Im}D$ & $B=-C^*$&$\mathrm{Im}B_3,C_3,B'_3$&$C_3=-D_3^*,B'_3=-C'_3$ \\
$\pi$-$(10)$-$0$-$0$ or $\pi$-$(10)$-$0$-$1$&$\underline{\mathrm{Im}A},B,\mathrm{Im}D$ & $B=-C^*$& $\mathrm{Im}B_3,C_3$ &$C_3=-D_3^*$ \\
\hline
0-$(11)$-$0$-$0$ or 0-$(11)$-$0$-$1$&$B$ & $B=-C$&$C_3,B'_3,\mathrm{Im}D'_3$ & $C_3=-D_3,B'_3=-C'^*_3$ \\
$\pi$-$(11)$-$0$-$0$ or $\pi$-$(11)$-$0$-$1$&$B$ & $B=-C$ & $C_3,\underline{\mathrm{Im}A'_3}$ & $C_3=-D_3$ \\
\hline
$\frac{\pi}{2}$-$(01)$-${1}$-$0$ or $\frac{\pi}{2}$-$(01)$-${1}$-$1$&$B$ & $B=-C$& $\mathrm{Im}C_3,\mathrm{Re}B'_3$ & $C_3\!=-\!D_3,\mathrm{Re}B'_3\!=\!-\mathrm{Im}B'_3,B'_3\!=\!-C'_3$ \\
\hline\hline
Chiral class $(I\mathcal{T},S)$ &{Nonzero 2nd NN}&\multirow{2}{*}{Constraint}&\multirow{2}{*}{Nonzero 3rd NN parameters}&\multirow{2}{*}{Constraints}\\
$\phi_1$-$(m_{\overline{C}_6}m_{S})$-${n}_{ST_1}$-$n_{\overline{C}_6S}$ &parameters&&&\\
\hline
0-$(00)$-$0$-$0$ or 0-$(00)$-$0$-$1$&$\underline{\mathrm{Im}A},B,\mathrm{Im}D$& $B=-C^*$&$\mathrm{Im}B_3,\mathrm{Im}C_3,\underline{\textcolor{RoyalBlue}{\mathrm{Re}A'_3}}^*,\mathrm{Im}B'_3$ & $C_3=-D^*_3,B'_3=-C'_3$\\
$\pi$-$(00)$-$0$-$0$ or $\pi$-$(00)$-$0$-$1$&$\underline{\mathrm{Im}A},B,\mathrm{Im}D$& $B=-C^*$& $\mathrm{Im}B_3,\mathrm{Im}C_3,\mathrm{Re}B'_3,\mathrm{Re}D'_3$ & $C_3=-D^*_3,B'_3=C'_3$ \\
\hline
0-$(01)$-$0$-$0$ or 0-$(01)$-$0$-$1$&$B$ & $B=-C$ & $\mathrm{Im}C_3,\underline{\textcolor{RoyalBlue}{\mathrm{Re}A'_3}^*},\mathrm{Im}B'_3,\mathrm{Im}D'_3$&$C_3=-D_3,B'_3=C'_3$ \\
$\pi$-$(01)$-$0$-$0$ or $\pi$-$(01)$-$0$-$1$&$B$ & $B=-C$ &$\mathrm{Im}C_3,\underline{\textcolor{Dandelion}{\mathrm{Im}A'_3}},\mathrm{Re}B'_3,\mathrm{Re}D'_3$ & $C_3=-D_3,B'_3=C'_3$ \\
\hline
0-$(10)$-$0$-$0$ or 0-$(10)$-$0$-$1$&$\underline{\mathrm{Im}A},B,\mathrm{Im}D$ & $B=-C^*$&$\underline{\mathrm{Im}A_3,\textcolor{RoyalBlue}{A'_3}^*}$&None\\
$\pi$-$(10)$-$0$-$0$ or $\pi$-$(10)$-$0$-$1$&$\underline{\mathrm{Im}A},B,\mathrm{Im}D$ & $B=-C^*$& $\underline{\mathrm{Im}A_3},B'_3,D'_3$&$B'_3=C'_3$ \\
\hline
0-$(11)$-$0$-$0$ or 0-$(11)$-$0$-$1$&$B$ & $B=-C$& $\underline{\textcolor{RoyalBlue}{\mathrm{Re}A'_3}^*}$ & None \\
$\pi$-$(11)$-$0$-$0$ or $\pi$-$(11)$-$0$-$1$&$B$ & $B=-C$ & $B'_3,\mathrm{Re}D'_3$ & $B'_3 = C'^*_3$ \\
\hline
\multirow{2}{*}{$\frac{\pi}{2}$-$(01)$-${1}$-$0$ or $\frac{\pi}{2}$-$(01)$-${1}$-$1$}&\multirow{2}{*}{$B$} & \multirow{2}{*}{$B=-C$}& \multirow{2}{*}{$\mathrm{Re}C_3,\underline{\textcolor{Plum}{\mathrm{Re}A'_3}},\mathrm{Re}B'_3,\mathrm{Re}D'_3$} & $C_3=-D_3,\underline{\textcolor{Plum}{\mathrm{Re}A'_3 =-\mathrm{Im}A'_3}},\mathrm{Re}B'_3 $ \\
&&&&$= \mathrm{Im}B'_3,B'_3=C'_3,\mathrm{Re}D'_3=\mathrm{Im}D'_3$\\
\hline\hline
Chiral class $(I,S\mathcal{T})$ &{Nonzero 2nd NN}&\multirow{2}{*}{Constraint}&\multirow{2}{*}{Nonzero 3rd NN parameters}&\multirow{2}{*}{Constraints}\\
 $\phi_1$-$(m_{\overline{C}_6}m_{S})$-${n}_{ST_1}$-$n_{\overline{C}_6S}$ &parameters&&&\\
\hline
0-$(00)$-$0$-$0$ or 0-$(00)$-$0$-$1$&$\underline{\mathrm{Re}A}$, $B$, $\mathrm{Re}D$& $B=C^*$ & $\mathrm{Re}B_3,\mathrm{Re}C_3,\mathrm{Re}B'_3,\mathrm{Re}D'_3$&$C_3=D_3,B'_3=C'_3$ \\
$\pi$-$(00)$-$0$-$0$ or $\pi$-$(00)$-$0$-$1$&$\underline{\mathrm{Re}A}$, $B$, $\mathrm{Re}D$& $B=C^*$& $\mathrm{Re}B_3,\mathrm{Re}C_3,\underline{\textcolor{BrickRed}{\mathrm{Re}A'_3}^\dag},\mathrm{Im}B'_3$ & $C_3=D_3,B'_3=-C'_3$\\
\hline
\multirow{2}{*}{0-$(01)$-$0$-$0$ or 0-$(01)$-$0$-$1$}&\multirow{2}{*}{$\underline{A},B,D$} & \multirow{2}{*}{$B=C$} & $\underline{\mathrm{Im}A_3},\mathrm{Re}B_3,\mathrm{Re}C_3$, & \multirow{2}{*}{$C_3=D_3,B'_3=C'_3$}\\
&&&$\underline{\textcolor{ForestGreen}{\mathrm{Im}A'_3}},\mathrm{Re}B'_3,\mathrm{Re}D'_3$&\\
\multirow{2}{*}{$\pi$-$(01)$-$0$-$0$ or $\pi$-$(01)$-$0$-$1$}&\multirow{2}{*}{$\underline{A},B,D$} & \multirow{2}{*}{$B=C$} & $\underline{\mathrm{Im}A_3},\mathrm{Re}B_3,\mathrm{Re}C_3$, & \multirow{2}{*}{$C_3=D_3,B'_3=C'_3$} \\
&&&$\underline{\textcolor{BrickRed}{\mathrm{Re}A'_3}^\dag},\mathrm{Im}B'_3,\mathrm{Im}D'_3$&\\
\hline
0-$(10)$-$0$-$0$ or 0-$(10)$-$0$-$1$&$\underline{\mathrm{Re}A}$, $B$, $\mathrm{Re}D$ & $B=C^*$ & $\mathrm{Re}B_3,C_3,B'_3,D'_3$ & $C_3=D^*_3,B'_3=C'_3$\\
$\pi$-$(10)$-$0$-$0$ or $\pi$-$(10)$-$0$-$1$&$\underline{\mathrm{Re}A}$, $B$, $\mathrm{Re}D$ & $B=C^*$ & $\mathrm{Re}B_3,C_3,\underline{\textcolor{BrickRed}{A'_3}^\dag}$ & $C_3=D^*_3$\\
\hline
0-$(11)$-$0$-$0$ or 0-$(11)$-$0$-$1$&$\underline{A},B,D$ & $B=C$ & $B_3,C_3,B'_3,\mathrm{Re}D'_3$ & $C_3=D_3,B'_3=C'^*_3$\\
$\pi$-$(11)$-$0$-$0$ or $\pi$-$(11)$-$0$-$1$&$\underline{A},B,D$ & $B=C$ & $B_3,C_3,\underline{\textcolor{BrickRed}{\mathrm{Re}A'_3}^\dag}$ & $C_3=D_3$\\
\hline
\multirow{2}{*}{$\frac{\pi}{2}$-$(01)$-${1}$-$0$ or $\frac{\pi}{2}$-$(01)$-${1}$-$1$}&\multirow{2}{*}{$\underline{A},B,D$} & \multirow{2}{*}{$B=C$} & {$\underline{\mathrm{Re}A_3},\mathrm{Im}B_3,\mathrm{Im}C_3,$} &
$C_3=D_3,\underline{\textcolor{TealBlue}{\mathrm{Re}A'_3=\mathrm{Im}A'_3}},\mathrm{Re}B'_3$\\
&&&$\underline{\textcolor{TealBlue}{\mathrm{Re}A'_3}},\mathrm{Re}B'_3,\mathrm{Re}D'_3$&$=-\mathrm{Im}B'_3,B'_3=C'_3,\mathrm{Re}D'_3 = - \mathrm{Im}D'_3$\\
\hline\hline
Chiral class $(I\mathcal{T},S\mathcal{T})$ &{Nonzero 2nd NN}&\multirow{2}{*}{Constraint}&\multirow{2}{*}{Nonzero 3rd NN parameters}&\multirow{2}{*}{Constraints}\\
 $\phi_1$-$(m_{\overline{C}_6}m_{S})$-${n}_{ST_1}$-$n_{\overline{C}_6S}$ &parameters&&&\\
\hline
0-$(00)$-$0$-$0$ or 0-$(00)$-$0$-$1$&$\underline{\mathrm{Re}A}, B, \mathrm{Re}D$& $B=C^*$&$\underline{\mathrm{Re}A_3},\mathrm{Im}C_3,\mathrm{Im}B'_3,\mathrm{Im}D'_3$ & $C_3=-D_3,B'_3=C'_3$\\
$\pi$-$(00)$-$0$-$0$ or $\pi$-$(00)$-$0$-$1$&$\underline{\mathrm{Re}A}, B, \mathrm{Re}D$& $B=C^*$&$\underline{\mathrm{Re}A_3},\mathrm{Im}C_3,\underline{\mathrm{Im}A'_3},\mathrm{Re}B'_3$ & $C_3=-D_3,B'_3=-C'_3$ \\
\hline
0-$(01)$-$0$-$0$ or 0-$(01)$-$0$-$1$&$\underline{A},B,D$ & $B=C$ & $\underline{\mathrm{Re}A_3},\mathrm{Im}B_3,\mathrm{Im}C_3,\mathrm{Im}B'_3$ & $C_3=D_3,B'_3=-C'_3$ \\
$\pi$-$(01)$-$0$-$0$ or $\pi$-$(01)$-$0$-$1$&$\underline{A},B,D$ & $B=C$ & $\underline{\mathrm{Re}A_3},\mathrm{Im}B_3,\mathrm{Im}C_3,\mathrm{Re}B'_3$ & $C_3=D_3,B'_3=-C'_3$\\
\hline
0-$(10)$-$0$-$0$ or 0-$(10)$-$0$-$1$&$\underline{\mathrm{Re}A}, B, \mathrm{Re}D$ & $B=C^*$ & $\underline{\mathrm{Re}A_3}$ & None\\
$\pi$-$(10)$-$0$-$0$ or $\pi$-$(10)$-$0$-$1$&$\underline{\mathrm{Re}A}, B, \mathrm{Re}D$ & $B=C^*$ & $\underline{\mathrm{Re}A_3},B'_3$ & $B'_3=-C'_3$\\
\hline
0-$(11)$-$0$-$0$ or 0-$(11)$-$0$-$1$&$\underline{A},B,D$ & $B=C$ &$\underline{A_3,\mathrm{Im}A'_3}$ & None\\
$\pi$-$(11)$-$0$-$0$ or $\pi$-$(11)$-$0$-$1$&$\underline{A},B,D$ & $B=C$& $\underline{A_3},B'_3,\mathrm{Im}D'_3$ & $B'_3 = -C'^*_3$ \\
\hline
\multirow{2}{*}{$\frac{\pi}{2}$-$(01)$-${1}$-$0$ or $\frac{\pi}{2}$-$(01)$-${1}$-$1$}&\multirow{2}{*}{$\underline{A},B,D$} & \multirow{2}{*}{$B=C$} & \multirow{2}{*}{$\underline{\mathrm{Im}A_3},\mathrm{Re}B_3,\mathrm{Re}C_3,\mathrm{Re}B'_3$} & \multirow{2}{*}{$C_3=D_3,\mathrm{Re}B'_3 = \mathrm{Im}B'_3,B'_3=-C'_3$}\\
&&&&\\
\hline\hline
\end{tabular}
\end{table*}

\begin{table*}
\caption{Independent 2nd NN mean-field parameters and constraints for the $(I,S)$, $(I\mathcal{T},S)$, $(I,S\mathcal{T})$, and $(I\mathcal{T},S\mathcal{T})$ $\mathbb{Z}_2$ chiral spin liquids.}\label{MFTparameters_z2_nnn}
\centering
\begin{tabular}{l|l|l}
\hline\hline
$(I,S)$ and $(I\mathcal{T},S)$, Class
&\multirow{2}{*}{Independent 2nd NN parameters}&\multirow{ 2}{*}{Constraints}\\
($\chi_1\chi_{ST_1}$)\---$(\chi_{\overline{C}_6S}\chi_{S\overline{C}_6}\chi_{\overline{C}_6})_j$
&&\\
\hline
{$(00)$\---  or $(\pi\pi)$\---$(000)$} & $\mathrm{Im}A_h$, $A_p$, $B_h$, $B_p$, $\mathrm{Im}D_h$, $D_p$  & $\mathrm{Im}C_h = \mathrm{Im}B_h$, $C_p=B_p$\\
{$(00)$\---  or $(\pi\pi)$\---$(\pi00)$} & $\mathrm{Im}A_h$, $A_p$, $B_h$, $B_p$, $\mathrm{Im}D_h$, $D_p$ & $\mathrm{Im}C_h = \mathrm{Im}B_h$, $C_p=B_p$ \\
{$(00)$\---  or $(\pi\pi)$\---$(0\pi\pi)$}& {$A_h$, $\mathrm{Im}A_p$, $B_h$, $B_p$, $D_h$, $\mathrm{Im}D_p$} &$C_h = B_h$, $C_p = - B^*_p$\\
{$(00)$\---  or $(\pi\pi)$\---$(\pi\pi\pi)$}& $A_h$, $\mathrm{Im}A_p$, $B_h$, $B_p$, $D_h$, $\mathrm{Im}D_p$& $C_h = B_h$, $C_p = - B^*_p$\\
\hline
\multirow{ 2}{*}{$(0\pi)$\---  or $(\pi0)$\---$(0\pi0)_1$} & \multirow{ 2}{*}{$\mathrm{Re}A_p$, $B_h$, $B_p$, $\mathrm{Re}D_p$} & $\mathrm{Im} A_p = -\sqrt{3}\mathrm{Re}A_p$, $C_h = -B_h$, \\
&& $C_p = e^{-\frac{2\pi}{3} i} B_p^*$, $\mathrm{Im}D_p = -\sqrt{3}\mathrm{Re} D_p$\\
{$(0\pi)$\---  or $(\pi0)$\---$(0\pi0)_3$} & {$\mathrm{Re}A_p$, $B_h$, $B_p$, $\mathrm{Re}D_p$} & $C_h =-B_h$, $C_p = B^*_p$\\
{$(0\pi)$\---  or $(\pi0)$\---$(\pi\pi0)_0$} & {$\mathrm{Re}A_p$, $B_h$, $B_p$, $\mathrm{Re}D_p$} & $C_h =-B_h$, $C_p = B^*_p$\\
\multirow{ 2}{*}{$(0\pi)$\---  or $(\pi0)$\---$(\pi\pi0)_2$}& \multirow{ 2}{*}{$\mathrm{Re}A_p$, $B_h$, $B_p$, $\mathrm{Re}D_p$} & 
$\mathrm{Im} A_p = \sqrt{3}\mathrm{Re}A_p$, $C_h = -B_h$, \\
&& $C_p= e^{\frac{2\pi}{3}i} B_p^*$, $\mathrm{Im}D_p = \sqrt{3}\mathrm{Re} D_p$\\
\hline
{$(0\pi)$\---  or $(\pi0)$\---$(00\pi)$} & {$\mathrm{Re}A_h$, $B_h$, $B_p$, $\mathrm{Re}D_h$} &  $C_h = B^*_h$, $C_p = - B_p$\\
{$(0\pi)$\---  or $(\pi0)$\---$(\pi0\pi)$} & {$\mathrm{Re}A_h$, $B_h$, $B_p$, $\mathrm{Re}D_h$} & $C_h = B^*_h$, $B_p = - C_p$\\
\hline
\multirow{ 2}{*}{$(0\pi)$\---  or $(\pi0)$\---$(0\pi\pi)_0$} & \multirow{ 2}{*}{$\mathrm{Re}A_p$, $B_h$, $B_p$, $\mathrm{Re}D_p$} & $\mathrm{Im}A_p = -\frac{1}{\sqrt{3}}\mathrm{Re}A_p$, $C_h=-B_h$,\\
&& $C_p = e^{-\frac{\pi}{3}i} B_p^*$, $\mathrm{Im}D_p = -\frac{1}{\sqrt{3}} \mathrm{Re} D_p$\\
\multirow{ 2}{*}{$(0\pi)$\---  or $(\pi0)$\---$(\pi\pi\pi)_0$} &\multirow{ 2}{*}{$\mathrm{Re}A_p$, $B_h$, $B_p$, $\mathrm{Re}D_p$} & $\mathrm{Im}A_p = -\frac{1}{\sqrt{3}}\mathrm{Re}A_p$, $C_h=-B_h$, \\
&&$C_p = e^{-\frac{\pi}{3}i}B_p^*$, $\mathrm{Im}D_p = -\frac{1}{\sqrt{3}} \mathrm{Re} D_p$\\
{$(0\pi)$\---  or $(\pi0)$\---$(0\pi\pi)_1$} & {$\mathrm{Im}A_p$, $B_h$, $B_p$, $\mathrm{Im}D_p$} & $C_h = -B_h$, $C_p = - B^*_p$\\
{$(0\pi)$\---  or $(\pi0)$\---$(\pi\pi\pi)_1$} & {$\mathrm{Im}A_p$, $B_h$, $B_p$, $\mathrm{Im}D_p$} & $C_h=-B_h$, $C_p = -B^*_p$\\
\hline
\hline
$(I,S\mathcal{T})$ and $(I\mathcal{T},S\mathcal{T})$, Class
&\multirow{2}{*}{Independent 2nd NN parameters}&\multirow{ 2}{*}{Constraints}\\
($\chi_1\chi_{ST_1}$)\---$(\chi_{\overline{C}_6S}\chi_{S\overline{C}_6}\chi_{\overline{C}_6})_j$
&&\\
\hline
{$(00)$\---  or $(\pi\pi)$\---$(000)$} & $\mathrm{Re}A_h,B_h,B_p,\mathrm{Re}D_h$  & $\mathrm{Im}C_h = -\mathrm{Im}B_h$, $C_p=-B_p$\\
{$(00)$\---  or $(\pi\pi)$\---$(\pi00)$} & $\mathrm{Re}A_h,B_h,B_p,\mathrm{Re}D_h$ & $\mathrm{Im}C_h = -\mathrm{Im}B_h$, $C_p=-B_p$ \\
{$(00)$\---  or $(\pi\pi)$\---$(0\pi\pi)$}& $\mathrm{Re}A_h,B_h,B_p,\mathrm{Re}D_h$ &$C_h = -B_h$, $C_p = B^*_p$\\
{$(00)$\---  or $(\pi\pi)$\---$(\pi\pi\pi)$}& $\mathrm{Re}A_h,B_h,B_p,\mathrm{Re}D_h$& $C_h = -B_h$, $C_p = B^*_p$\\
\hline
\multirow{ 2}{*}{$(0\pi)$\---  or $(\pi0)$\---$(0\pi0)_1$} & \multirow{ 2}{*}{$A_h,\mathrm{Re}A_p,B_h,B_p, D_h, \mathrm{Re}D_p$} & $\mathrm{Im} A_p = \frac{1}{\sqrt{3}}\mathrm{Re}A_p$, $C_h = B_h$, \\
&& $C_p = -e^{-\frac{2\pi}{3} i} B_p^*$, $\mathrm{Im}D_p = \frac{1}{\sqrt{3}}\mathrm{Re} D_p$\\
{$(0\pi)$\---  or $(\pi0)$\---$(0\pi0)_3$} & {$A_h,\mathrm{Im}A_p,B_h,B_p, D_h, \mathrm{Im}D_p$} & $C_h =B_h$, $C_p = -B^*_p$\\
{$(0\pi)$\---  or $(\pi0)$\---$(\pi\pi0)_0$} & {$A_h,\mathrm{Im}A_p,B_h,B_p, D_h, \mathrm{Im}D_p$} & $C_h =B_h$, $C_p = -B^*_p$\\
\multirow{ 2}{*}{$(0\pi)$\---  or $(\pi0)$\---$(\pi\pi0)_2$}& \multirow{ 2}{*}{$A_h,\mathrm{Re}A_p,B_h,B_p, D_h, \mathrm{Re}D_p$} & 
$\mathrm{Im} A_p = -\frac{1}{\sqrt{3}}\mathrm{Re}A_p$, $C_h = B_h$, \\
&& $C_p= -e^{\frac{2\pi}{3}i} B_p^*$, $\mathrm{Im}D_p = -\frac{1}{\sqrt{3}}\mathrm{Re} D_p$\\
\hline
{$(0\pi)$\---  or $(\pi0)$\---$(00\pi)$} & {$\mathrm{Im}A_h,A_p,B_h,B_p,\mathrm{Im}D_h, D_p$} &  $C_h = -B^*_h$, $C_p =  B_p$\\
{$(0\pi)$\---  or $(\pi0)$\---$(\pi0\pi)$} & {$\mathrm{Im}A_h,A_p,B_h,B_p,\mathrm{Im}D_h, D_p$} & $C_h = -B^*_h$, $B_p =  C_p$\\
\hline
\multirow{ 2}{*}{$(0\pi)$\---  or $(\pi0)$\---$(0\pi\pi)_0$} & \multirow{ 2}{*}{$A_h,\mathrm{Re}A_p,B_h,B_p,D_h,\mathrm{Re}D_p$} & $\mathrm{Im}A_p = {\sqrt{3}}\mathrm{Re}A_p$, $C_h=B_h$,\\
&& $C_p = -e^{-\frac{\pi}{3}i} B_p^*$, $\mathrm{Im}D_p = {\sqrt{3}} \mathrm{Re} D_p$\\

\multirow{ 2}{*}{$(0\pi)$\---  or $(\pi0)$\---$(\pi\pi\pi)_0$} &\multirow{ 2}{*}{$A_h,\mathrm{Re}A_p,B_h,B_p,D_h,\mathrm{Re}D_p$} & $\mathrm{Im}A_p = {\sqrt{3}}\mathrm{Re}A_p$, $C_h=B_h$, \\
&&$C_p = -e^{-\frac{\pi}{3}i}B_p^*$, $\mathrm{Im}D_p = {\sqrt{3}} \mathrm{Re} D_p$\\
{$(0\pi)$\---  or $(\pi0)$\---$(0\pi\pi)_1$} & {$A_h,\mathrm{Re}A_p,B_h,B_p,D_h,\mathrm{Re}D_p$} & $C_h=B_h$, $C_p=B^*_p$\\
{$(0\pi)$\---  or $(\pi0)$\---$(\pi\pi\pi)_1$} & {$A_h,\mathrm{Re}A_p,B_h,B_p,D_h,\mathrm{Re}D_p$} & $C_h=B_h$, $C_p=B^*_p$\\
\hline\hline
\end{tabular}
\end{table*}

\twocolumngrid

\section{Unprojected equal-time spin structure factor for $\mathrm{U(1)}$ spin-singlet {\it Ans\"atze}}\label{app:unprojected_SSF}

\begin{figure*}
  \centering
  \includegraphics[width=1.0\linewidth]{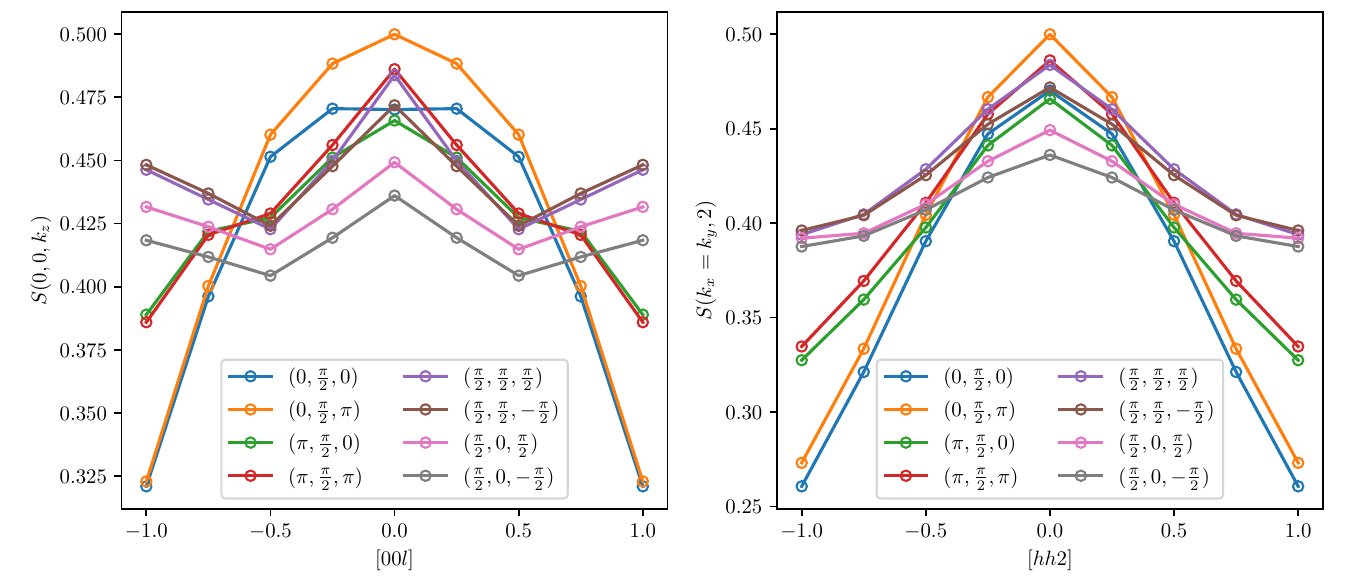}
  \caption{One-dimensional cuts of the equal-time spin structure factors for the eight \emph{unprojected} $\mathrm{U}(1)$ chiral spin-liquid {\it Ans\"atze}, taken along the two principal directions through the pinch point and along a representative high-symmetry path. The data are normalized such that $\sum_{\mathbf q} S(\mathbf q)=1$.}
  \label{fig:sq_unprojected}
\end{figure*}

Below we rewrite the equal-time spin structure factor, Eq.~\eqref{etssf}, in terms of Abrikosov partons:

\begin{widetext}
\begin{equation}
\begin{aligned}
&S(\mathbf{q})\\
&=\frac{1}{N}\sum\limits_{\mathbf{r}_\mu,\mathbf{r}'_\nu}
e^{-i\mathbf{q}\cdot (\mathbf{r}_\mu-\mathbf{r}'_\nu)} 
\left\langle
\hat{\mathbf{S}}_{\mathbf{r}_\mu}\cdot
\hat{\mathbf{S}}_{\mathbf{r}'_\nu}
\right\rangle\\
&=
\frac{1}{4N^3} \sum\limits_{\mathbf{r}_\mu,\mathbf{r}'_\nu} \sum\limits_{\mathbf{k}_1,\mathbf{k}_2,\mathbf{k}_3,\mathbf{k}_4}
e^{-i\mathbf{q}\cdot (\mathbf{r}_\mu-\mathbf{r}'_\nu)} e^{i(\mathbf{k}_2-\mathbf{k}_1)\cdot\mathbf{r}_\mu}e^{i(\mathbf{k}_4-\mathbf{k}_3)\cdot \mathbf{r}'_\nu} \sum\limits_{\alpha,\sigma_1,\sigma_2,\sigma_3,\sigma_4}\left(\sigma^\alpha\right)_{\sigma_1,\sigma_2}\left(\sigma^\alpha\right)_{\sigma_3,\sigma_4}\left\langle f^\dag_{\mathbf{k}_1,\mu\sigma_1} f_{\mathbf{k}_2,\mu\sigma_2}f^\dag_{\mathbf{k}_3,\nu\sigma_3}f_{\mathbf{k}_4,\nu\sigma_4}\right\rangle\\
&=\frac{1}{4N}\sum\limits_{\mathbf{k}_1,\mathbf{k}_3}\sum\limits_{\mu,\nu}e^{-i\mathbf{q}\cdot(\hat{\varepsilon}_\mu-\hat{\varepsilon}_\nu)}\sum\limits_{\alpha,\sigma_1,\sigma_2,\sigma_3,\sigma_4}
\left(\sigma^\alpha\right)_{\sigma_1,\sigma_2}\left(\sigma^\alpha\right)_{\sigma_3,\sigma_4}\left\langle f^\dag_{\mathbf{k}_1,\mu\sigma_1} f_{\mathbf{k}_1-\mathbf{q},\mu\sigma_2}f^\dag_{\mathbf{k}_3,\nu\sigma_3}f_{\mathbf{k}_3+\mathbf{q},\nu\sigma_4}\right\rangle\,.
\end{aligned}
\end{equation}
\end{widetext}
where $\mathbf{r}$ labels the ($0$-flux, $\pi$-flux, or $\frac{\pi}{2}$-flux) parton primitive cell, Greek symbols such as $\mu,\nu$ label the sublattices in the unit cell, and $\hat{\varepsilon}_\mu$ labels the sublattice displacements in the primitive cell. $\sigma_{1,2,3,4} = \uparrow,\downarrow$, and $\alpha=x,y,z$.
We write the fermions in the eigen-energy basis:
\begin{equation}
f_{\mathbf{k},\mu\sigma} = \sum\limits_{\rho} \sum\limits_{\tau=\uparrow,\downarrow}
\left( U(\mathbf{k})\right)_{\mu\sigma,\rho\tau}\widetilde{f}_{\mathbf{k},\rho\tau}\,.
\end{equation}

Using Wick's theorem, we have
\begin{widetext}
\begin{equation}
\begin{aligned}
&\left\langle f^\dag_{\mathbf{k}_1,\mu\sigma_1} f_{\mathbf{k}_1-\mathbf{q},\mu\sigma_2}f^\dag_{\mathbf{k}_3,\nu\sigma_3}f_{\mathbf{k}_3+\mathbf{q},\nu\sigma_4}\right\rangle\\
&=
\sum\limits_{\rho_1,\rho_2}\sum\limits_{\tau_1,\tau_2}
\delta_{\mathbf{k}_1,\mathbf{k}_3+\mathbf{q}}\left(U(\mathbf{k}_1)\right)^*_{\mu\sigma_1,\rho_1\tau_1}
\left(U(\mathbf{k}_1-\mathbf{q})\right)_{\mu\sigma_2,\rho_2\tau_2}
\left(U(\mathbf{k}_1-\mathbf{q})\right)^*_{\nu\sigma_3,\rho_2\tau_2}
\left(U(\mathbf{k}_1)\right)_{\nu\sigma_4,\rho_1\tau_1}
n_{\mathbf{k}_1,\rho_1\tau_1} (1-n_{\mathbf{k}_1-\mathbf{q},\rho_2\tau_2})\\
&\qquad\qquad+
\delta_{\mathbf{q},\mathbf{0}}
\left(U(\mathbf{k}_1)\right)^*_{\mu\sigma_1,\rho_1\tau_1}
\left(U(\mathbf{k}_1)\right)_{\mu\sigma_2,\rho_1\tau_1}\left(U(\mathbf{k}_3)\right)^*_{\nu\sigma_3,\rho_2\tau_2}\left(U(\mathbf{k}_3)\right)_{\nu\sigma_4,\rho_2\tau_2} n_{\mathbf{k}_1,\rho_1\tau_1}n_{\mathbf{k}_3,\rho_2\tau_2}\,,
\end{aligned}
\end{equation}
\end{widetext}
the 2nd term comes from the disconnected part $\left\langle \hat{S}^\alpha_{\mathbf{r}_\mu}\right\rangle\left\langle\hat{S}^\alpha_{\mathbf{r}'_\nu}\right\rangle$ and only contributes to the momentum $\mathbf{q}=\mathbf{0}$. As we will focus on $\mathbf{q}\neq \mathbf{0}$, we will drop this term.

Now, since the parton Hamiltonian only contains spin singlet hoppings, $U$ does not carry spin index $\sigma$ or $\tau$, then we can perform the sum $\sum_\alpha (\sigma^\alpha)_{\sigma_1,\sigma_2} (\sigma^\alpha)_{\sigma_3,\sigma_4} = 2\delta_{\sigma_1 \sigma_4} \delta_{\sigma_2 \sigma_3}- \delta_{\sigma_1\sigma_2}\delta_{\sigma_3\sigma_4}$, and note that even for the term $\delta_{\sigma_1\sigma_2}\delta_{\sigma_3\sigma_4}$ we will implicitly have the requirement $\delta_{\sigma_1\sigma_4}\delta_{\sigma_2 \sigma_3}$ due to the SU(2) symmetry of the $U$ matrix, so that when summing over $\sigma_{1,2,3,4}$ we get a factor of $2\cdot 2\cdot 2 - 2 = 6$. Now, we have
\begin{widetext}
\begin{equation}
\begin{aligned}
S(\mathbf{q})&=\frac{3}{2N}\sum\limits_{\mu,\nu}e^{-i\mathbf{q}\cdot(\hat{\varepsilon}_\mu-\hat{\varepsilon}_\nu)}\sum\limits_{\rho_1,\rho_2}
\sum\limits_{\mathbf{k}_1}\left(U(\mathbf{k}_1)\right)^*_{\mu,\rho_1}
\left(U(\mathbf{k}_1-\mathbf{q})\right)_{\mu,\rho_2}
\left(U(\mathbf{k}_1-\mathbf{q})\right)^*_{\nu,\rho_2}
\left(U(\mathbf{k}_1)\right)_{\nu,\rho_1} n_{\mathbf{k}_1,\rho_1} (1-n_{\mathbf{k}_1-\mathbf{q},\rho_2})\\
&=\frac{3}{2N}
\sum_{\rho_1,\rho_2}\sum_{\mathbf{k}_1}
|[U^\dag(\mathbf{k}_1)\Lambda(\mathbf{q}) U(\mathbf{k}_1-\mathbf{q})]_{\rho_1,\rho_2}|^2 n_{\mathbf{k}_1,\rho_1} (1-n_{\mathbf{k}_1-\mathbf{q},\rho_2})\,,\\
\end{aligned}
\end{equation}
\end{widetext}
where we defined the diagonal matrix $\Lambda(\mathbf{q}) = \mathrm{diag}(e^{-i\mathbf{q}\cdot \hat{\varepsilon}_\mu})$; note that $\Lambda(\mathbf{q})$ and $U$ are $4\varepsilon^2\times 4\varepsilon^2$ matrices, where the enlargement factor $\varepsilon = 1,2,4$ for $0$-flux, $\pi$-flux, and $\pi/2$-flux states. $n_{\mathbf{k},\rho}$ is the Fermi--Dirac distribution function for the eigen energy $E_{\mathbf{k},\rho}$ of the parton Hamiltonian at an appropriately chosen temperature.

In Fig.~\ref{fig:sq_unprojected}, we plot the equal-time spin structure factor $S(\mathbf{q})$ for the eight \emph{unprojected} $\mathrm{U(1)}$ chiral spin liquid {\it Ans\"atze} along directions $[00l]$ and $[hh2]$. The mean-field {\it Ans\"atze} assume a particular gauge fixing, and the structure factor $S(\mathbf{q})$ for these unprojected mean-field states only contains contribution from the matter field (spinons) and not the $\mathrm{U(1)}$ gauge field. The results are obtained on a $32\times 32\times 32$ meshing of momenta in the Brillouin zone. We have verified that changing the momentum meshing does not qualitatively alter the results. This indicates that the unprojected spinon contribution does not exhibit any obvious divergent or pinch-point-like singularity on the cuts shown here. While weak nonanalyticities associated with gapless spinons cannot be completely excluded on the basis of these numerics alone, they are subleading on the present scales and do not account for the pronounced pinch-point structure observed in the projected data.

\twocolumngrid

\bibliography{pyro_chiral}

\end{document}